\documentclass[a4paper,11pt]{article}
\usepackage{jheppub} % for details on the use of the package, please
\usepackage{graphicx,here}

\usepackage{amsmath}
\usepackage[utf8]{inputenc}

\usepackage{amssymb}
\usepackage{amsfonts}
\usepackage{amsbsy}
\usepackage{url}
\usepackage{enumerate}
\usepackage{caption}
\usepackage{subcaption}
\usepackage{color}
\usepackage{bm}
\usepackage{bbm}
\usepackage{comment}
\usepackage{multirow,bigdelim}
\usepackage[head1]{DotArrow}
\usepackage[normalem]{ulem}

\newcommand{\be}{\begin{eqnarray}}
\newcommand{\ee}{\end{eqnarray}}
\newcommand{\p}{\partial}

\makeatletter
\@addtoreset{equation}{section}
\makeatother

\newcommand{\bee}{\begin{equation}}
\newcommand{\eee}{(\end{equation})}

\newcommand\rsout{\bgroup\markoverwith{\textcolor{red}{\rule[0.5ex]{2pt}{0.4pt}}}\ULon}

\newcommand{\yh}[1]{\textcolor{magenta}{[YH]: \textit{#1}}}

\graphicspath{{./figs/}}
\newbox{\ORCIDicon}
\sbox{\ORCIDicon}{\large \includegraphics[width=0.8em]{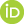}}

\title{
Composite topological solitons consisting of domain walls, strings, and monopoles in $O(N)$ models
}

\author[a,b]{Minoru Eto,\,\href{https://orcid.org/0000-0002-2554-1888}{\usebox{\ORCIDicon}}}
\emailAdd{meto@sci.kj.yamagata-u.ac.jp}
\affiliation[a]{Department of Physics, Yamagata University, Kojirakawa-machi 1-4-12, Yamagata, Yamagata 990-8560, Japan}
\affiliation[b]{Research and Education Center for Natural Sciences, Keio University, 4-1-1 Hiyoshi, Yokohama, Kanagawa 223-8521, Japan}

\author[b,c]{Yu Hamada,\,\href{https://orcid.org/0000-0002-0227-5919}{\usebox{\ORCIDicon}}}
\emailAdd{yuhamada@post.kek.jp}
\affiliation[c]{KEK Theory Center, Tsukuba 305-0801, Japan}

\author[b,d,e]{and Muneto Nitta,\,\href{https://orcid.org/0000-0002-3851-9305	}{\usebox{\ORCIDicon}}}
\emailAdd{nitta@phys-h.keio.ac.jp}
\affiliation[d]{Department of Physics, Keio University, 4-1-1 Hiyoshi, Kanagawa 223-8521, Japan}
\affiliation[e]{
International Institute for Sustainability with Knotted Chiral Meta Matter(SKCM$^2$), Hiroshima University, 1-3-2 Kagamiyama, Higashi-Hiroshima, Hiroshima 739-8511, Japan
}
\abstract{
We study various composites of global solitons consisting of domain walls,
strings, and monopoles in 
linear $O(N)$ models with $N=2$ and $3$.
Spontaneous symmetry breaking (SSB) 
of the $O(N)$ symmetry down to $O(N-1)$ 
results in the vacuum manifold 
$S^{N-1}$, 
together with a perturbed scalar potential
in the presence of a small explicit symmetry breaking (ESB) interaction.
The $O(2)$ model is equivalent to the axion model 
admitting topological global (axion) strings  %which are relevant for the axion cosmology.
attached by $N_{\rm DW}$ domain walls.
\if0 
While in the conventional explanation
SSB is $\mathbb{Z}_{N_{\rm DW}} \to 1$ 
supporting axion domain walls, 
we point out that the exact SSB for $N_{\rm DW} = 1$ and $2$ are $\mathbb{Z}_2\to \mathbb{Z}_2$ and
$(\mathbb{Z}_2)^2 \to \mathbb{Z}_2$, respectively. This is crucial especially for $N_{\rm DW}=2$, 
\fi 
We point out for the 
$N_{\rm DW} = 2$ case 
that 
the topological stability of the string with two domain walls is ensured by sequential SSBs $(\mathbb{Z}_2)^2 \to \mathbb{Z}_2 \to 1$, 
where the first SSB occurs in the vacuum leading 
to the topological domain wall as a mother soliton,
only inside which the second SSB occurs 
giving rise to a subsequent  
kink inside the mother wall.
From the bulk viewpoint, this kink is identical to a global string as a daughter soliton. 
This observation can be naturally extended to the $O(3)$ model, 
where a global monopole as a daughter soliton appears as a kink in a mother string or as a vortex on a mother domain wall, depending on ESB interactions. 
In the most generic case, 
the stability of the composite system consisting of the monopole, string, and domain wall is understood by the SSB $(\mathbb{Z}_2)^3 \to (\mathbb{Z}_2)^2 \to \mathbb{Z}_2 \to 1$,
in which the first SSB at the vacuum gives rise to the domain wall triggering the second one,
so that the daughter string appears as a domain wall inside the mother wall triggering the third SSB,
which leads to a granddaughter 
monopole as a kink inside the daughter vortex.
We demonstrate numerical simulations for the dynamical evolution of the composite solitons.
}

\preprint{YGHP-23-02, KEK-TH-2515}

\begin{document}

\maketitle

%%%%%%%%%%%%%%%%% I N T R O D U C T I O N %%%%%%%%%%%%%%%%%%

\section{Introduction}

Topological solitons such as monopoles, vortices 
and domain walls ubiquitously appear in nature and play crucial roles. 
Not only in quantum field theories 
\cite{Rajaraman:1987,Manton:2004tk, Shnir:2005vvi,Vachaspati:2006zz,Dunajski:2010zz,Weinberg:2012pjx,Shnir:2018yzp}
including supersymmetric gauge theories 
\cite{Tong:2005un,Tong:2008qd,Eto:2006pg,Shifman:2007ce,Shifman:2009zz} 
and quantum chromodynamics(QCD) 
\cite{Eto:2013hoa},
they also appear 
in cosmology 
\cite{Kibble:1976sj,Kibble:1980mv,Vilenkin:1984ib,Hindmarsh:1994re,Vachaspati:2015cma,Vilenkin:2000jqa} 
and various condensed matter systems 
\cite{Mermin:1979zz}: 
superfluids \cite{Volovik:2003fe,
Svistunov:2015},
Josephson junctions 
\cite{Ustinov2015}, 
quantum Hall effects 
\cite{Ezawa},
%Bose-Einstein condensates (BECs) of 
ultracold atomic gasses 
\cite{Kawaguchi:2012ii}, 
nonlinear media 
\cite{Pismen,Bunkov:2000},
liquid crystals
\cite{RevModPhys.84.497,Smalyukh:2020zin,Smalyukh:2022} and active matter 
\cite{Shankar2022}.

%In particular, 
%composite solitons composed of 
%different topological solitons 
%attract attentions 
%in field theory \cite{Nitta:2022ahj} 
%(see also references therein), 
%cosmology \cite{Vilenkin:2000jqa},
%domain-wall Skyrmions in QCD \cite{Eto:2023lyo},
%and condensed matter physics 
%such as 
%various composite solitons in $^3$He superfluids 
%\cite{Volovik:2019goo,Volovik:2020zqc} 
%and 
%domain-wall Skyrmions in chiral magnets \cite{PhysRevB.99.184412,PhysRevB.102.094402,Nagase:2020imn,Yang:2021,Ross:2022vsa} 
%(see also \cite{Kim:2017lsi}). 
In particular, 
composite solitons composed of 
different topological solitons 
attract attentions 
in field theory \cite{Nitta:2022ahj} 
(see also references therein), 
cosmology \cite{Vilenkin:2000jqa},
and condensed matter physics.
For instance, domain-wall Skyrmions in QCD \cite{Eto:2023lyo} and in chiral magnets~\cite{PhysRevB.99.184412,PhysRevB.102.094402,Nagase:2020imn,Yang:2021,Ross:2022vsa} 
(see also \cite{Kim:2017lsi}),
and composite solitons in $^3$He superfluids 
\cite{Volovik:2019goo,Volovik:2020zqc}.
On the theoretical side,
by using the most general composite 
soliton consisting of 
kinks, vortices, monopoles and Yang-Mills instantons, all possible relations among topological solitons are exhausted~\cite{Nitta:2022ahj}. 
On the experimental side, 
domain-wall Skyrmions in chiral magnets 
are expected to be useful for a racetrack memory.
In addition, in cosmology, cosmic strings (vortex strings) attached by domain walls \cite{Kibble:1982dd,Everett:1982nm}, such as axion strings with axion domain walls~\cite{Vilenkin:1982ks,Vilenkin:2000jqa,Kawasaki:2013ae}, exhibit non-trivial dynamics driven by the tensions of the strings 
and domain walls and have been studied for decades.

In this paper, 
we focus on 
composite solitons consisting of 
global monopoles, global vortices and domain walls,
motivated by the fact that most of topological solitons in condensed matter physics are in fact global solitons.
Monopoles can be attached by vortices, 
or can be immersed into a domain wall. 
The former and latter are called 
vortex monopoles and domain-wall monopoles, respectively. 
As a lower dimensional analogue,
a vortex can be attached by 
 one (two  or more) domain wall(s), 
 as axion strings.
This can be called domain-wall vortices. 
All previously known 
domain-wall vortices, 
vortex monopoles and domain-wall monopoles 
are summarized in Appendix \ref{sec:summary-composites}.
More specifically, here 
we study composite solitons in linear $O(N)$ models ($N=2,3$)  with 
a spontaneous symmetry breaking (SSB) 
$O(N) \to O(N-1)$
together with small explicit symmetry breaking (ESB) terms that are linear or quadratic with respect to the fields.

In the $O(2)$ model, we consider composites consisting of domain walls and vortices,
which are the same as the so-called axion strings attached with the $N_{\rm DW}=1$ or $2$ domain walls.
In the literature, such composites have been understood in terms of 
an approximate $SO(2)$ and exact $\mathbb{Z}_{N_{\rm DW}}$ symmetries spontaneously broken into $\mathbbm{1}$. 
We however present another understanding of these composites.
Firstly we find that the proper exact SSB at the vacuum is $\mathbb{Z}_2\times \mathbb{Z}_2 \to \mathbb{Z}_2$ for $N_{\rm DW}=2$.
This correct understanding for the SSB leads to a novel picture that the composite soliton is stabilized by two sequent SSBs; 
$G=\mathbb{Z}_2\times \mathbb{Z}_2 \to H=\mathbb{Z}_2$ at the vacuum producing the topological domain wall as a mother soliton,
inside which $H=\mathbb{Z}_2$ is spontaneously broken further into $\mathbbm{1}$, resulting in production of a subsequent kink in the mother domain wall.
This can be regarded as a daughter vortex of the mother domain wall from the bulk perspective. 
This composite soliton is a global counterpart of an 
Abrikosov-Nielsen-Olesen (ANO) vortex
\cite{Abrikosov:1956sx,Nielsen:1973cs} 
inside a domain wall  
\cite{Auzzi:2006ju,Nitta:2012xq,Kobayashi:2013ju,Nitta:2015mma,Fujimori:2016tmw}, 
or local non-Abelian vortex  \cite{Hanany:2003hp,Auzzi:2003fs,Eto:2005yh,Eto:2006cx} 
realized as 
non-Abelian sine-Gordon solitons 
\cite{Nitta:2014rxa,Eto:2015uqa} 
inside a non-Abelian domain wall \cite{Shifman:2003uh,Eto:2005cc,Eto:2008dm}.

We also consider global-monopole analogues 
of vortex monopoles and domain wall monopoles in the linear $O(3)$ model.
Depending on the ESB terms and the sign of their coefficients, 
there can be a global monopole as a daughter soliton attached with one or two strings and 
a daughter monopole localized on a domain wall. 
While the monopole attached with one string is not stable, that with two strings is topologically stable,
as can be understood by sequential SSBs: $O(2)\times \mathbb{Z}_2 \to \mathbb{Z}_2 \times \mathbb{Z}_2$ in the vacuum, producing the topological strings, and $\mathbb{Z}_2 \times \mathbb{Z}_2 \to \mathbbm{1}$ inside the strings.
This monopole is regarded as a kink connecting the two strings.
This composite soliton 
is a global counterpart of 
a local monopole (`t Hooft-Polyakov monopole 
\cite{tHooft:1974kcl,Polyakov:1974ek}) 
attached by two 
non-Abelian vortex strings 
\cite{Tong:2003pz,Shifman:2004dr,Hanany:2004ea,Nitta:2010nd,Eto:2011cv}. 

On the other hand, a monopole localized on the wall is associated with $O(2)\times \mathbb{Z}_2 \to O(2)$ at the vacuum, producing the topological domain wall, and $O(2) \to \mathbb{Z}_2$ inside the domain wall,
which implies that the daughter monopole is 
a vortex inside the mother domain wall.
Again, this composite soliton is a global counterpart of 
a local (`t Hooft-Polyakov) 
monopole immersed into a non-Abelian domain wall 
\cite{Nitta:2015mxa}.
These composite solitons appear only when the subsequent (second-step) SSB takes place inside the mother solitons.
We also observe the phase transition between the broken and symmetric phases varying the coefficients of the ESB terms.
This phase transition inside the mother solitons can provide interesting phenomena as discussed in Sec.~\ref{sec:summary}.

%\textcolor{red}{
%One of interesting findings 
%is that 
%a phase transition 
%of the SSB in the these 
%mother solitons.
%In the symmetric phase, 
%the SSB does not occur 
%while it occurs 
%in the broken phase. 
%}

Furthermore, when the ESB terms in the potential of the $O(3)$ model take more general forms,
a composite soliton consisting of the domain wall, vortex string and monopole can exist.
This topological stability can be understood by the sequential SSBs of 
$(\mathbb{Z}_2)^3 \to(\mathbb{Z}_2)^2 $ at the vacuum leading to the domain wall,
$(\mathbb{Z}_2)^2 \to(\mathbb{Z}_2)^1 $ inside the wall leading to a wall (daughter string from the bulk perspective) on the mother wall,
and $(\mathbb{Z}_2)^1 \to \mathbbm{1} $ inside the daughter wall,
leading to a kink inside the daughter vortex.
This kink is nothing but a monopole as a granddaughter soliton from the bulk perspective.
We also investigate the real-time dynamics of these composite solitons.

This paper is organized as follows.
In Sec.~\ref{sec:string-wall}, we study the topological strings attached with one or two domain walls in the $O(2)$ model.
In Sec.~\ref{sec:monopole}, we study, in the $O(3)$ model, a global monopole attached with one (or two) string(s), and a monopole localized on a domain wall.
In sec.~\ref{sec:monopole-string-wall},
we study the composite soliton consisting of the domain wall, vortex string and monopole.
We present the summary and discussion in Sec.~\ref{sec:summary}.
In Appendix~\ref{sec:summary-composites}, 
we classify composite solitons and summarize known composite solitons 
composed of vortices and domain walls, 
and those composed of 
monopoles and vortices.
In Appendix~\ref{sec:appendix}, we give a comment on the effective Lagrangian approach.

%Definition in intro: 
%spontaneous symmetry breaking (SSB).

%\newpage

%%%%%%%%%%%%%%%%% model %%%%%%%%%%%%%%%%%%

\section{String-wall composites}% in axion models}
\label{sec:string-wall}
In this section, 
we study string-domain wall composites in the linear 
$O(2)$ model with the small ESB potential term.

\subsection{Linear $O(2)$ model}

Let us consider an axion model or
an $O(2)$ linear sigma model consisting of 
a two-component real scalar field ${\bm \phi} = (\phi_1,\phi_2)$ in $(2+1)$ and $(3+1)$ dimensions in this section.
The Lagrangian we will study is  
\be
{\cal L}_N = \frac{1}{2}(\p_\mu{\bm \phi})^2 - \frac{\lambda}{4}\left({\bm \phi}^2 - v^2\right)^2 -  \alpha V_N\, ,
\label{eq:O3_lag}
\ee
where the ESB potential $V_N$ is given by
\be
V_N = (\phi_2)^N
\label{eq:VN}
\ee
with $N\leq 4$ for stable vacua.
In this work we will focus on $N=0$, $1$, and $2$.
A (global) symmetry $G$ of the model and an unbroken subgroup $H$ in the vacuum depend on $N$.

\paragraph{$N=0$ case:} 
The symmetries are $G=O(2)$ and $H=\mathbb{Z}_2$. For example, 
$\left(\begin{smallmatrix}1 &0\\0&-1\end{smallmatrix}\right) \in O(2)$ is unbroken for a specific vacuum ${\bm \phi}=(v,0)$.
Thus, the vacuum manifold is $G/H = S^1$. 

\begin{figure}[ht]
\centering
\includegraphics[width=12cm]{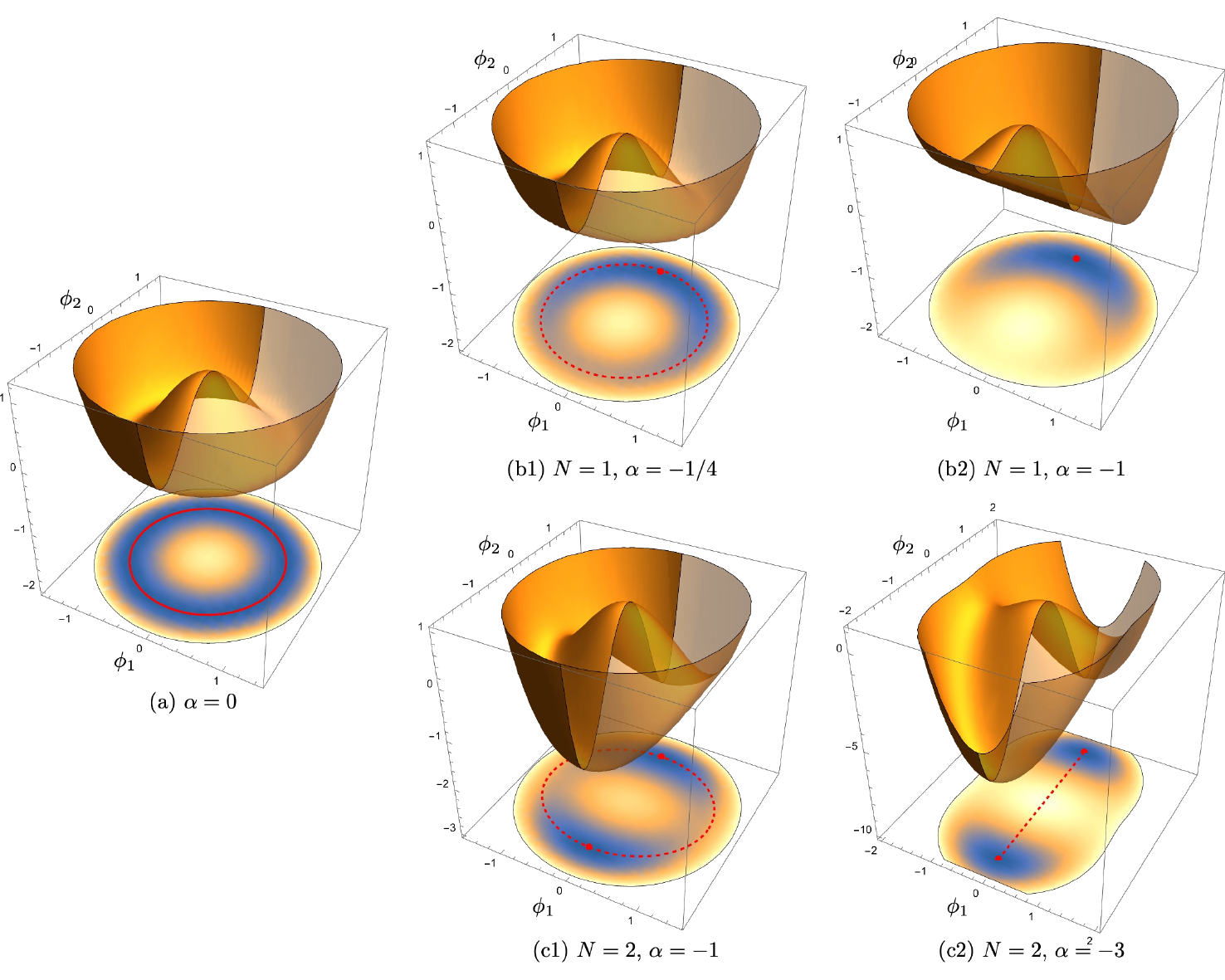}
\caption{The total scalar potential including $V_N$ as a function of $(\phi_1,\phi_2)$. 
The parameters are set as $(\lambda,v) = (4,1)$ for concreteness. }
\label{fig:pot01}
\end{figure}

\paragraph{$N=1$ case:}
$O(2)$ is explicitly broken into $G = \mathbb{Z}_2^{(1)}$ which acts only on $\phi_1$ as 
$(\phi_1,\phi_2) \to (\pm \phi_1,\phi_2)$.
We set $\alpha < 0$ without loss of generality. There is a unique
vacuum on the positive side of the $\phi_2$ axis, as shown in Fig.~\ref{fig:pot01} (b1) and (b2).
The symmetry $G$ is not spontaneously broken, so that we have $H = \mathbb{Z}_2^{(1)}$. 

\paragraph{$N=2$ case:}
We have 
$G = \mathbb{Z}_2^{(1)} \times \mathbb{Z}_2^{(2)}$ where $\mathbb{Z}_2^{(2)}$ acts as $(\phi_1,\phi_2) \to (\phi_1,\pm  \phi_2)$.
Again we set $\alpha<0$.
The vacuum manifold consists of two discrete points on the $\phi_2$ axis as shown in Fig.~\ref{fig:pot01} (c1) and (c2).
Only $\mathbb{Z}_2^{(2)}$ is spontaneously broken, so that we have $H = \mathbb{Z}_2^{(1)}$. 
The vacuum manifold is $G/H \simeq {\mathbb Z}_2$.
These symmetry breaking patterns are summarized in Table.~\ref{tab:1}.
The rows of ``single soliton'' and ``composite soliton'' will be discussed in more detail in the following sections.
\\
%%%%%%%%%%%%%
%%%%%%%%%%%%%

%%%%%%%%%%%%%%%%%
\begin{table}[ht]
\centering
\begin{tabular}{c|ccc}
\hline
& $N=0$ & $N=1$ & $N=2$ \\
\hline
$G$ & $O(2)$ & $\mathbb{Z}_2^{(1)}$ & $\mathbb{Z}_2^{(1)}\times \mathbb{Z}_2^{(2)}$\\
$H$ & $\mathbb{Z}_2$ & $\mathbb{Z}_2^{(1)}$ & $\mathbb{Z}_2^{(1)}$\\
vacuum manifold & $S^1$ & 1 point & 2 points \\
single soliton & vortex & none & domain wall \\
composite soliton & none & VW & VW$^2$ \\
\hline
\end{tabular}
\caption{Solitons in the linear $O(2)$ model.
The description ``VW'' (``VW$^2$'') represents the composite solitons consisting of one vortex and one (two) domain wall(s).
}
\label{tab:1}
\end{table}
%%%%%%%%%%%%%%%

This model \eqref{eq:O3_lag} is identical to the well known axion(-like) model with the ``domain wall number'' $N_\mathrm{DW}=N$,
\be
{\cal L}^{(\text{axion})} = \frac{1}{2}|\p_\mu\varphi|^2 
- \frac{\lambda}{4}\left(|\varphi|^2 - v^2\right)^2 
- \frac{2\alpha v^2}{N^2}\left(\cos N\Theta - 1\right),
\label{eq:L_axion}
\ee
with a complex scalar $\varphi = v f e^{i\Theta}$.
The Lagrangian (\ref{eq:L_axion}) is essentially equivalent to our Lagrangian (\ref{eq:O3_lag}) with $V_N = (\phi_2)^N$ (up to some overall and additive constants), as is understood by rewriting $\varphi = \phi_2 + i \phi_1$.
If one takes the limit of $\lambda \to \infty$, the model reduces to a sine-Gordon model with the $2\pi/N$ periodicity for $N\neq 0$ or the XY model for $N=0$.

In this paper, we are mostly interested in the case that the ESB term is small,
\be
\lambda v^4 \gg |\alpha| v^N.
\label{eq:wall_weak}
\ee
(Note that the mass dimension of $\alpha$ is $d+1 - N(d-1)/2$ in $d+1$ dimensions.)
This condition ensures that the structure of the minimum of the potential for $N=0$ is not significantly deformed by $\alpha V_N$ ($N=1,2$). 
%Namely, the original $S^1$ vacuum manifold does not completely disappear but the $S^1$
%still remains in the low-lying structure of the potential.
Namely,  
the potential $V_N$ does not largely affect the original 
($N=0$) vacuum manifold, which consists of continuously degenerated minima forming $S^1$, 
but slightly resolves the continuous degeneracy leaving $N$ points as the true minima.
Thus, the original ($N=0$) vacuum manifold $S^1$ is still approximately realized as the minimum of the deformed potential,
and hence we call this $S^1$ the quasi-vacuum manifold.
This quasi-vacuum manifold is 
parameterized by 
a pseudo-Nambu-Goldstone boson 
appearing as a consequence of 
the SSB of 
the approximate symmetry $O(2)$ into ${\mathbb Z}_2$.

\subsection{Topological solitons for $N=0$}

\begin{figure}[ht]
\begin{center}
\includegraphics[height=5cm]{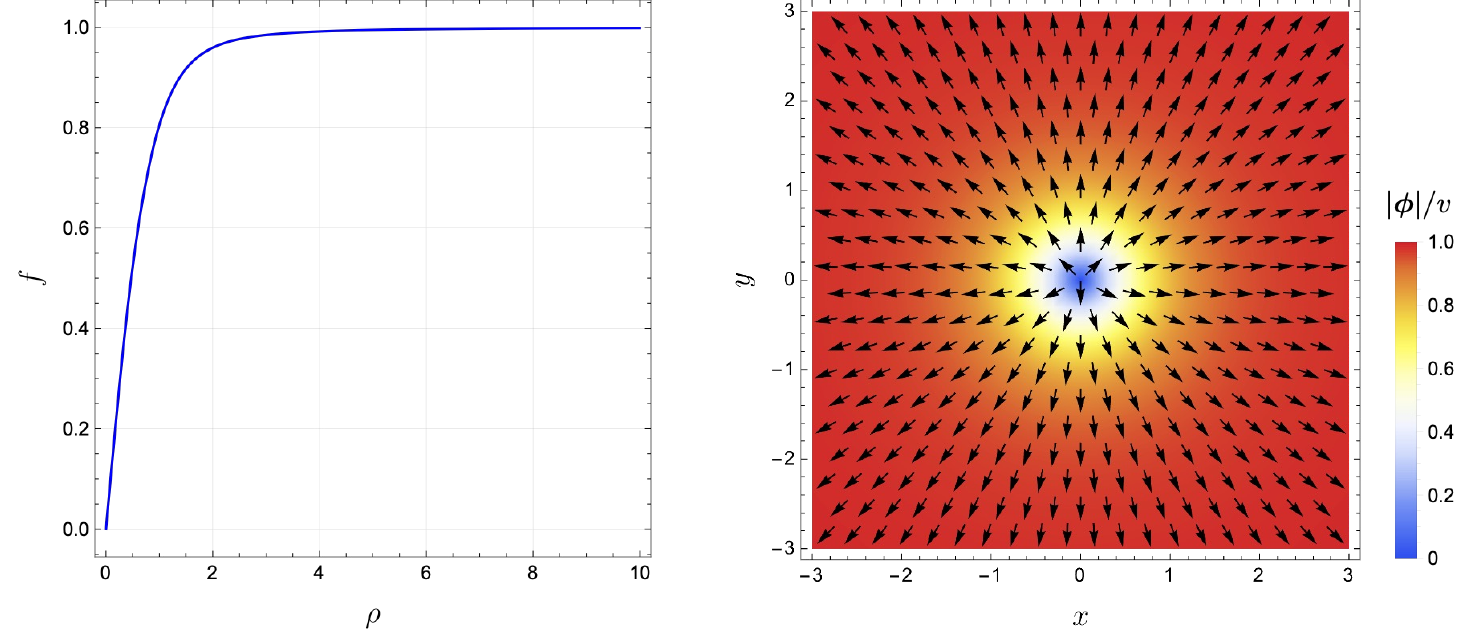}
\caption{The numerical solution of a global vortex for $v=1$ and $\lambda = 4$.
The left figure shows the amplitude $f(\rho)$. In the right figure
the color density corresponds to the amplitude $|{\bm \phi}|/v$ and the arrows are the vector plot of
the normalized field $\hat {\bm \phi}$.}
\label{fig:2d_string}
\end{center}
\end{figure}

In the case of $N=0$, the vacuum manifold is $S^1$ of the radius $v$, which is not simply connected.
Therefore it gives rise to topological solitons characterized by the fundamental homotopy group
$\pi_1(S^1) = \mathbb{Z}$, the so-called global vortices.
Since no analytic solutions even for a single winding vortex are available,  
we need a numerical analysis to obtain the solutions.
A suitable ansatz for a static vortex solution that is axially symmetric about the $z$ axis is given by
\be
{\bm \phi} = v f(\rho) \hat{\bm \rho},
\label{eq:vortex_ansatz}
\ee
with $\rho = \sqrt{x^2+y^2}$, and $\hat{\bm \rho}=(x,y)/\rho$.
The equation of motion (EOM) reads
\be
\frac{d^2 f}{d\rho^2} + \frac{1}{\rho}\frac{df}{d\rho} - \frac{f}{\rho^2} - \lambda v^2 f \left(f^2- 1\right) = 0.
\label{eq:EOM_global_string}
\ee
The profile function $f(\rho)$ should satisfy the boundary condition
\be
f(0) = 0,\quad f(\infty) = 1.
\ee
Fig.~\ref{fig:2d_string} shows a numerical solution for $\lambda v^2 = 4$.

\subsection{Non-topological string-wall composite for $N=1$}
%\subsubsection{Symmetry breaking by additional potentials}
\label{sec:SB_AV_two}

Here we briefly explain a vortex-wall composite for $N=1$. As explained previously,
the model has only $\mathbb{Z}_2^{(1)}$ symmetry which
is not spontaneously broken. Consequently, there is the unique vacuum.
Therefore, there are no topologically stable objects under the presence of $V_1$.
Nevertheless, if the effect of the additional potential $V_1$ is sufficiently small, namely the condition (\ref{eq:wall_weak}) is satisfied, 
the quasi-vacuum manifold $S^1$ shown in Fig.~\ref{fig:pot01}(b1) allows solitonic objects, that is global vortices 
winding the quasi-vacuum manifold $S^1$. The soliton looks like a global vortex on one side, but at the same time
it also looks like a domain wall on the other side. This is because the $S^1$ is not the flat bottom of the potential
but the sine-Gordon-like potential appears, $V_1 \simeq v \cos\Theta$, where $\Theta$ is the phase of $\phi_1 + i \phi_2$.
Hence, when one goes around the vortex, one inevitably passes the sine-Gordon-like potential once. This explains that
the global vortex is necessarily attached by the domain wall.
This kind of ``soliton-attached solitons'' are called composite solitons in this paper, and 
the description ``VW'' in Table.~\ref{tab:1} represents the composite soliton consisting of the one vortex and one domain wall.

\begin{figure}[h]
\centering
\includegraphics[height=5cm]{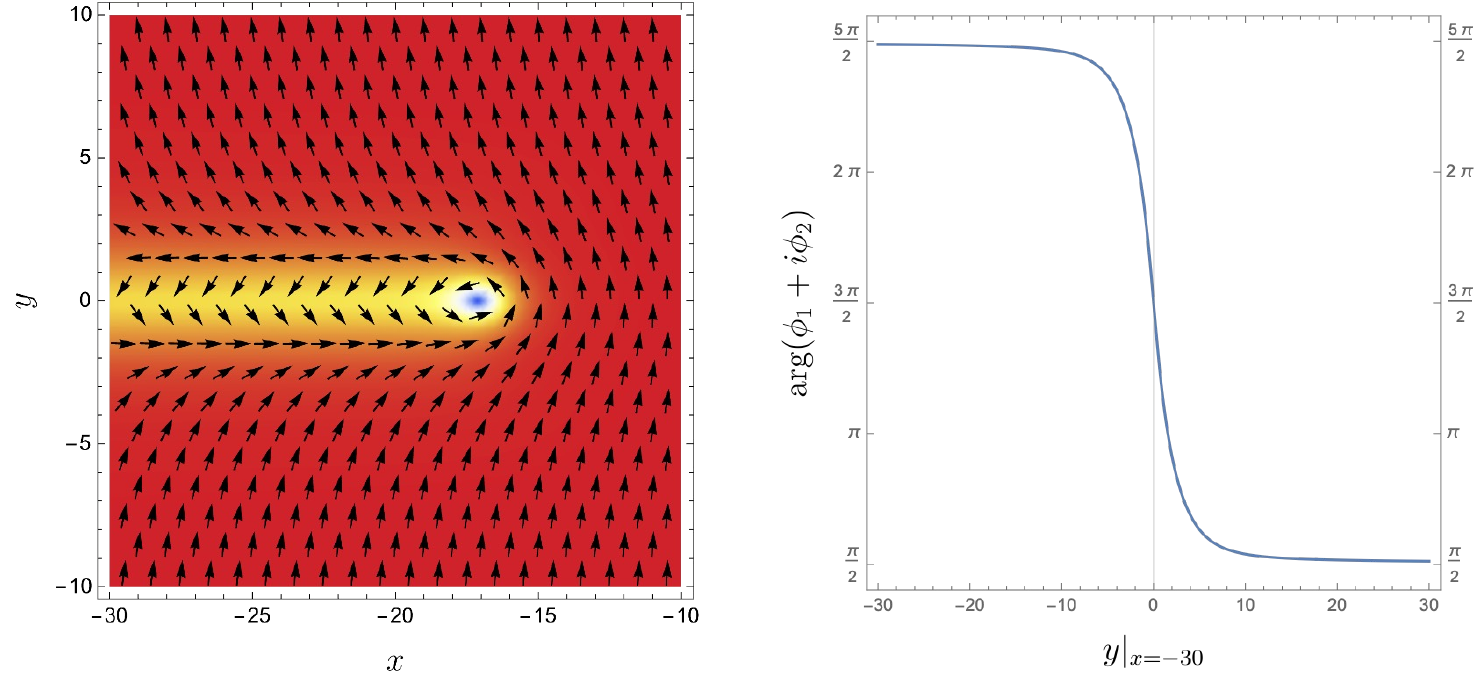}
\caption{
The vortex-wall composite in $N=1$ model. 
The color shows the amplitude and the vectors show the phase of ${\bm \phi}$ in the left panel.
The right panel shows the sine-Gordon-like soliton in the phase $\Theta = \arg(\phi_1+i\phi_2)$ on the slice $x=-30$.
}
\label{fig:wall_vortex_n1_half}
\end{figure}

In order to construct a numerical solution of the vortex-wall composite in two dimensional plane, we set the boundary
condition as follows. At an edge of box, say $x = -L$, we set $\Theta(L,y)$ to be an anti-sine-Gordon-like soliton,
$\Theta(L,y) \simeq 4\arctan\exp (-\sqrt{\alpha} y) + \pi/2$
behaving 
$\Theta(L,y\to -L) \to \pi/2$ and 
$\Theta(L,y\to L) \to -3\pi/2$.  On the remaining three edges, we set $\Theta$ as a constant $\Theta = \pi/2$.
A numerical solution obtained by making use of a standard relaxation technique is shown in Fig.~\ref{fig:wall_vortex_n1_half}.
The solution is not topologically stable in the sense that it cannot be static since the domain wall continuously pulls the vortex.

%%%%%%%%%%%%%
%%%%%%%%%%%%%%

\begin{figure}[h]
\begin{center}
\includegraphics[height=5cm]{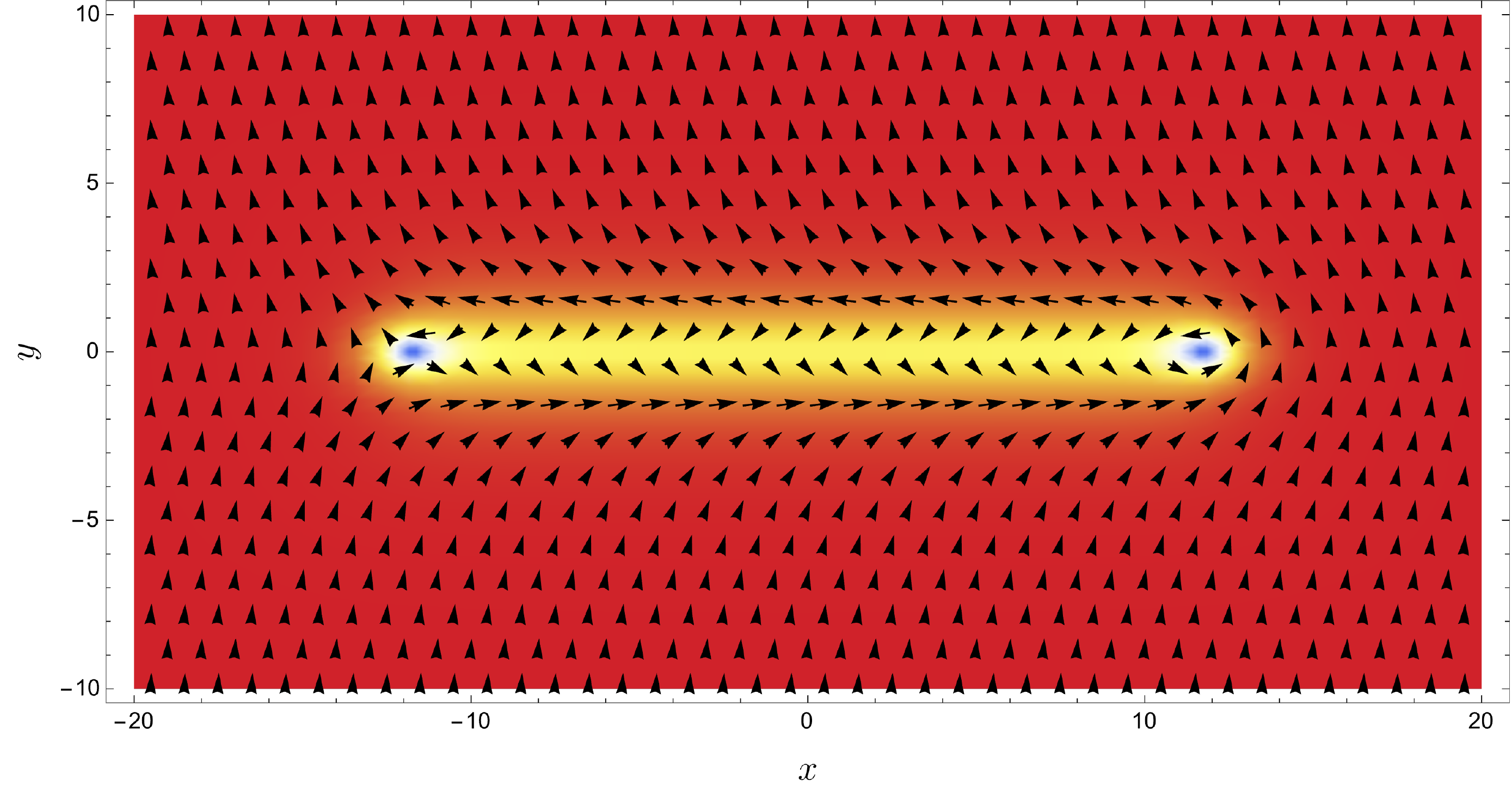}
\caption{
The confined pair of vortex and anti-vortex connected by the sine-Gordon-like domain wall.
}
\label{fig:wall_vortex_n1_pair}
\end{center}
\end{figure}

The sine-Gordon domain wall can also terminate with an anti-global vortex at the opposite end.
The resulting configuration is a sort of confined pair of vortex and anti vortex connected by  the domain wall.
A numerical solution is shown in Fig.~\ref{fig:wall_vortex_n1_pair}.
Again, this is not stable. Because of the domain wall tension, 
the vortex and anti-vortex are pulled and 
eventually pair annihilate to decay into the vacuum.

\subsection{Topological string-wall composite for $N=2$}

As stated above, our model is identical to the axion(-like) model \eqref{eq:L_axion}. % with $N_\mathrm{DW}(=N)=2$.
However, there is a point that is not fully emphasized in the literature.
It is said that in the axion model, a global $U(1)$ symmetry associated with the phase rotation of $\varphi$ (called the Peccei-Quinn symmetry) is explicitly broken into $\mathbb{Z}_N$ due to the cosine potential,
$G={\mathbb Z}_N:\, \varphi \to e^{i2\pi k/N}\varphi$ with $k=0,1,\cdots,N-1$.
In particular, $G=\mathbb{Z}_2$ for $N=2$. 
However, we emphasize that the Lagrangian \eqref{eq:L_axion} is also invariant under the CP symmetry, $\varphi\to \varphi^\ast$,
and hence the true symmetry is $G = \mathbb{Z}_2\times \mathbb{Z}_2^\mathrm{CP}$, 
where the action of the former $\mathbb{Z}_2 $ is equivalent to $\mathbb{Z}_2^{(1+2)}: (\phi_1,\phi_2) \to - (\phi_1,\phi_2)$ in our model
while the latter $\mathbb{Z}_2^\mathrm{CP}$ acts as $\mathbb{Z}_2^{(1)}$.
%It is essentially identical to our Lagrangian (\ref{eq:O3_lag}) with $V_2 = (\phi_2)^2$.
Hence, the genuine symmetry is $\mathbb{Z}_2^{(1)} \times \mathbb{Z}_2^{(2)}$ instead of $\mathbb{Z}_2^{(1+2)}$.
The presence of two $\mathbb{Z}_2$'s is crucial for the arguments below.

\subsubsection{Topological domain wall with $\mathbb{Z}_2^{(2)}$ charge}
\label{sec:tdw_z2}

We first explain topological domain walls in the case of $N=2$. 
As is shown in Table~\ref{tab:1},
the $\mathbb{Z}_2^{(2)}$ symmetry is spontaneously broken,
and it gives rise to topologically stable domain walls.
Note that there are two phases for the domain walls corresponding to whether or not the $\mathbb{Z}_2^{(1)}$ symmetry is further broken \textit{inside the walls}.
If the condition (\ref{eq:wall_weak}) is satisfied, the $S^1$ structure remains as the quasi vacuum manifold
as indicated by the red-dashed curve in Fig.~\ref{fig:pot01} (c1). 
Therefore, for the field configuration connecting $(0,v)$ and $(0,-v)$, the straight path in the field space
is shorter but energetically higher than detours around the potential barrier at $|{\bm \phi}|= 0$.
The field configuration tends to lie along the low-lying $S^1$ locus. 
Now, there are two possibilities, the field configuration goes around either clockwise or counterclockwise. 
These can be distinguished by whether $\phi_1$ takes positive or negative value on the way.
In other words, $\phi_1$ positively or negatively condenses near the core of the domain wall.
Namely, the $\mathbb{Z}_2^{(1)}$ symmetry is spontaneously broken near the domain wall core,
which is schematically expressed as
\be
\mathbb{Z}_2^{(1)} \times \mathbb{Z}_2^{(2)}
\xrightarrow[]{\ \text{vac}\ }
\mathbb{Z}_2^{(1)}
%\xrightarrow[]{\ \text{wall}\ }%(|\alpha| v^2/\lambda)^{1/4}\ }
\dotarrow{\ \text{wall}\ }
1.
\label{eq:z2xz2toz2to1}
\ee
%The first symmetry breaking (the solid-line arrow) gives rise to the domain wall, and it triggers the second one (the dashed-line arrow).
%Hence, the domain wall has the $\mathbb{Z}_2^{(1)}$ charge.
%Notice that the spontaneous symmetry breaking (SSB) occurring at the vacuum is the first SSB corresponding to the solid-line arrow.
%The second SSB of the dashed-line arrow takes place only inside the domain wall. One should carefully distinguish the solid and dashed arrows.
The SSB at the vacuum corresponds to the first SSB indicated by the solid-line arrow.
It gives rise to the domain wall, inside which the second SSB indicated by the dashed-line arrow takes place.
Hence, the domain wall has the $\mathbb{Z}_2^{(1)}$ charge 
(or  the $\mathbb{Z}_2^{(1)}$ modulus like Ising spin).
One should carefully distinguish the solid and dashed arrows.

\begin{figure}[tbp]
\centering
\includegraphics[width=15cm]{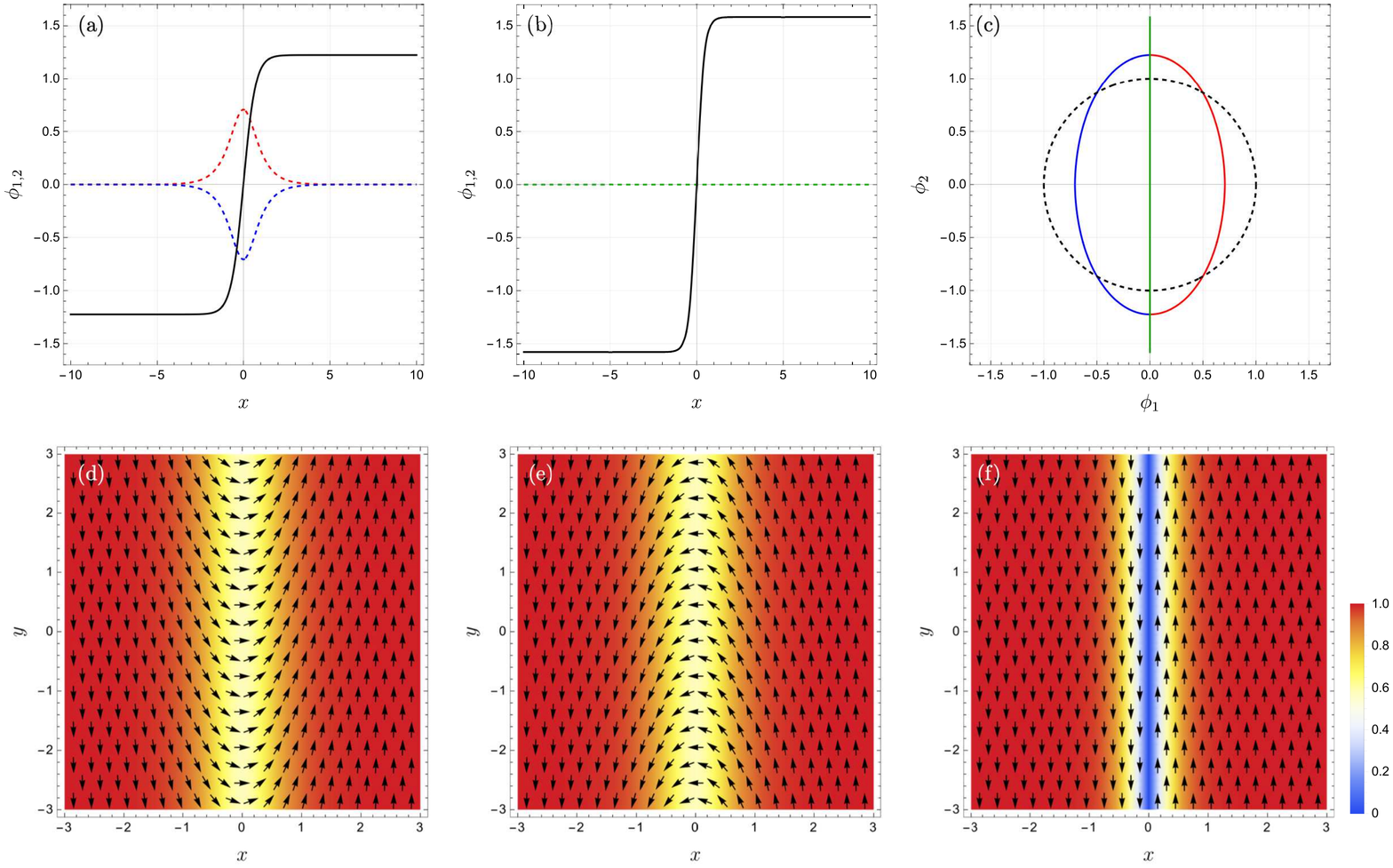}
\caption{The topological domain wall solutions in the case with $N=2$ in the linear $O(2)$ model. The parameter combinations are chosen as
$(\lambda,v) = (4,1)$. 
(a): The $\mathbb{Z}_2^{(1)}$ charged domain wall solutions for $\alpha=-1$. 
The black curve stands for $\phi_2(x)$, and the dashed curves show $\phi_1(x)$.
The color (red and blue) of the dashed curves indicates the $\mathbb{Z}_2^{(1)}$ charge.
(b): The $\mathbb{Z}_2^{(1)}$ neutral domain wall for $\alpha = - 3$. 
$\phi_2$ is the black curve and the green-dashed line stands for $\phi_1=0$.
(c): The orbits corresponding to the solutions in (a) and (b) expressed in the $\phi_1\phi_2$ plane.
(d), (e), and (f) show the vector plot of ${\bm \phi}$ and the density plot of $|{\bm \phi}|/v_{\rm max}$
in the $xy$ plane for the domain wall solutions of (a) with the blue-dashed, red-dashed,
and (b), respectively ($v_{\rm max}$ is the maximum value of $|{\bm \phi}|$).
}
\label{fig:wall_n2}
\end{figure}

The topological nature of the domain wall is associated with the first SSB,
\begin{equation}
\pi_0\left((\mathbb{Z}_2^{(1)} \times \mathbb{Z}_2^{(2)})/\mathbb{Z}_2^{(1)}\right) 
\simeq
\mathbb{Z}_2 \neq 1.
\end{equation} 
We show numerical solutions for the topological domain wall perpendicular to the $x$ axis in Fig.~\ref{fig:wall_n2}.
The panel (a) shows the scalar fields $\phi_{1,2}$ as the function of $x$, and (c) shows the corresponding parametric plot in the field space. The orbits are half ovals reflecting the small deformation by $\alpha V_2$.
%Due to the additional potential $\alpha V_2$, the orbits are not a half circle but a half oval. \textcolor{red}{comment in tex souce}
% Due to additional は変。なぜならそれがないとそもそもdomain wallがないから。
The panels (d) and (e) show $|\bm{\phi}|/v_\mathrm{max}$ and the phase of $\phi_2+i\phi_1$ in the $xy$ plane.
While the genuine topological charge is associated with the spontaneously broken $\mathbb{Z}_2^{(2)}$, the
domain walls can also be characterized by a quasi-topological quantity, the winding number of the low-lying $S^1$ (oval) locus. Clearly, this winding number is quantized in a half integer in comparison with
the topological global vortices for $N=0$.

There is another type of domain wall, when the additional potential $V_2$ has such a large coupling constant $|\alpha|$ as the 
condition (\ref{eq:wall_weak}) is not met. Since the original $S^1$ structure completely disappears,
the field configuration corresponding to the domain wall forms an orbit in the $\phi_1\phi_2$ plane as a straight segment connecting $(0,\pm\tilde v)$,
the red-dashed line in Fig.~\ref{fig:pot01}(c2).
For this domain wall, $\phi_1$ is kept to be zero everywhere, so that
the $\mathbb{Z}_2^{(1)}$ symmetry is unbroken everywhere. 
As a result, there is only a unique domain wall without the $\mathbb{Z}_2^{(1)}$ charge, which means that it is neutral under the $\mathbb{Z}_2^{(1)}$ symmetry.
The relevant symmetry breaking in this case is
\be
\mathbb{Z}_2^{(1)} \times \mathbb{Z}_2^{(2)}
\xrightarrow[]{\ \text{vac}\ }
\mathbb{Z}_2^{(1)}.
\ee
The numerical solution is shown in Fig.~\ref{fig:wall_n2}(b) and (f).
%The relevant topological current for this is neither ${\cal W}_\alpha$ nor $\hat{\cal W}_\alpha$ since the former 
%vanishes due to $\phi_1 = 0$ and the latter is not only zero but is singular at the domain wall center $\phi_1 = \phi_2=0$. 
%By the same reason the dual  ``electric field'' vanishes ${\bm E} = {\bm 0}$.
%The genuine topological current for the domain wall  perpendicular to the $x^1$ axis is given by
%\be
%w_\alpha = \frac{1}{2}\epsilon_{\alpha\beta}\p^\beta \phi_2,\qquad (\alpha = 0,1).
%\ee
%This is a conserved current since it trivially satisfies $\p^\alpha w_\alpha = 0$.
%Note that $w_\alpha$ can be used for the domain walls with the symmetry breaking (\ref{eq:z2xz2toz2to1})
%but it cannot distinguish the $\mathbb{Z}_2^{(1)}$ charge. Therefore, we prefer to use
%$\hat W_\alpha$ for the $\mathbb{Z}_2^{(1)}$ charged domain walls.

%%%%%%%%%%%%%%%%
\begin{figure}[tbp]
\begin{center}
\includegraphics[width=8cm]{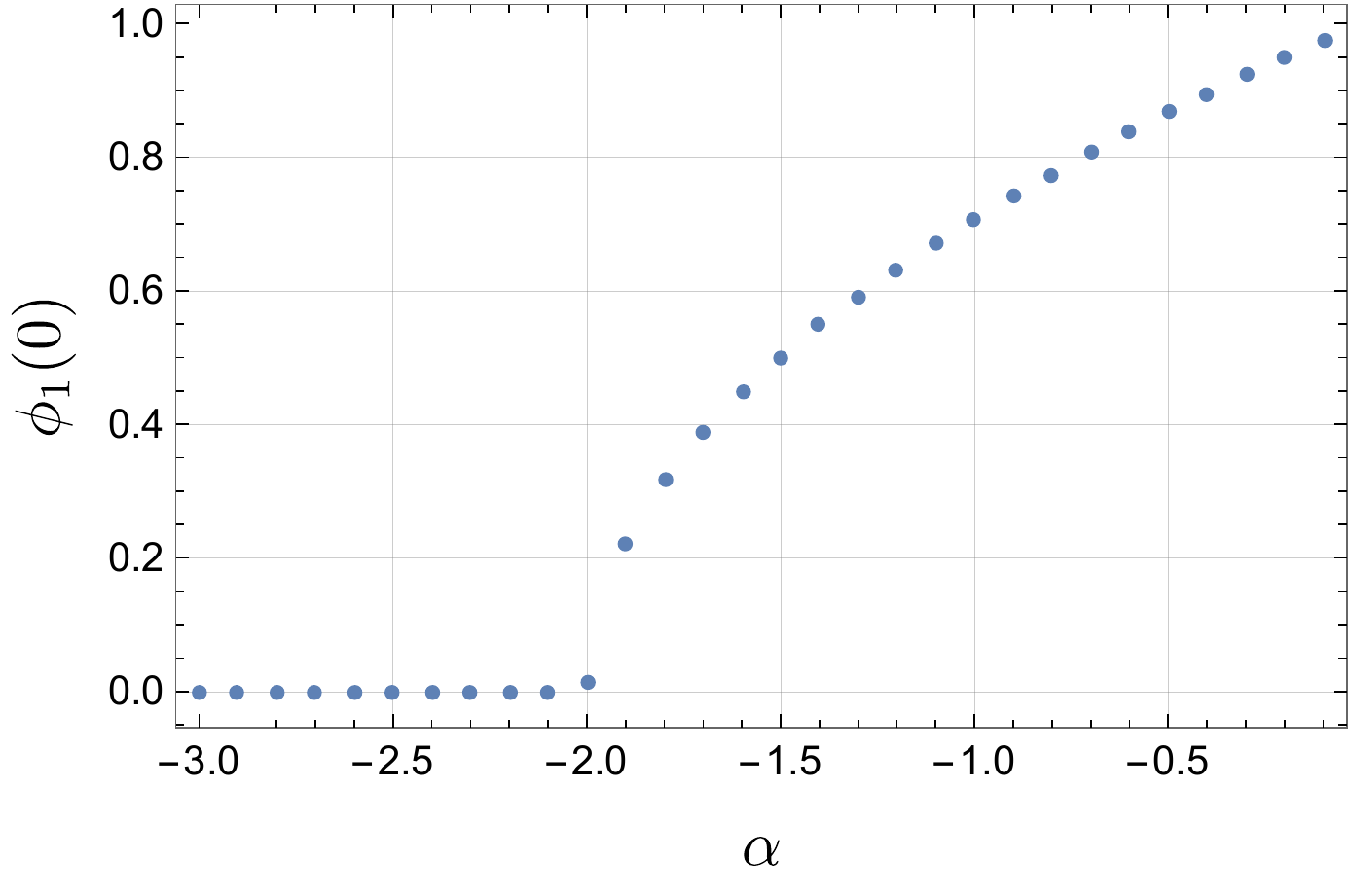}
\caption{The transition of the $\mathbb{Z}_2^{(1)}$ neutral and charged domain wall for the case of $N=2$.
The domain wall is $\mathbb{Z}_2^{(1)}$ charged for $\phi_1(0) \neq 0$ while is neutral for $\phi_1(0) = 0$.
The parameters are taken as $(\lambda,v) = (4,1)$.
}
\label{fig:wall_transition}
\end{center}
\end{figure}
%%%%%%%%%%%%%%%%

The transition between the $\mathbb{Z}_2^{(1)}$ neutral domain wall and charged one is shown in Fig.~\ref{fig:wall_transition}.
We plot the value of $\phi_1$ at the center of domain wall ($x=0$) which is an order parameter for the $\mathbb{Z}_2^{(1)}$
symmetry as a function of $\alpha$ while the other parameters
are fixed as $(\lambda,v) = (4,1)$. The domain wall is neutral if $\phi_1(0)=0$ while charged if $\phi_1(0) \neq 0$.

%\subsubsection{The topological domain wall with a global vortex: domain wall in domain wall}
%\label{sec:string-wall}

%In Sec.~\ref{sec:topological_wall}, 
%we have learned that the spontaneously broken $\mathbb{Z}_2^{(2)}$ symmetry gives rise to the topological domain walls,
%and the further spontaneously broken $\mathbb{Z}_2^{(1)}$ symmetry 
%endows the $\mathbb{Z}_2^{(1)}$ charges to them. 

\subsubsection{Topological wall-string composite: a kink in a domain wall}

Let us next  consider composite solitons for the small $\alpha$ satisfying the condition (\ref{eq:wall_weak}).
Hereafter, we will refer the $\mathbb{Z}_2^{(1)}$-charged domain wall associated with the 
spontaneously broken $\mathbb{Z}_2^{(2)}$ to the mother domain wall.
The composite soliton we are going to study consists of the global vortex 
attached by the two domain walls.

When the $\mathbb{Z}_2^{(2)}$ symmetry is spontaneously broken in the vacuum, it gives rise to the
mother domain walls in the vacuum. Similarly, when the $\mathbb{Z}_2^{(1)}$ symmetry
is spontaneously broken in the mother domain wall, it gives rise to a kink inside the mother domain wall.
%We will call this a daughter kink.
Namely, %the daughter 
this kink connects two domain walls with different $\mathbb{Z}_2^{(1)}$ charges.
Below we will identify this kink 
as a global vortex from the bulk point of view, 
which we call a daughter vortex.

A concrete field configuration which illustrates the above properties is realized by a product ansatz
\be
{\bm \phi} = v \left(
\begin{array}{c}
\tanh\alpha y \cos \Theta(x)\\
\gamma \sin\Theta(x)
\end{array}
\right),\quad
\Theta(x) = 2 \arctan e^{-\beta x} + \frac{\pi}{2},
\label{eq:PA_sw}
\ee
with $\alpha,\beta$ and $\gamma$ are constants. $\Theta$ is the sine-Gordon soliton with a half period which embodies the
mother domain wall, and $\tanh \alpha y$
corresponds to the inner kink (daughter vortex). In Appendix \ref{sec:appendix} we derive this ansatz by making use of an effective theory approach.
Indeed, taking $y \gg 0$ or $y \ll 0$, 
(i.e., far from the inner kink,)
it reduces to the half-sine Gordon soliton going around the left or right half of the quasi-vacuum manifold $S^1$ as
\be
{\bm \phi}\big|_{y\gg0} = v \left(
\begin{array}{c}
\cos \Theta(x)\\
\gamma\sin\Theta(x)
\end{array}
\right),\quad
{\bm \phi}\big|_{y\ll0} = v \left(
\begin{array}{c}
-\cos \Theta(x)\\
\gamma\sin\Theta(x)
\end{array}
\right).
\ee
On the mother domain wall at $x=0$, we have $\Theta = \pi$ and the configuration reduces to
\be
{\bm \phi}\big|_{x=0} = v \left(
\begin{array}{c}
-\tanh\alpha y\\
0
\end{array}
\right).
\ee
This exhibits the inner kink at which the $\phi_1$ condensation flips its sign.

\begin{figure}[tbp]
\begin{center}
\includegraphics[height=6cm]{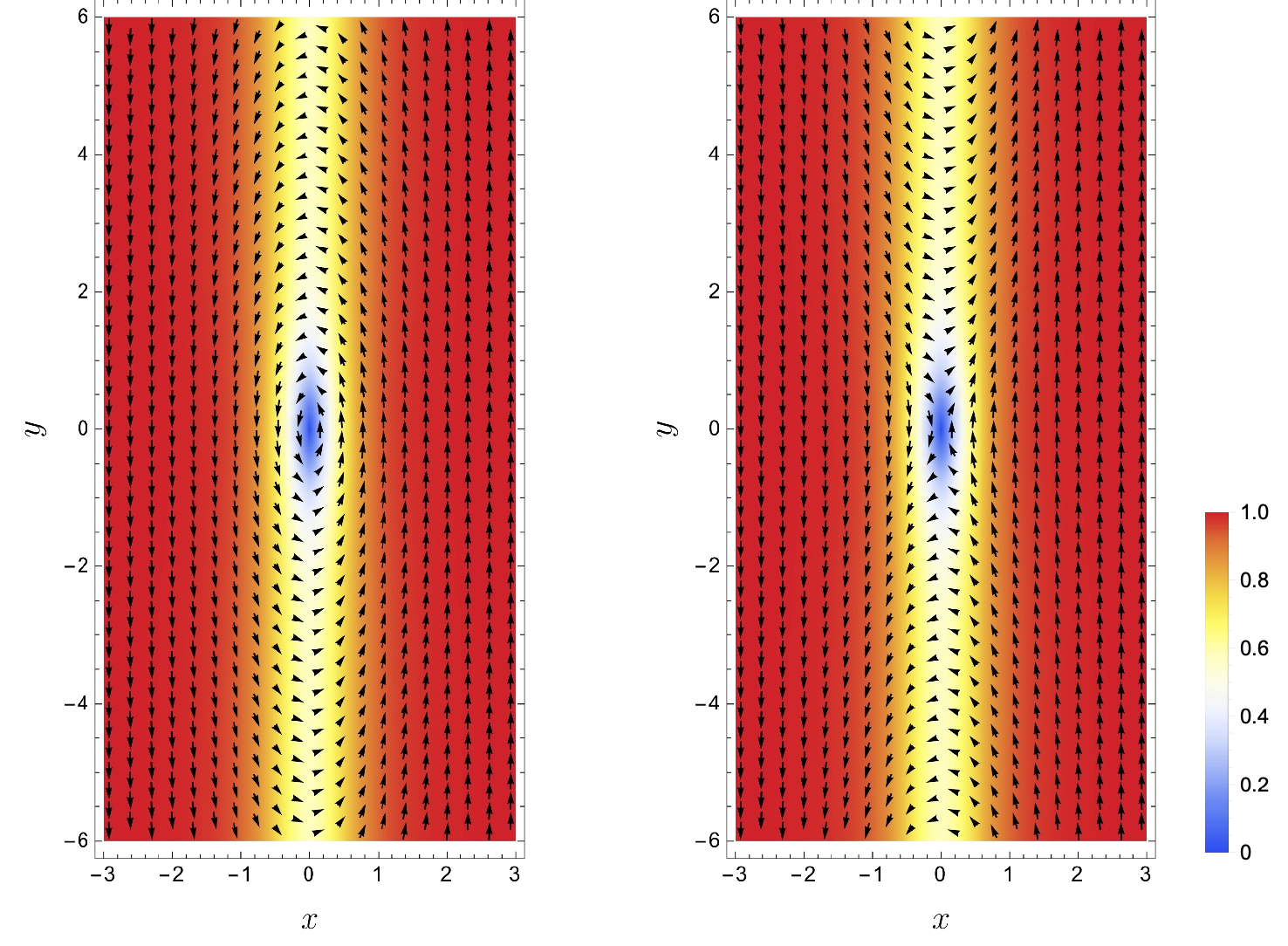}
\caption{
Topological vortex-wall composites in the $N=2$ model. 
The upper part of the mother wall in the left (right) panel has a negative (positive) $\phi_1$ condensation whereas that for the lower part is opposite. Hence, the inner kinks correspond to the vortex and the anti-vortex on the left and right panels, respectively.
}
\label{fig:wall_vortex_n2}
\end{center}
\end{figure}

%In addition to endowing the $\mathbb{Z}_2^{(1)}$ charges,
%the symmetry breaking of $\mathbb{Z}_2^{(1)}$ can also give rise another domain walls,
%just as the symmetry breaking of $\mathbb{Z}_2^{(2)}$ gives rise to the host domain walls.
%However, it occurs only inside the host domain walls, the guest domain walls appear 
%in the host domain walls.
%Hence, they are  point-like solitons.
%In other words,
%the two host domain walls with different $\mathbb{Z}_2^{(1)}$ charges given in Fig.~\ref{fig:wall_n2}(d) and (e) 
%can be connected by the guest domain walls.
The product ansatz (\ref{eq:PA_sw}) is also useful for constructing a numerical solution by a relaxation method.
We can adopt it as an initial configuration, and it quickly converges. 
Fig.~\ref{fig:wall_vortex_n2} shows numerical solutions.  Clearly, these are made by joining the two ${\mathbb Z}_2$ charged domain walls given in the panels (d) and (e) of Fig.~\ref{fig:wall_n2}.
The junction point, corresponding to the guest domain wall, can also be seen as the global vortex.
Indeed, one observes that the black arrow counterclockwise (clockwise) rotates once as we counterclockwise go around the boundary
of the left (right) panel of Fig.~\ref{fig:wall_vortex_n2}. 
The phase $\Theta = \arg(\phi_1+i\phi_2)$ rotates by $\pi$ only on the upper and lower edges while it is constant at the left and right edges.
In total, the phase rotate by $2\pi$.
Moreover, $|{\bm \phi}|$ vanishes at the junction point, which is necessary to avoid a singularity. 

%%%%%%%
%%%%%%%%

%The topological origin of the composite system can be understood in two different ways.
%One is the way given by Preskil-Vilenkin's work.
%For $\alpha=0$, the model has a symmetry $O(2)$, which spontaneously breaks into $\mathbb{Z}_2$ due to the VEV of $\bm{\phi}$.
%This symmetry breaking gives rise to the global vortex associated with the first homotopy group $\pi_1(S^1)\neq 1$.
%Switching on $\alpha\neq 0$ with small $|\alpha|$, 
%the original $O(2)$ symmetry is only an approximate symmetry and explicitly broken into $\mathbb{Z}_2^{(1)} \times \mathbb{Z}_2^{(2)}$,
%leading to the non-trivial zeroth homotopy group of the vacuum manifold $\pi_0(S^0)=\mathbb{Z}_2$. 
%\textcolor{red}{How does $S^0$ appear suddenly?}
%As a result, the global vortex becomes attached by two domain walls, and the system is topologically stable due to the non-trivial $\pi_0$.

The topological origin of the composite system can be understood in the following way. Firstly, we quickly review the conventional way given by Preskil-Vilenkin's work~\cite{Preskill:1992ck}. For $\alpha=0$, the model has a symmetry $SO(2)$, which spontaneously breaks due to the VEV of $\bm{\phi}$. This symmetry breaking gives rise to the global vortex associated with the first homotopy group $\pi_1(S^1) \simeq \mathbb{Z}$.
Switching on $\alpha\neq 0$ with small $|\alpha|$, 
the original $SO(2)$ symmetry is only an approximate symmetry and the exact symmetry is $\mathbb{Z}_2$. It is spontaneously spontaneously broken, 
leading to the non-trivial zeroth homotopy group of the vacuum manifold and hence the topological domain wall. In Ref.~\cite{Preskill:1992ck}, this is expressed schematically as follows:
\begin{alignat}{2}
G_{\rm approx} &= ~SO(2) & \qquad \to \qquad & 1 \nonumber\\
&\cup & & \cup \\
G_{\rm exact} &= ~\mathbb{Z}_2 & \qquad \to \qquad & 1 \nonumber
\end{alignat}
The approximate SSB in the top row implies the global vortex while
the exact SSB in the bottom row ensures the topological domain walls.

%
%On the other hand, we can give another interpretation of the composite system: a wall in a wall.
%As stated above, the SSB of $\mathbb{Z}_2^{(2)}$ in the vacuum gives rise to the
%host domain walls in the vacuum. 
%In the host domain walls, the $\mathbb{Z}_2^{(1)}$ symmetry is spontaneously broken, leading to the guest domain wall inside the host.
%Therefore, the topological stability of the composite soliton owes the SSBs summarized as follows:
%\be
%\mathbb{Z}_2^{(1)} \times \mathbb{Z}_2^{(2)}
%\xrightarrow[\ \text{host wall}\ ]{\ \text{vac}\ }
%\mathbb{Z}_2^{(1)}
%\xrightarrow[\ \text{guest wall}\ ]{\ \text{host wall}\ }%(|\alpha| v^2/\lambda)^{1/4}\ }
%1.
%\label{eq:2comp_sequence}
%\ee
%The words above the arrows correspond to the origins of the SSBs, and
%those below to the topological solitons generated by 
%the corresponding SSBs.
%It endows the genuine topologically nontrivial property to
%the composite solitons.
%This is a clear distinction of 
%the topological composite soliton for $N=2$ 
%from the non-topological composite soliton for $N=1$.

We can improve this sequence by correcting the symmetries, leading to another interpretation of the composite system: a wall in a wall.
As stated above, the SSB of $\mathbb{Z}_2^{(2)}$ in the vacuum gives rise to the
host domain walls in the vacuum. 
In the host domain walls, the $\mathbb{Z}_2^{(1)}$ symmetry is spontaneously broken, leading to the guest domain wall inside the host.
Therefore, the topological stability of the composite soliton owes the SSBs summarized as follows:
\begin{alignat}{3}
G_{\rm approx}= & O(2) \quad &\xrightarrow[\ \text{vortex}\ ]{\ \text{vac}\ }\quad &\mathbb{Z}_2 & \nonumber\\
&\cup & & \cup & \nonumber\\
G_{\rm exact}=
\mathbb{Z}_2^{(1)} &\times \mathbb{Z}_2^{(2)}
\quad &\xrightarrow[\ \text{mother wall}\ ]{\ \text{vac}\ }\quad 
&\mathbb{Z}_2^{(1)}
& \quad \xrightarrow[\ \text{kink}\ ]{\ \text{mother wall}\ }\quad %(|\alpha| v^2/\lambda)^{1/4}\ }
1. 
\label{eq:2comp_sequence}
\end{alignat}
The words above the arrows correspond to the triggers of the SSBs, and
those below to the topological solitons generated by 
the corresponding SSBs.
It endows the genuine topologically nontrivial property to
the composite solitons.
This is a clear distinction of 
the topological composite soliton for $N=2$ 
from the non-topological composite soliton for $N=1$.

\begin{figure}[tbp]
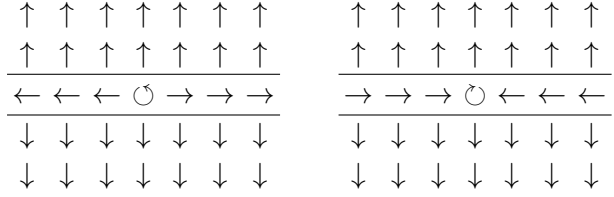

\begin{eqnarray*}
\begin{array}{ccccccc}
%\uparrow & \uparrow & \uparrow & \uparrow & \uparrow & \uparrow\\
\uparrow & \uparrow & \uparrow & \uparrow & \uparrow & \uparrow & \uparrow\\
\uparrow & \uparrow & \uparrow & \uparrow & \uparrow & \uparrow & \uparrow\\
\hline
\leftarrow & \leftarrow & \leftarrow & \circlearrowleft & \rightarrow & \rightarrow & \rightarrow\\
\hline
\downarrow & \downarrow & \downarrow & \downarrow & \downarrow & \downarrow & \downarrow\\
\downarrow & \downarrow & \downarrow & \downarrow & \downarrow & \downarrow & \downarrow\\
%\downarrow & \downarrow & \downarrow & \downarrow & \downarrow & \downarrow
\end{array}\qquad
\begin{array}{ccccccc}
%\uparrow & \uparrow & \uparrow & \uparrow & \uparrow & \uparrow\\
\uparrow & \uparrow & \uparrow & \uparrow & \uparrow & \uparrow & \uparrow\\
\uparrow & \uparrow & \uparrow & \uparrow & \uparrow & \uparrow & \uparrow\\
\hline
\rightarrow & \rightarrow & \rightarrow & \circlearrowright & \leftarrow & \leftarrow & \leftarrow\\
\hline
\downarrow & \downarrow & \downarrow & \downarrow & \downarrow & \downarrow & \downarrow\\
\downarrow & \downarrow & \downarrow & \downarrow & \downarrow & \downarrow & \downarrow\\
%\downarrow & \downarrow & \downarrow & \downarrow & \downarrow & \downarrow
\end{array}
%\quad
%\begin{array}{cccccc}
%\downarrow & \downarrow & \downarrow & \downarrow & \downarrow & \downarrow \\
%\downarrow & \downarrow & \downarrow & \downarrow & \downarrow & \downarrow \\
%%\downarrow & \downarrow & \downarrow & \downarrow & \downarrow & \downarrow
%\hline
%\leftarrow & \leftarrow & \leftarrow & \rightarrow & \rightarrow & \rightarrow\\
%\hline
%%\uparrow & \uparrow & \uparrow & \uparrow & \uparrow & \uparrow\\
%\uparrow & \uparrow & \uparrow & \uparrow & \uparrow & \uparrow\\
%\uparrow & \uparrow & \uparrow & \uparrow & \uparrow & \uparrow
%\end{array}\quad
%\begin{array}{cccccc}
%\downarrow & \downarrow & \downarrow & \downarrow & \downarrow & \downarrow \\
%\downarrow & \downarrow & \downarrow & \downarrow & \downarrow & \downarrow \\
%%\downarrow & \downarrow & \downarrow & \downarrow & \downarrow & \downarrow
%\hline
%\rightarrow & \rightarrow & \rightarrow & \leftarrow & \leftarrow & \leftarrow\\
%\hline
%%\uparrow & \uparrow & \uparrow & \uparrow & \uparrow & \uparrow\\
%\uparrow & \uparrow & \uparrow & \uparrow & \uparrow & \uparrow\\
%\uparrow & \uparrow & \uparrow & \uparrow & \uparrow & \uparrow
%\end{array}
\end{eqnarray*}

\caption{Schematic pictures of the composite solitons of a domain wall and vortices.
    The small arrows represent configurations of ${\bm \phi}$. 
    The horizontal lines denote a domain wall. 
    The left (right) represent a (anti-)kink which is regarded as a (anti)vortex. 
\label{fig:composite_2solitons0}}
\end{figure}

We here give an intuitive explanation which can connect 
the original global vortex for the $O(2)$ symmetry and
the string-wall composite for the $\mathbb{Z}_2^{(1)} \times \mathbb{Z}_2^{(2)}$. The $SO(2)$ symmetry rotates a two dimensional arrow ``$\to$''
which corresponds to a vector $\bm{\phi}=(\phi_1,\phi_2)$ in the field space. For example, when we go around the string, the
arrow spins along the path 
as $\rightarrow \nearrow \uparrow \nwarrow \leftarrow \swarrow
\downarrow \searrow \rightarrow$. How many times the arrow rotates is
counted by $\pi_1[SO(2)] \simeq {\mathbb Z}$.
On the other hand, the $\mathbb{Z}_2^{(2)}$ symmetry flips $\uparrow$ into $\downarrow$ and \textit{vice versa}, and it gives rise to the mother domain wall. Inside the mother domain wall, the 
arrow points either $\leftarrow$ or $\rightarrow$ associated with the
$\mathbb{Z}_2^{(1)}$ charge. Thus, when the %daughter (host) 
inner kink is generated,
the distribution of the arrows on the $xy$ plane is either of those depicted in 
Fig.~\ref{fig:composite_2solitons0}.
%%%
%Although these are associated with $\mathbb{Z}_2^{(1)} \times \mathbb{Z}_2^{(2)}$,
When we go around the boundary counterclockwise, the arrows rotates counterclockwise 
on the left figure 
and clockwise on the right figure.
This clearly explains that the string-wall composite with respect to $\mathbb{Z}_2^{(1)} \times \mathbb{Z}_2^{(2)}$ has the non-trivial topology and is homotopic to the genuine $SO(2)$ global vortex characterized by $\pi_1[SO(2)]$.

A local-soliton  counterpart of 
this composite soliton is  
an ANO vortex inside a domain wall  
\cite{Auzzi:2006ju,Nitta:2012xq,Nitta:2015mma} 
in which the ESB was introduced 
for a global symmetry.
A texture version is a domain-wall Skyrmions in 2+1 dimensions \cite{Nitta:2012xq,Kobayashi:2013ju} (see also \cite{
Sutcliffe:1992he,
Stratopoulos:1992hq,
Kudryavtsev:1997nw,Jennings:2013aea,Bychkov:2016cwc})
and also those 
in chiral magnets \cite{PhysRevB.99.184412,PhysRevB.102.094402,Nagase:2020imn,Yang:2021,Ross:2022vsa} 
(see also~\cite{Kim:2017lsi}).

%%%
\subsection{Dynamics of vortex-wall composites}

\begin{figure}[tbp]
\begin{center}
\includegraphics[width=10cm]{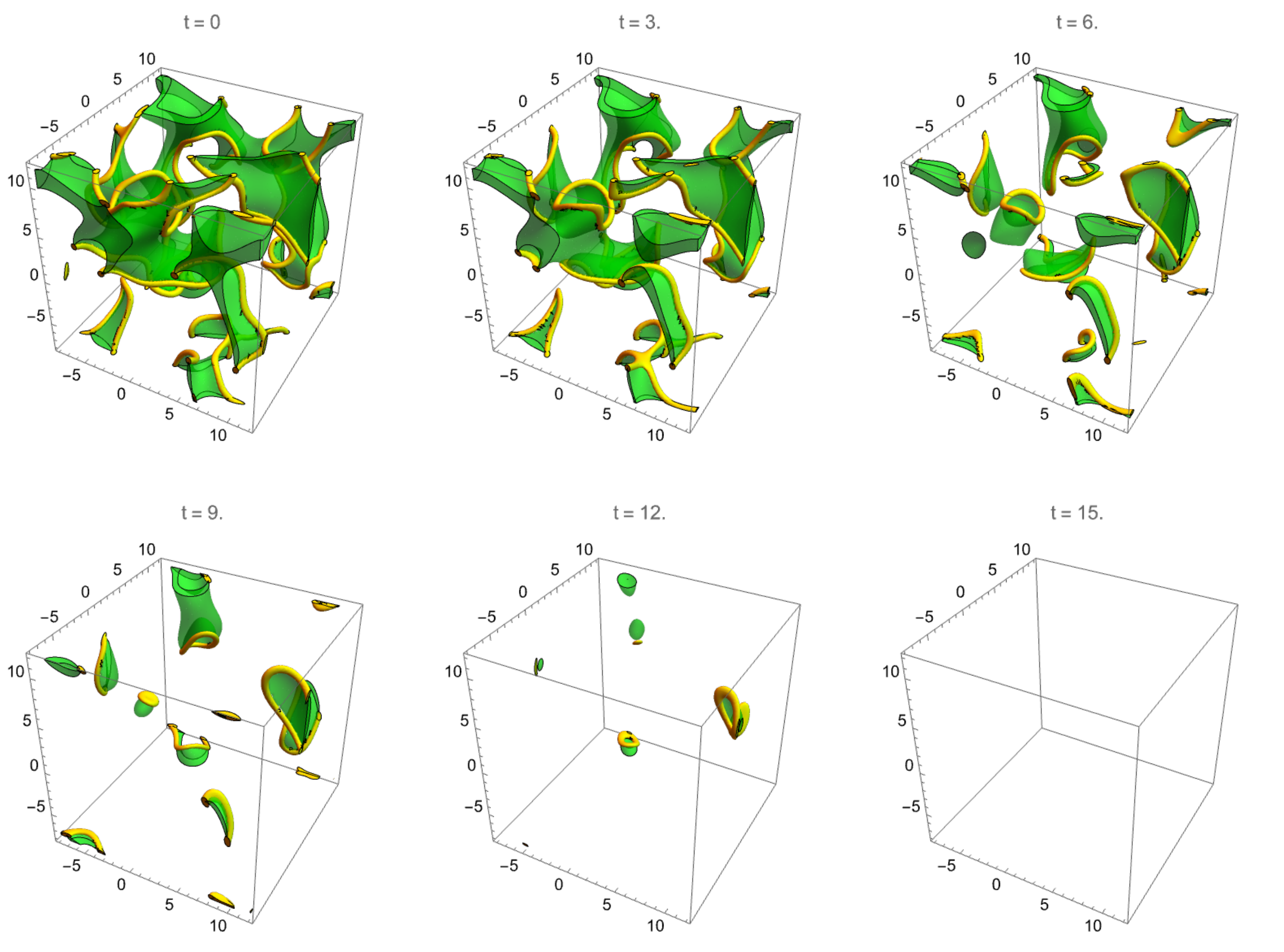}
\caption{
A numerical evolution of $N=1$ vortex-wall composites.
%of random initial configuration of the string-wall composites in $N=1$ model
%with the periodic boundary conditions. 
%The green objects stand for
%the domain walls, and the yellow staff stand for the global strings. 
%The composites quickly decay because they are topologically unstable.
}
\label{fig:wall_string_snaps_N2}
\end{center}
\end{figure}

We numerically investigate dynamics of both the $N=1$ non-topological composite solitons and the $N=2$ topological composite solitons in $(3+1)$ dimensions. 
In this dimensionality, vortices are string 
and domain walls are sheets.
A domain-wall sheet can stretch inside a closed string. 
See, e.g., Ref.~\cite{Eto:2022lhu} 
for three-dimensional configurations of 
a domain-wall disk 
bounded by a circular vortex loop.
We firstly show a dynamical evolution of a randomly chosen initial configuration ($t=0$) in the model of $N=1$ 
in Fig.~\ref{fig:wall_string_snaps_N2}.
The global strings (yellow) are attached by single green domain walls. The composite decays quite fast in comparison with
the $N=2$ case shown below.

%%%%%%%%%%%%%%
%%%%%%%%%%%%%%

\begin{figure}[th]
\begin{center}
\includegraphics[width=12cm]{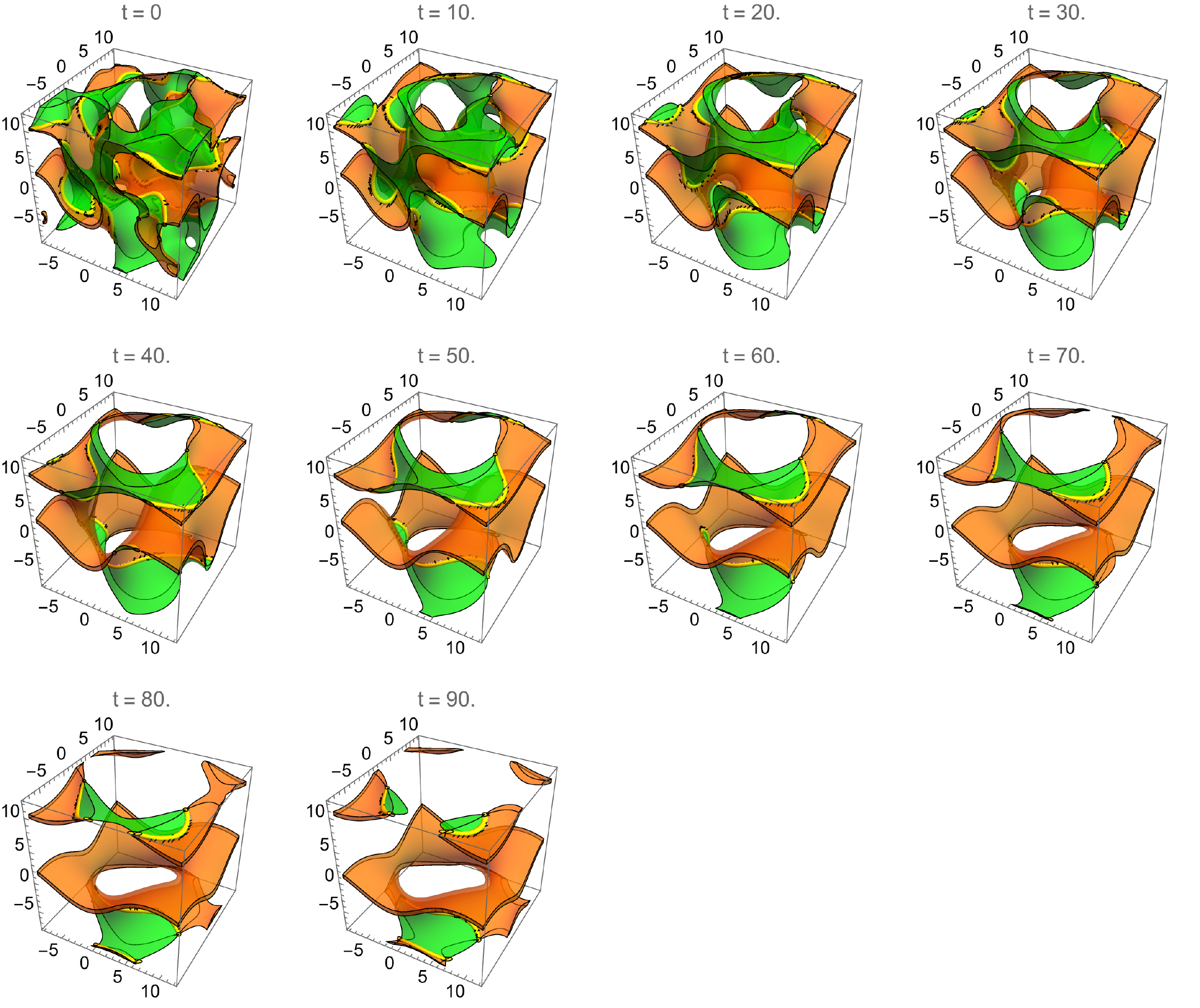}
\caption{
A numerical evolution of $N=2$ vortex-wall composites.
%A numerical evolution of random initial configuration of the string-wall composites in $N=2$ model
%with the periodic boundary conditions. 
%The green and orange region stand for
%the host domain walls with different $\mathbb{Z}_2^{(1)}$ charges. The yellow strings 
%corresponding the guest domain walls which appear
%at the boundaries between the green and orange domain walls. 
%The domain walls are long-lived because they are topologically stable.
}
\label{fig:wall_string_snaps}
\end{center}
\end{figure}

Next, we show time evolution
with a randomly chosen initial configuration ($t=0$) in the $N=2$ model in Fig.~\ref{fig:wall_string_snaps}.
The global strings (the guest domain walls corresponding the yellow objects ) are
attached by a pair of green and orange host domain walls. 
The topological composite solitons are quite long-lived in comparison with the non-topological one.

In cosmological contexts, long-lived domain walls should be ruled out because they are
inconsistent with cosmological and astrophysical observations.
Hence, the $N=2$ model would have the domain wall problem whereas
the $N=1$ case would be safe, 
as is well known in the context of 
the axion strings and domain walls.

%\clearpage

\section{Monopole-string composites 
and monopole-wall composites}
\label{sec:monopole}

\subsection{Linear $O(3)$ model}

In this section, we are going to investigate composite solitons in the model with a three-component scalar field
${\bm \phi} = (\phi_1,\phi_2,\phi_3)$ in $(3+1)$ dimensions. We will study the Lagrangian 
\be
{\cal L} = \frac{1}{2}(\p_\mu{\bm \phi})^2 - \frac{\lambda}{4}\left({\bm \phi}^2 - v^2\right)^2 - \alpha V_N\,, 
\label{eq:O3_lag_4}
\ee
where $V_N$ is an ESB term given by 
\be
V_N = (\phi_3)^N.
\ee
This Lagrangian has a similar form to the one given in Eq.~(\ref{eq:O3_lag}).
In the limit of  $\lambda \to \infty$, 
the model reduces to the  $O(3)$ nonlinear sigma model.

\begin{figure}[tbp]
\begin{center}
\includegraphics[width=13cm]{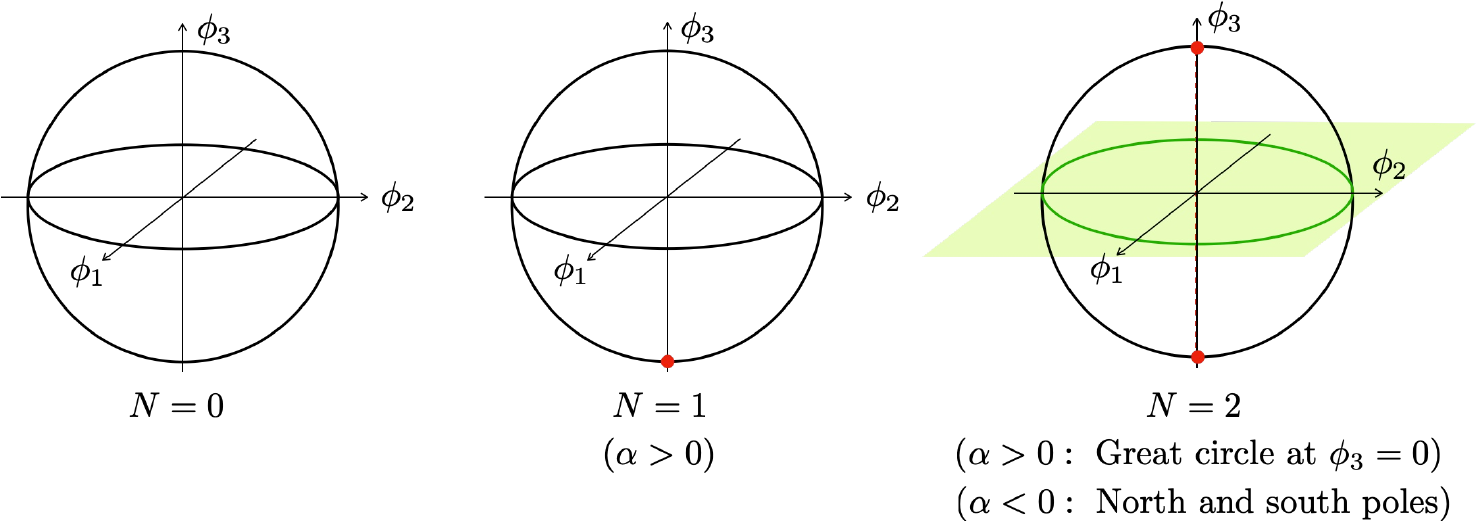}
\caption{The vacua of the three-component model.}
\label{fig:pot_N3}
\end{center}
\end{figure}

As before, we will study $N=0,1,2$.
The symmetry $G$ of the Lagrangian and $H$ of the vacua depend on $N$.
\paragraph{The $N=0$ case:}
The symmetries are $G=O(3)$ and $H=O(2)$, which leads to the vacuum manifold $G/H \simeq S^2$;
\paragraph{The $N=1$ case:}
We have $G = O(2)^{(12)}$ which acts on the first two components
of  $(\phi_1,\phi_2,\phi_3)$.
We set, without loss of generality, $\alpha > 0$. There is a unique
vacuum on the negative side of the $\phi_3$ axis, as shown in 
the middle panel of Fig.~\ref{fig:pot_N3}.
The symmetry $G$ is unbroken in the vacuum, leading to $H = O(2)^{(12)}$. 
%%%%%%%%%%
%%%%%%%%%%

\paragraph{The $N=2$ case:}
We have 
$G = O(2)^{(12)} \times \mathbb{Z}_2^{(3)}$ where $\mathbb{Z}_2^{(3)}$ acts as $\phi_3 \to - \phi_3$ while
$\phi_{1,2}$ is not changed.
The vacuum manifold and the unbroken subgroup $H$ depend on the sign of $\alpha$ 
as shown in the rightmost panel in Fig.~\ref{fig:pot_N3}.
For $\alpha > 0$, the vacuum manifold is a great circle with $\phi_3 = 0$ 
(called the easy-plane potential), and $H=\mathbb{Z}_2^{(12)} \times \mathbb{Z}_2^{(3)}$.
On the other hand, for $\alpha < 0$, the vacuum manifold consists of
two discrete points on the $\phi_3$ axis (called the easy-axis potential)  and we have
$H = O(2)^{(12)}$.
These are summarized in Table.~\ref{tab:2}.

%%%%%%%%%%%%%%
\begin{table}[tbp]
\begin{center}
\begin{tabular}{c||c|c|c|c}
\hline
& $N=0$ & $N=1$ &  \multicolumn{2}{c}{$N=2$}\\
\hline
\hline
$G$ & $O(3)$ & $O(2)^{(12)}$ &  \multicolumn{2}{c}{$O(2)^{(12)}\times\mathbb{Z}_2^{(3)}$}\\
\hline
& & & $\alpha > 0$ & $\alpha<0$\\
\cline{4-5}
$H$ & $O(2)$ & $O(2)^{(12)}$ & $\mathbb{Z}_2^{(12)}\times\mathbb{Z}_2^{(3)}$ & $O(2)^{(12)}$\\
\hline
Vacuum manifold & $S^2$ & 1 point & $S^1$ & 2 points\\
\hline
Single soliton & monopole & none & vortex & domain wall \\
\hline
Composite soliton & none & MS & MS$^2$ & WM \\
\hline
\end{tabular}
\caption{
The patterns of the symmetry breaking and possible solitons in the linear $O(3)$ models.
The composite solitons ``MS'' and ``MS$^2$'' indicate a global monopole attached by one and two strings, respectively,
while ``MW'' indicates a monopole localized on a domain wall.
}
\label{tab:2}
\end{center}
\end{table}
%%%%%%%%%%%

Again, we assume that the ESB potential $V_N$ in 
\eqref{eq:wall_weak} is small.
Thus, the vacuum manifold $S^2$ in the case of $N=0$ is not largely deformed even for $N=1,2$ but is approximately realized as the quasi-vacuum manifold.

\subsection{Topological solitons for $N=0$}

%\subsection{Global monopoles}

\begin{figure}[tbp]
\begin{center}
\includegraphics[height=5cm]{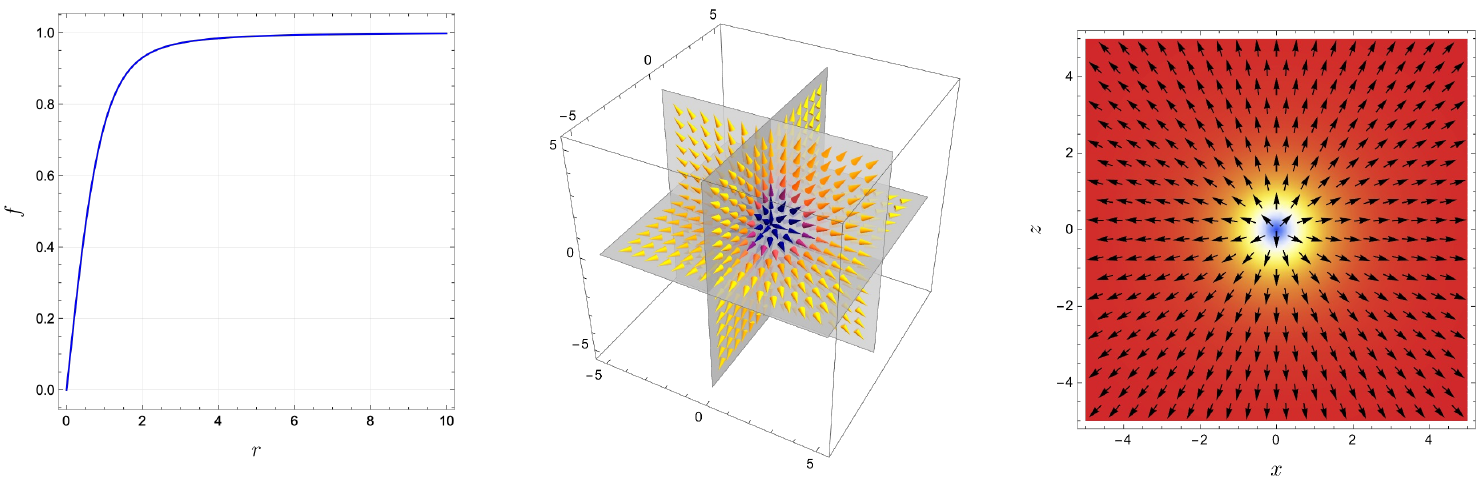}
\caption{
The numerical solution of a spherical global monopole for $(\lambda,v) = (4,1)$.
In the middle and right panels, the arrows and the colors indicate the vector plot for $\bm{\phi}$ and the amplitude $|\bm{\phi}|/v=f(r)$, respectively.
}
\label{fig:monopole}
\end{center}
\end{figure}

Let us begin with the simplest case of $N=0$ where the vacuum manifold is $SO(3)/SO(2) \simeq S^2$. 
The nontrivial second homotopy group $\pi_2(S^2)\simeq {\mathbb Z}$
ensures the existence of the global monopoles. Similarly to the global vortices, 
no analytic solutions even for a single winding monopole have been obtained, so that we 
need a numerical analysis to have a concrete solution.
The hedgehog ansatz for an radially symmetric monopole solution is given by
\be
{\bm \phi} = v f(r) \hat{\bm r},
\label{eq:hedgehog}
\ee
with $r = \sqrt{x^2+y^2+z^2}$, and $\hat{\bm r}=(x,y,z)/r$.
The EOM for $f(r)$ reads
\be
\frac{d^2 f}{dr^2} + \frac{2}{r}\frac{df}{dr} - \frac{2f}{r^2} - \lambda v^2 f \left(f^2- 1\right) = 0.
\ee
The amplitude $f(r)$ should satisfy the boundary condition
\be
f(0) = 0,\quad f(\infty) = 1.
\ee
We show a typical numerical solution in Fig.~\ref{fig:monopole}.

%%%%%%%%%%%%%

For later convenience, let us introduce the monopole charge density by
\be
{\cal M}_0 = \frac{1}{3!} \epsilon_{ijk} \epsilon_{abc} \p_i \phi_a \p_j\phi_b\p_k\phi_c,\quad(i,j,k=1,2,3;\ a,b,c = 1,2,3).
\ee
We will use this to detect the monopoles in the subsequent sections.
The topological charge of monopole is then given by
\be
M = \int d^3x\,{\cal M}_0 = \frac{4\pi v^3}{3} n,
\ee
with $n \in \pi_2(S^2) \simeq \mathbb{Z}$.

\subsection{Non-topological monopole-string composite soliton for $N=1$}

%\subsection{Monopole-antimonopole}

%We now deform the $O(3)$ symmetric Lagrangian (\ref{eq:O3_lag_4}) by adding the following potential\footnote{
%Non-static and non-topological configurations studied in this section can be thought of as a four dimensional version 
%of those found in the linearly deformed $O(2)$ model in three dimensions in Sec.~\ref{sec:nontop_vortex_wall_3d}.}
%\be
%V_1 = \alpha \phi_3.
%\ee
%We can assume $\alpha >0$ without loss of generality because the sign of $\alpha$ can be absorbed by reparametrization of $\phi_3$.
%The added term explicitly breaks $O(3)$ down to $O(2)^{(12)}$ which is the symmetry for $\phi_1$ and $\phi_2$.
%%(Remember that the remaining symmetry for the two component model with $V_1$ was $\mathbb{Z}_2^{(1)}$, see Sec.~\ref{sec:SB_AV_two}.)

We next study the $N=1$ case with $\alpha > 0$. 
The global minimum of the potential reads
$
{\bm \phi} = (0,0, -\tilde v),
$
where $\tilde v$ $(> 0)$ is the solution of
$
\lambda \tilde v (\tilde v^2 - v^2) = \alpha.
%\label{eq:v_tilde}
$
The vacuum corresponds to the south pole of the quasi vacuum manifold $S^2$, see Fig.~\ref{fig:pot_N3} (middle).
The exact symmetry 
$G = O(2)^{(12)}$ of the Lagrangian is not spontaneously broken and the above vacuum is topologically trivial. 
Therefore, there is no stable topological soliton.
However, when the smallness condition \eqref{eq:wall_weak} is satisfied, 
the $S^2$ structure remains as the quasi vacuum manifold as mentioned.
Then, there exist quasi-topological composite solitons which winds the quasi vacuum manifold $S^2$.

\begin{figure}[t]
\begin{center}
\includegraphics[width=13cm]{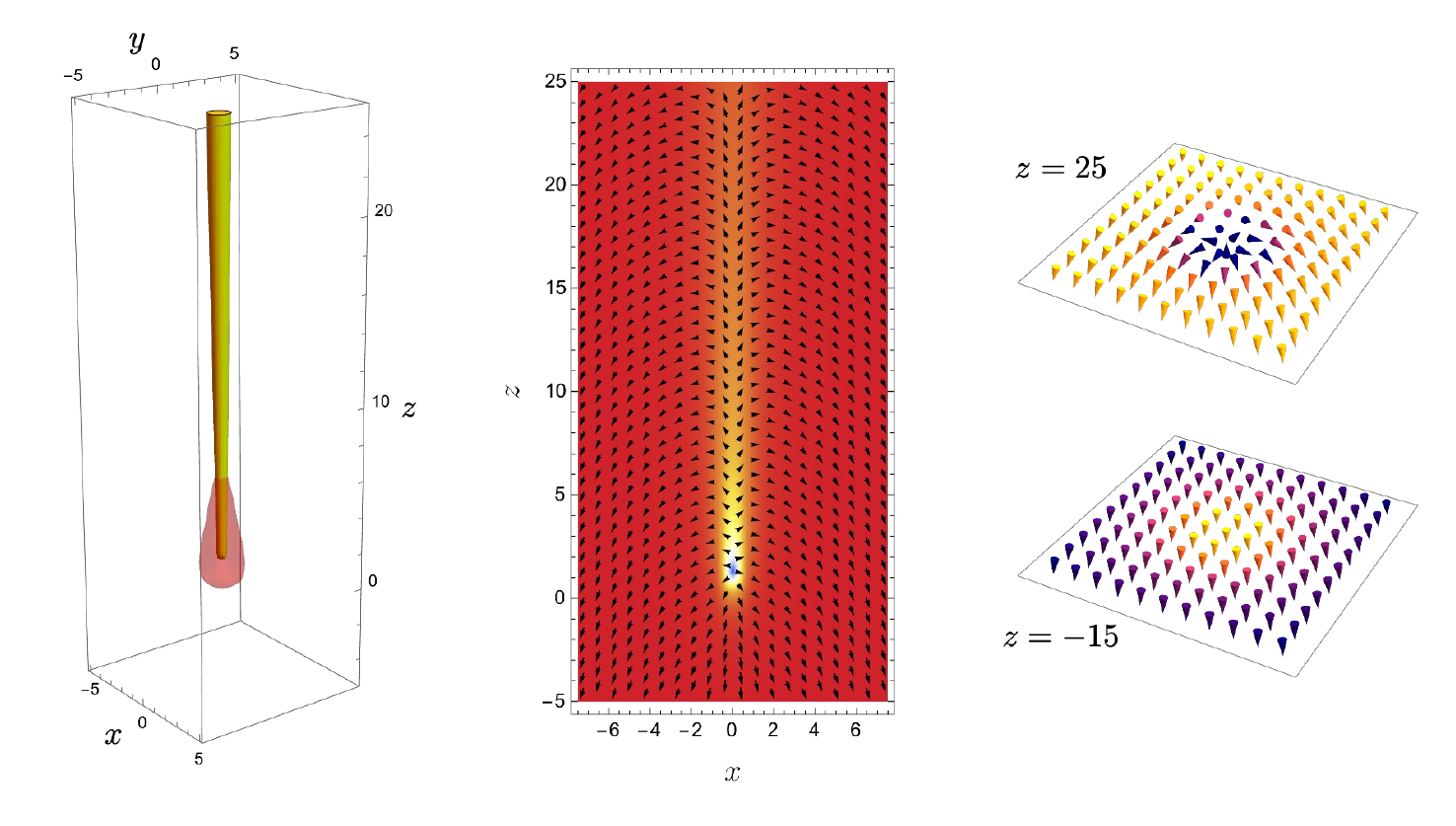}
\caption{A numerical solution of the non-topological string-monopole composite soliton
for $(v,\lambda,\alpha) = (1,4,1/4)$.
(left) The red surface shows the isosurface of monopole charge density ${\cal M}_0= 1/18$
and the yellow surface shows the isosurface of $|\phi_1+i\phi_2| = 1/2$.
(middle) The vector plot of ${\bm \phi}$ at $y=0$.
(right) The vector plots of ${\bm \phi}$ at $z = 25$ and  $z=-15$ for $x \in [-15,15]$ and $y\in [-15,15]$.}
\label{fig:1monopole_1string}
\end{center}
\end{figure}

We construct a numerical solution in a box of the size $(2L)^3$.
We should set ${\bm \phi} = (0,0,-\tilde v)$ on the five faces of the numerical box. 
On the remaining boundary, say the top face at $z = L$, we should set ${\bm \phi}(x,y,L)$ to be an $O(3)$ lump-like configuration,
\be
{\bm \phi}\big|_{z=L} = v\left( \frac{a (x+iy)+a^*(x-iy)}{\rho^2+|a|^2},\ 
\frac{a(x+iy)-a^*(x-iy)}{i(\rho^2+|a|^2)},\ \gamma\frac{|a|^2 -x^2-y^2}{\rho^2+|a|^2}\right),
\ee
with $a$ and $\gamma$ are constants. This satisfies the algebraic equation for an ellipsoid 
$
(\phi_1)^2  + (\phi_2)^2 + \gamma^{-2}(\phi_3)^2 = v^2.
$
It is, however, not easy to prepare a suitable initial configuration for the standard relaxation scheme. Therefore,
we adopt the spherical hedgehog configuration as the initial configuration and evolve it with the Neumann boundary condition
by the relaxation scheme. The result is shown in Fig.~\ref{fig:1monopole_1string}. The monopole can easily be detected by
drawing the charge density ${\cal M}_0$. It is largely deformed to be a droplet shape and attached by a single string on the upper side.
We emphasize that the string is not similar to a global vortex but to a $O(3)$ sigma model lump. Indeed, on the plane normal to the sting, 
${\bm \phi}$ points down at the boundary of the plane whereas it points up at the center of the string, 
see the right-upper panel of Fig.~\ref{fig:1monopole_1string}.
This is of course not stable since 
the string pulls the monopole.

\begin{figure}[tbp]
\begin{center}
\includegraphics[width=8cm]{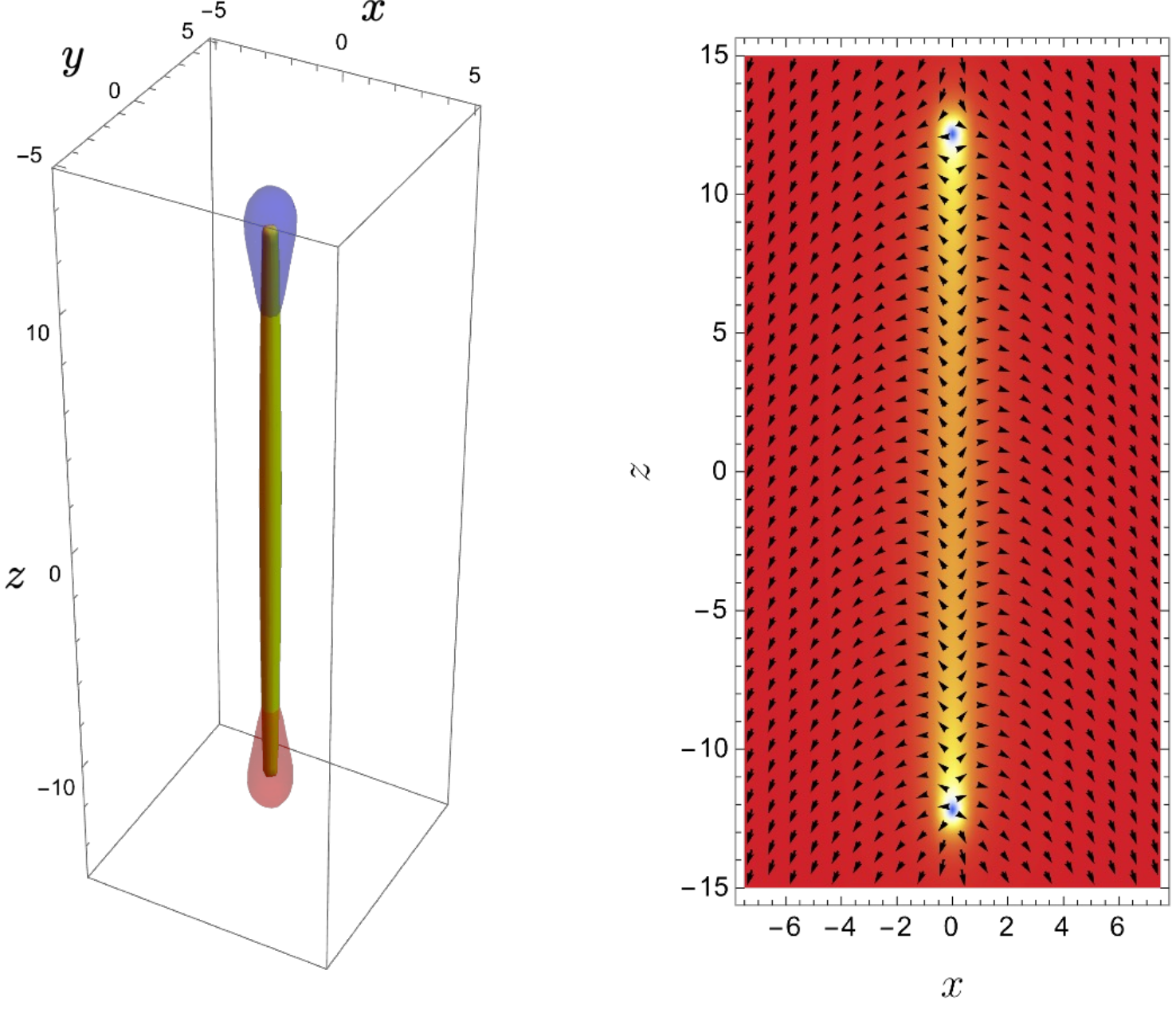}
\caption{The non-topological antimonopole-string-monopole composite soliton
for $(v,\lambda,\alpha) = (1,4,1/4)$.}
\label{fig:1monopole_1string_0}
\end{center}
\end{figure}

The string can also be terminated by anti-monopole.
We show a pair of monopole and anti-monopole connected by the string, 
a dumbbell, in Fig.~\ref{fig:1monopole_1string_0}.
Again, this is not stable. 
The monopole and anti-monopole are pulled by the string and eventually annihilate in pair. 
This configuration is called a toron 
in condensed matter physics \cite{Smalyukh:2022}.

%%%%%%%%%%
\subsubsection{Dynamical simulations for monopole-string composites for $N=1$}

We numerically simulate the real time dynamics of the monopole-string composites in the $N=1$ case.
We prepare random and periodic configurations as initial configurations for the real time dynamics 
(we smear the random configuration a little by using the relaxation scheme).
The field ${\bm \phi}$ evolves from the initial configurations with zero velocity $\dot{\bm \phi}(t=0)={\bm 0}$.
We impose the periodic boundary conditions in the $x$, $y$, and $z$ directions. 
Therefore, the numbers of the monopoles and anti-monopoles in the initial states are precisely identical. 

\begin{figure}[tbp]
\begin{center}
\includegraphics[width=10cm]{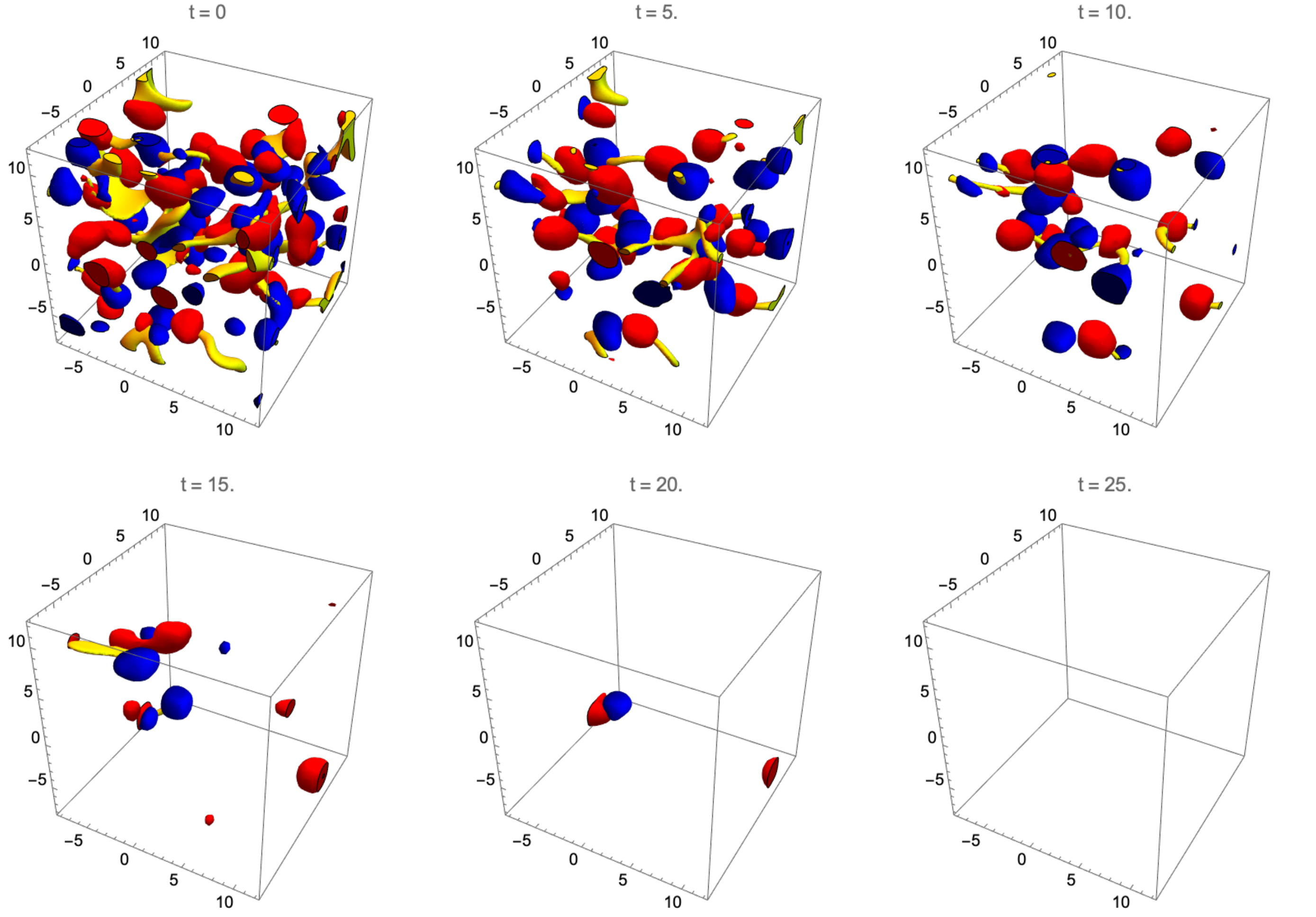}
\caption{A real time dynamics of the $N=1$ non-topological monopole-string composites 
for  $(v,\lambda,\alpha) = (1,4,1/4)$.
The red (blue) surface shows the isosurface of the monopole charge density ${\cal M}_0 = 1/24$ ($-1/24$), 
representing (anti-)monopoles.
The yellow surface shows the isosurface of $|\phi_1+i\phi_2| = 1/2$, representing non-topological strings. The initial configuration is shown in the top-leftmost panel,
and we show the five snapshots at every $\delta t = 5$ time with the unit $v$.}
\label{fig:rtd_1monopole_1string}
\end{center}
\end{figure}

A typical evolution of the $N=1$ non-topological monopole-string composites is shown in Fig.~\ref{fig:rtd_1monopole_1string}.
The top-leftmost panel shows
the initial state which is quite cluttered with many monopoles (red lumps) and anti-monopoles (blue lumps).
The strings correspond to the yellow objects but they are not initially so stringy. As soon as the time evolves, however,
the yellow objects transform into the stringy stuffs connecting one (red) monopole and one (blue) anti-monopoles. The pairs of monopole
and anti-monopole with the string annihilate very fast, and soon disappear from the numerical box.
The cosmological monopole problem
does not exist in this case.

\subsection{Topological monopole-string composite for $N=2$ with $\alpha > 0$}\label{sec:N=2,alpha>0}

%We next deform the $O(3)$ symmetric Lagrangian (\ref{eq:O3_lag_4}) by adding the quadratic term of $\phi_3$
%\be
%V_2 = \alpha_2 (\phi_3)^2.
%\label{eq:V2}
%\ee
%This term explicitly breaks $O(3)$ symmetry down to  
%$O(2)^{(12)} \times \mathbb{Z}_2^{(3)}$ where the $O(2)^{(12)}$ is rotation among $\phi_1$ and $\phi_2$, and
%$\mathbb{Z}_2^{(3)}$ flips the sign of $\phi_3$. 
%(This is a straightforward extension of the remaining symmetry 
%for the two component model with $V_2$ was $\mathbb{Z}_2^{(1)} \times \mathbb{Z}_2^{(2)}$, see Sec.~\ref{sec:SB_AV_two}.)
%Hence, no topologically stable global monopoles exist as in the previous section with the linear perturbation.
%
%As we will shortly show, the vacuum structures of the three component model for $\alpha_2 > 0$ and $\alpha_2 < 0$ are 
%significantly different from those for the two component model studied in Sec.~\ref{sec:SB_AV_two}.
%Therefore, we need to deal with $\alpha_2 > 0$ and $<0$, separately.
%
%
%\subsection{Topological global strings for $\alpha_2 > 0$}
In this subsection, we study the 
$N=2$ cases with $\alpha > 0$.

\subsubsection{Topological string with $\mathbb{Z}_2^{(3)}$ charge}

We first focus on topological strings in the $N=2$ case with $\alpha > 0$ (the easy-plane potential). 
The vacuum manifold is $S^1$ of the radius $v$ in the $\phi_1\phi_2$ 
plane which corresponds to the equator of the unperturbed $S^2$ with 
$\alpha = 0$, see Fig.~\ref{fig:pot_N3} (right panel with $\alpha>0$).
%\begin{figure}[t]
%\begin{center}
%\includegraphics[width=10cm]{deformed_V_2}
%\caption{The vacuum manifold under the presence of $V_2$.}
%\label{fig:deformed_V_2}
%\end{center}
%\end{figure}
As is summarized in Table~\ref{tab:2}, the exact symmetry of the Lagrangian $G=O(2)^{(12)} \times \mathbb{Z}_2^{(3)}$
is spontaneously broken to $H = \mathbb{Z}_2^{(12)} \times \mathbb{Z}_2^{(3)}$.
Therefore, it gives rise to a sort of global strings characterized by
$\pi_1(G/H) = \pi_1(SO(2)^{(12)}) = \mathbb{Z}$.

Let us consider a straight global string extending along the $z$ axis.
Namely, we impose $\p_0 \phi_{1,2,3} = \p_3 \phi_{1,2,3}= 0$ and set the following boundary condition 
\be
\phi_1 + i \phi_2 \to v e^{i\theta},\quad \phi_3 \to 0,\qquad \text{as}\quad \rho \to \infty,
\label{eq:bc_string_3comp}
\ee
%It turns out that there are qualitatively different two string solutions depending 
%on  $|\alpha_2|$.
%In order to understand this point, we recall deformations of the scalar potential 
%given in Fig.~\ref{fig:pot01}(c1) and (c2) for the two component field ${\bm \phi} = (\phi_1,\phi_2)$ in the previous section.
%There we have observed that  the quasi $S^1$ structure remains at the bottom of scalar potential.
%Similarly, the quasi $S^2$ structure remains at the bottom of scalar potential for the three component field 
%if the additional potential $V_2$ is sufficiently weak as
%\be
%|\alpha_2| \ll \lambda v^2.
%\label{eq:smallness2}
%\ee
which describes a map with the winding number unity from a loop on the $xy$ plane, $\theta \in [0,2\pi)$, to a loop in the field internal space, $\mathrm{arg} \, (\phi_1 + i\phi_2) \in [0,2\pi)$.
On the other hand, as approaching the origin of the $xy$ plane,  $\rho=0$,
the loop in the field space must be contracted for regularity, i.e., $\phi_1 = \phi_2=0$.
Since the $S^2$ structure remains as the quasi vacuum manifold of the potential due to the small coupling condition (\ref{eq:wall_weak}),
${\bm \phi}$ energetically prefers to leave off from the $\phi_3=0$ plane at $\rho=0$,
resulting in that the contraction can take place at two points: the north pole ($\phi_3>0$) or the south pole ($\phi_3 <0 $) of $S^2$.
For the former (latter) case, as $\rho$ varying from $\rho \to \infty$ to $\rho = 0$,
the loop in the internal space sweeps the upper (lower) hemisphere of the quasi vacuum manifold $S^2$.
%(old version:)
%Since we are assuming the small coupling regime (\ref{eq:wall_weak}), the $S^2$ structure remains as the quasi vacuum manifold of the scalar potential.
%Hence, for the configuration satisfying the boundary condition (\ref{eq:bc_string_3comp}), 
%the winding of the $\phi_1 + i \phi_2$ tends to 
%${\bm \phi}$ prefers to leave off from the $\phi_3=0$ plane and taking a detour through the quasi $S^2$.
%Now, there are two choices for the detour: closing the vacuum manifold $S^1$ by the upper or lower hemisphere.
%
In other words, the third component $\phi_3$ positively or negatively condenses inside the global string.
Consequently, the $\mathbb{Z}_2^{(3)}$ symmetry which is unbroken in the vacuum  is spontaneously broken by the global string as
\be
O(2)^{(12)} \times \mathbb{Z}_2^{(3)} 
\xrightarrow[]{\ \ \text{vac}\ \ }
\mathbb{Z}_2^{(12)} \times \mathbb{Z}_2^{(3)}
%\xrightarrow[]{\ \ \text{string}\ \ }
\dotarrow{\ \ \text{string}\ \ }
1.
\label{eq:O(2)xz2toz2to1}
\ee 
Hence, the topological string in the weak coupling regime has the $\mathbb{Z}_2^{(3)}$ charge.

On the contrary, in a regime where $|\alpha|$ is so large that the condition (\ref{eq:wall_weak}) is not satisfied, 
the $S^2$ structure completely disappears in the scalar potential. 
Then the contraction of the loop can occur without leaving the $\phi_3=0$ plane.
Therefore, $\mathbb{Z}_2^{(3)}$ is unbroken and the global string with large $|\alpha|$ is neutral 
under the $\mathbb{Z}_2^{(3)}$ transformation.
%% old version:
%Then ${\bm \phi}$ satisfying the boundary condition (\ref{eq:bc_string_3comp}) closes the equator with the flat disk at $\phi_3=0$.
%Therefore, $\mathbb{Z}_2^{(3)}$ is unbroken and the global string in the strong coupling regime is neutral 
%under the $\mathbb{Z}_2^{(3)}$ transformation.
%%

Let us make an appropriate ansatz for the axially symmetric string with the minimal winding number 
\be
\phi_1 + i \phi_2 = v f(\rho) e^{i\theta},\quad \phi_3 = v g(\rho).
\label{eq:ansatz_global_string}
\ee
Then, the EOMs read
\be
\frac{d^2f}{d\rho^2} + \frac{1}{\rho}\frac{df}{d\rho} - \frac{f}{\rho^2} - \lambda v^2 f(f^2+g^2-1) = 0,\\
\frac{d^2g}{d\rho^2} + \frac{1}{\rho}\frac{dg}{d\rho} - \lambda v^2 g(f^2+g^2-1) - 2\alpha_2 g = 0.
\ee
Note that $g=0$ solves the second equation, and the first one with $g=0$ is identical to
that we studied in Eq.~(\ref{eq:EOM_global_string}) for the global vortices.
However, of course, we should not assume $g=0$ from the beginning when  it is an unstable solution. 
We numerically solve these with the boundary condition
\be
f(0) = 0,\quad f(\infty) = 1,\quad \frac{dg}{d\rho}\bigg|_{\rho=0} = 0,\quad g(\infty) = 0.
\ee

\begin{figure}[h]
\begin{center}
\includegraphics[width=12cm]{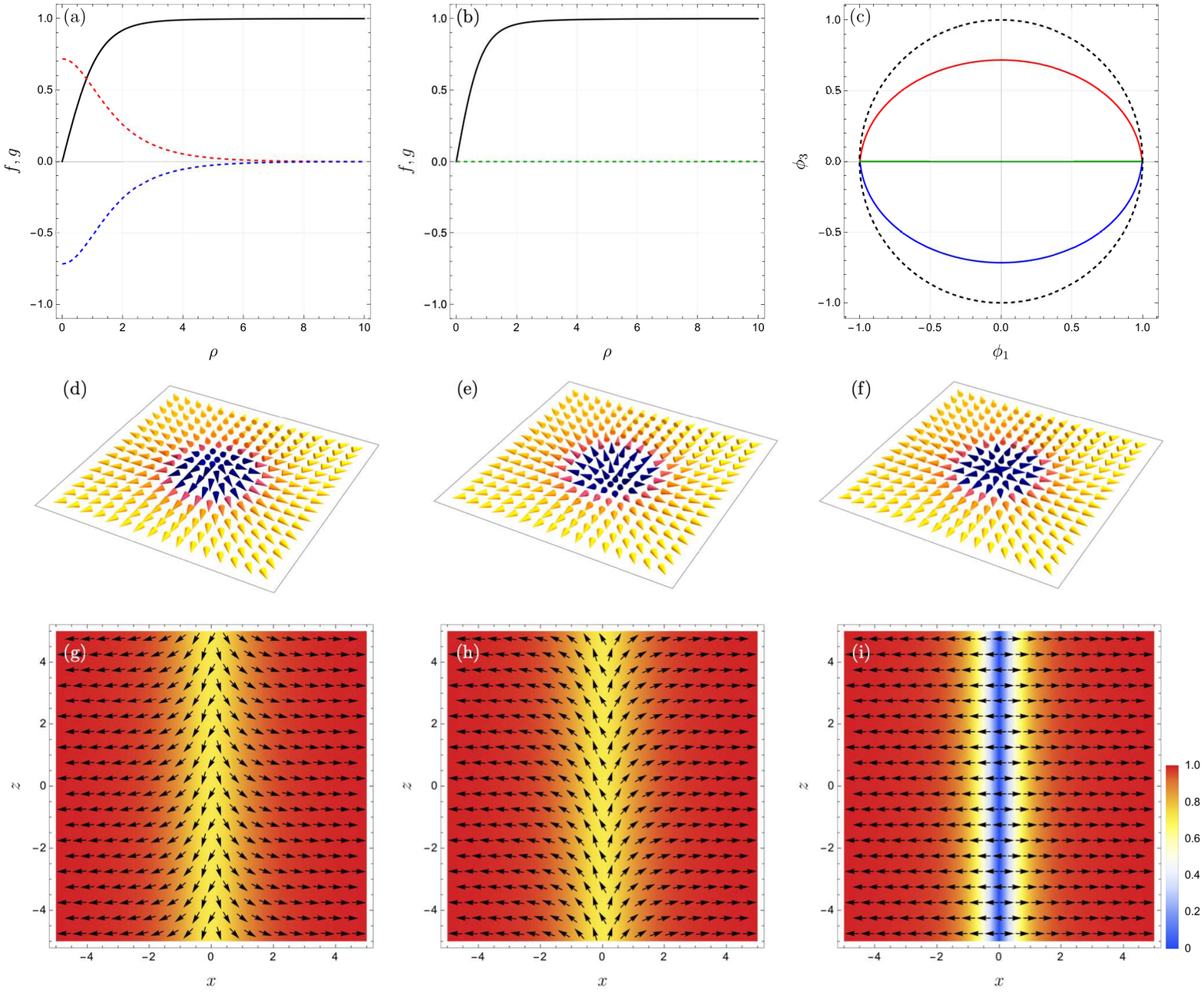}
\caption{The axially symmetric string solutions for $(\lambda,v) = (4,1)$ and $\alpha=1/4$ for (a) and $\alpha = 1$ for (b). 
(a) The $\mathbb{Z}_2^{(3)}$ charged global strings. The black-solid curve shows  
$f(\rho)$ and the red-dashed and blue-dashed curves corresponds to $g(\rho)$.
(b) The $\mathbb{Z}_2^{(3)}$ neutral global string. The black-solid curve corresponds to $f(\rho)$ and the green-dashed
line shows $g(\rho)=0$.
(c) The parametric plots of the solutions given in (a) and (b) in the $\phi_1\phi_3$ plane ($\phi_2=0$). 
The black-dashed curve indicates the unperturbed order parameter space $S^2$ at $\alpha=0$.
(d), (e), (f) show the 3d vector plot of ${\bm \phi}$ on the $xy$ plane for the sting solutions of 
(a) with the blue-dashed curve, red-dashed curve,
and (b), respectively. (g) [(h)] shows the 3d vector plot on the slices $x=0$, $y=0$, and $z=0$ corresponding
to (d) [(e)].}
\label{fig:string_z2}
\end{center}
\end{figure}
Fig.~\ref{fig:string_z2}(a) shows numerical solutions in the weak coupling regime,
%The black-solid curve shows $f(\rho)$ whereas the red-dashed (blue-dashed)
%curve corresponds to  $g(\rho)$. 
and (b) shows the $\mathbb{Z}_2^{(3)}$ neutral string in the strong coupling regime. 
(c) shows the solutions represented in the $\phi_1\phi_3$ plane.
%The colors of curve match with those of the dashed curves in (a) and (b). 
The red (blue) curve does not lie on the
upper (lower) hemisphere (black-dashed curve) 
but it lies on a slightly squashed upper (lower) hemisphere, namely the quasi vacuum manifold $S^2$.
On the contrary, the green curve corresponds to the solution with large $|\alpha|$ and has $\phi_3=0$ everywhere.  
As can be seen in (d), (e), (f),  
it is common among the three string solutions that ${\bm \phi}$ radially spreads as the 2 dimensional hedgehog in asymptotic far region,
namely $(\phi_1,\phi_2) \to v (\cos\theta,\sin\theta)$ with $\phi_3 \to 0$. This ensures the winding number $n=1$
around the vacuum manifold $S^1$.
On the other hand, they show different behaviors near the center of string: $\bm{\phi}$ points down and up for (d) and (e)
whereas $\phi_3$ for (f) is everywhere zero. 
This reflects the $\mathbb{Z}_2^{(3)}$ charge of the strings; (d) and (e) are charged whereas (f) is neutral.

The $\mathbb{Z}_2^{(3)}$ charged strings can be also interpreted as textures (half-lumps) wrapping the quasi vacuum manifold $S^2$ whose winding number is half quantized
because they cover a half of the quasi vacuum manifold $S^2$. 
Such half lumps are also called 
merons in condensed matter physics.
However, note that the wrapping number as the texture is not a conserved topological charge 
but an approximate charge.
The topological nature comes from $\pi_1(S^1) \simeq \mathbb{Z}$ together with the $\mathbb{Z}_2^{(3)}$ charge.
%Finally, the panels (g), (h) and (i) show the two dimensional vector plots 
%of ${\bm \phi}$ on the slice at $y=0$.

\begin{figure}[tbp]
\begin{center}
\includegraphics[width=6.5cm]{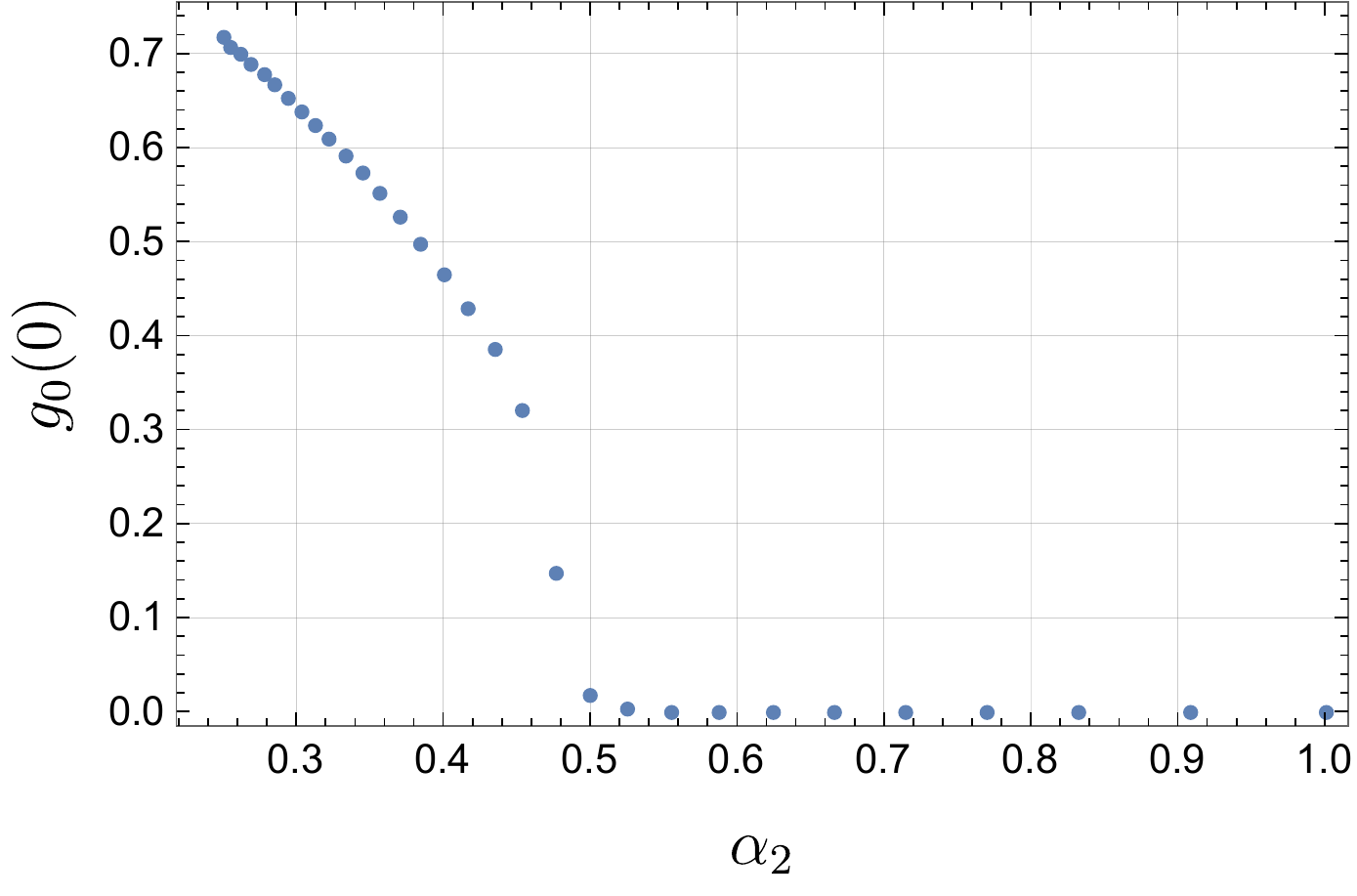}
\caption{The value of $g_0(0)$ as the function of $\alpha_2$. The string is coreless in the sense of $|\bm{\phi}|\neq 0$ if $g_0(0) \neq 0$, otherwise it has a core.
The other parameters are fixed as $(v,\lambda) = (1,4)$.}
\label{fig:core_vs_coreless}
\end{center}
\end{figure}

In order to determine when the global strings become $\mathbb{Z}_2^{(3)}$ charged, 
we measure $g|_{\rho =0}$ by varying $\alpha$ while $v$ and $\lambda$ are fixed. The result is shown in Fig.~\ref{fig:core_vs_coreless}.
As is consistent with Eq.~(\ref{eq:wall_weak}), the $\mathbb{Z}_2^{(3)}$ charged string appears in the weak coupling regime.
Therefore this figure shows  
the phase transition (of Ising spins)
inside the vortex.

%%%%%%%%%%
%%%%%%%%%%%

In the limit of  $\lambda \to \infty$, 
the model reduces to the  $O(3)$ nonlinear sigma model 
with an easy-plane potential.
In this limit,  
a lump charge of $\pi_2(S^2) \simeq {\mathbb Z}$ becomes 
 exact in contrast to 
the linear model for which the lump 
charge is only approximate. 
\if0 %%%
However, lumps are unstable to shrink and become singular from the Derick's scaling argument 
\cite{Derrick:1964ww}. 
In order to avoid it,
\fi %%%
If we add 
four-derivative Skyrme terms, 
the model is a baby-Skyrme model 
with an easy-plane potential, 
admitting stable 
half-baby Skyrmions (merons) 
\cite{Jaykka:2010bq,Kobayashi:2013aja,Kobayashi:2013wra}.

\subsubsection{Topological monopole-string composite: a kink inside a string}

Now we are ready to study the global monopoles in the presence of 
a perturbation $V_2$ with positive and small $\alpha$.
An important clue is that the spontaneous breaking of $\mathbb{Z}_2^{(3)}$ inside the string.
Just as a usual $\mathbb{Z}_2$ breaking in $1+1$ dimensions gives rise to a topological domain wall, the domain wall is spontaneously generated on the string. 
This domain wall is a point-like object from the perspective of 3+1 dimensions in the bulk, 
and it is a junction point which connects the two strings with the different
$\mathbb{Z}_2^{(3)}$ charges.
It is nothing but a global monopole, which is deformed by $V_2$,
attached by the two strings on the opposite sides. 
Notice that $\pi_2(S^2)$ is  only approximately meaningful for characterizing the composite solitons since $S^2$ is not the exact vacuum manifold. 
Alternatively, the combination of $\pi_1(S^1) \simeq \mathbb{Z}$ for the mother strings  and $\mathbb{Z}_2^{(3)}$ for the domain wall inside it is more appropriate to characterize the composite soliton.
This is summarized by the sequential symmetry breaking
%\me{
\begin{alignat}{3}
G_{\rm approx}=&O(3) \quad &\xrightarrow[\ \ \text{monopole}\ \ ]{\ \ \text{vac}\ \ }\quad &O(2) & \nonumber \\
&\cup & &\cup & \nonumber \\
G_{\rm exct}=&
O(2)^{(12)} \times \mathbb{Z}_2^{(3)} 
\quad &\xrightarrow[\ \ \text{mother string}\ \ ]{\ \ \text{vac}\ \ }\quad
&\mathbb{Z}_2^{(12)} \times \mathbb{Z}_2^{(3)}
\quad &\xrightarrow[\ \ \text{kink}\ \ ]{\ \ \text{mother string}\ \ }\quad
1.
\label{eq:SSB_sequence_monopole_string}
%\label{eq:O(2)xz2toz2to1}
\end{alignat}

With the picture that the monopole is the kink in the string at hand, we are lead to a natural product ansatz
\be
{\bm \phi} = v\left(
\begin{array}{c}
\frac{x}{\sqrt{\rho^2 + a^2}}  \\
\frac{y}{\sqrt{\rho^2 + a^2}} \\
\frac{c a}{\sqrt{\rho^2 + a^2}} \tanh b z 
\end{array}
\right),\label{eq:kink-on-string}
\ee
with $a$, $b$ and $c$ are constants.
See Appendix~\ref{sec:appendix} for the validity of this ansatz. 
The $z$ dependence comes from the kink while the $x,y$ dependence exhibits the $\mathbb{Z}_2^{(3)}$ charged string.
In fact,
for $z \gg 0$ and $z \ll 0$, we have
\be
{\bm \phi}\big|_{z\gg0} = v\left(
\begin{array}{c}
\frac{x}{\sqrt{\rho^2 + a^2}}  \\
\frac{y}{\sqrt{\rho^2 + a^2}} \\
\frac{c a}{\sqrt{\rho^2 + a^2}}
\end{array}
\right),\quad
{\bm \phi}\big|_{z\ll0} = v\left(
\begin{array}{c}
\frac{x}{\sqrt{\rho^2 + a^2}}  \\
\frac{y}{\sqrt{\rho^2 + a^2}} \\
-\frac{c a}{\sqrt{\rho^2 + a^2}}
\end{array}
\right).
\ee
These are nothing but the $\mathbb{Z}_2^{(2)}$ charged mother strings which
wrap the upper- and lower-half hemispheres of the quasi-vacuum manifold $S^2$ (ellipsoid).
On the other hand, at the center of string $(x=y=0)$ we have
\be
{\bm \phi}\big|_{x=y=0} = v\left(
\begin{array}{c}
0\\
0 \\
\frac{c a}{\sqrt{\rho^2 + a^2}} \tanh b z 
\end{array}
\right).
\ee
This is clearly the inner kink which embodies the flip of the $\phi_3$ condensation.

\begin{figure}[tbp]
\begin{center}
\includegraphics[width=12cm]{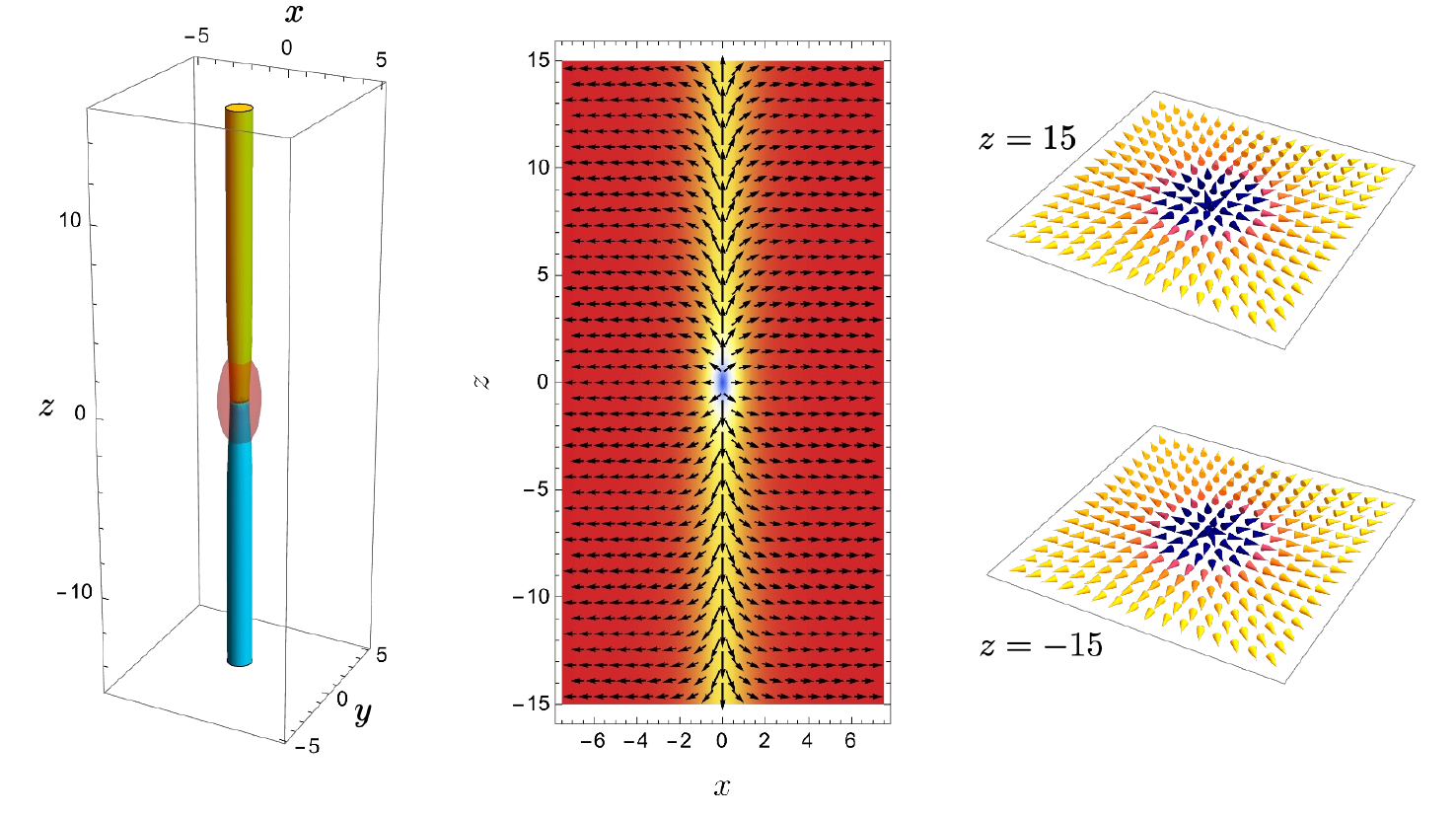}
\caption{A numerical solution of the topological monopole-string composite for $(v,\lambda,\alpha_2) = (1,4,1/4)$.
(left) The red surface is the isosurface of ${\cal M}_0 = 1/24$.
The yellow and cyan surfaces show $|\phi_1+i\phi_2| = 1/2$ with $\phi_3 > 0$
and $\phi_3 < 0$, respectively.
(middle) The vector plot of ${\bm \phi}$ at $y=0$.
(right) The vector plot of ${\bm \phi}$ at $z=\pm 15$.}
\label{fig:1monopole_2string}
\end{center}
\end{figure}
We can make use of the above ansatz as an initial configuration for the relaxation scheme.
The numerical solution obtained at the convergence of relaxation scheme is shown in Fig.~\ref{fig:1monopole_2string}.
Note that the final state is independent of the parameters of the initial configurations.

\begin{figure}[tbp]
\begin{center}
\includegraphics[width=3.5cm]{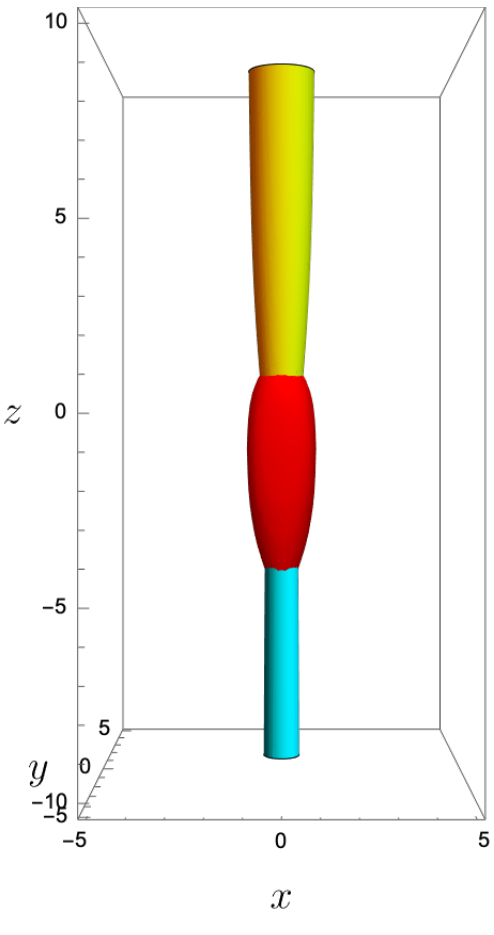}
\caption{Asymmetric monopole-string composite.
The tension of strings are different. This configuration is unstable since the string with larger tension continuously pulls the monopole.
}
\label{fig:ms_asym}
\end{center}
\end{figure}

So far, we have considered the ESB potentials
which includes only linear or quadratic terms. The linear potential leads to
the non-topological composite solitons while the quadratic one allows the topologically
nontrivial composite solitons. If we mix the linear and quadratic terms, some symmetries
are explicitly broken and it leads to imbalance of tension of the constituent solitons.
Here, we give only one example:
\be
V_{1+2} = \alpha_1 \phi_3 + \alpha_2 (\phi_3)^2,
%= \alpha_2\left(\phi_3- \frac{\alpha_1}{2\alpha_2}\right)^2
%- \frac{\alpha_1^2}{4\alpha_2},
\label{eq:V12}
\ee
For the parameter $(v,\lambda,\alpha_1,\alpha_2) = (1,4,-1/10,1/4)$,
we have found the monopole-string composite but the two strings have different tensions, see Fig.~\ref{fig:ms_asym}. 
Thus, the monopole cannot be static.
This resembles the unstable Nambu monopole in the two Higgs doublet model~\cite{Eto:2020hjb}.
%%%%%%
%%%%%%%
If we consider more generic ESB  potential 
beyond the quadratic order 
in the fields, more complicated composite solitons should emerge.
We will report the details elsewhere.

As mentioned in introduction, 
a local-soliton version of this composite soliton 
is a local monopole (`t Hooft-Polyakov monopole 
\cite{tHooft:1974kcl,Polyakov:1974ek}) 
attached by two 
non-Abelian vortex strings, 
in which 
monopoles are realized 
as kinks on the vortex 
\cite{Tong:2003pz,Shifman:2004dr,Hanany:2004ea,Nitta:2010nd,Eto:2011cv}. 
This is realized 
by introducing an ESB term for 
a global symmetry that 
is locked with the gauge symmetry in the vacuum.
As a version of textures,
Skyrmions inside a vortex string 
as sine-Gordon solitons 
were also studied in Refs.~\cite{Gudnason:2014hsa,Gudnason:2016yix}.

%%%%%%%%%%
\subsubsection{Dynamical simulations for monopole-string composites for $N=2$}

We numerically simulate the real time dynamics of the monopole-string composites in the $N=2$ case.
As the case of $N=1$, 
We prepare random and periodic configurations as initial configurations for the real time dynamics and smear it by using the relaxation scheme.
Again,
We impose the periodic boundary conditions in the $x$, $y$, and $z$ directions.

\begin{figure}[tbp]
\begin{center}
\includegraphics[width=12cm]{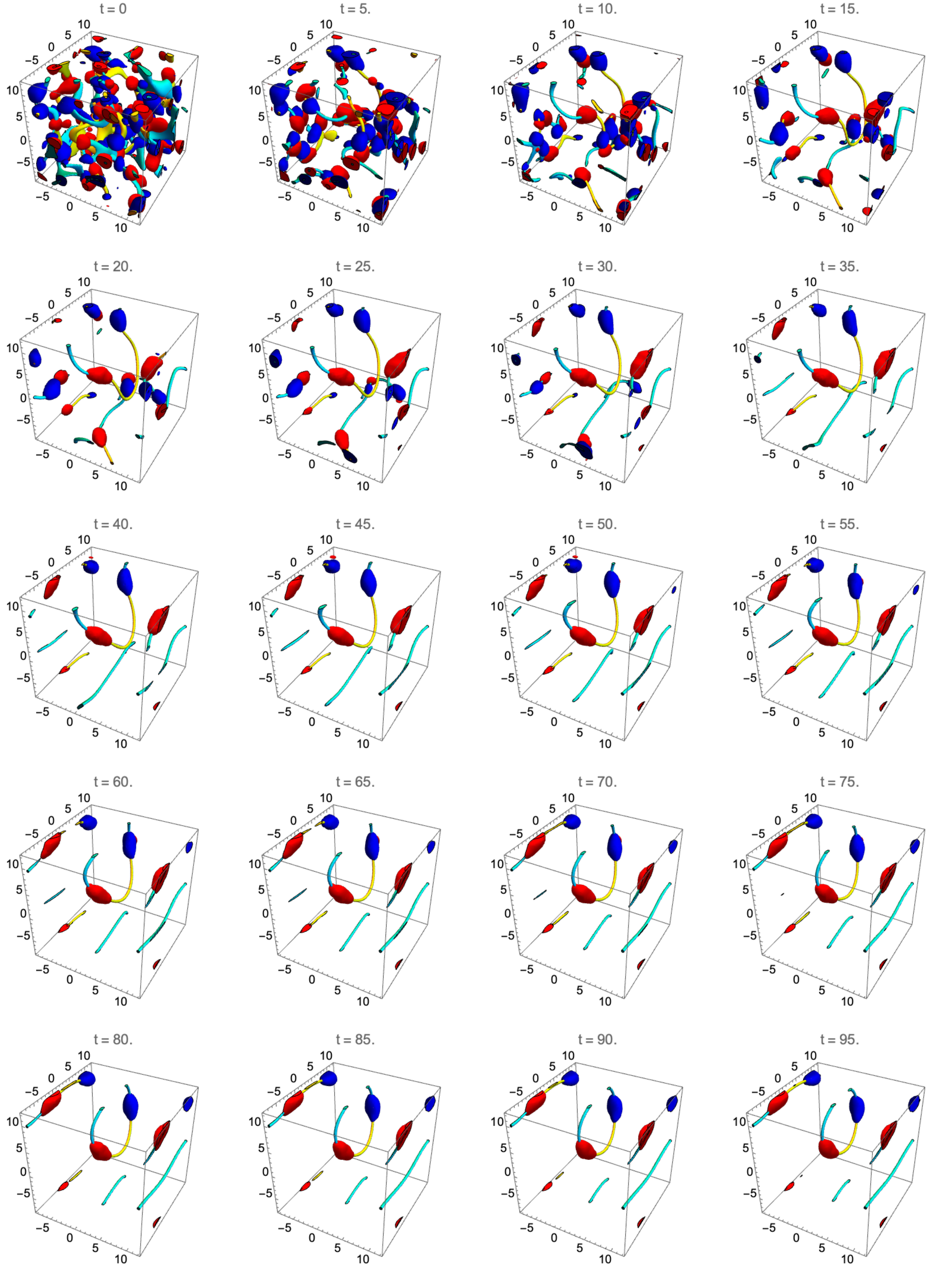}
\caption{$(v,\lambda,\alpha) = (1,4,1/4)$.
The red (blue) surface shows the isosurface of the monopole charge density ${\cal M}_0 = 1/24$ ($-1/24$), 
representing (anti-)monopoles.
The yellow and cyan surfaces show the isosurface of $|\phi_1+i\phi_2| = 1/2$ with $\phi_3 > 0$
and $\phi_3 < 0$, respectively, 
representing topological strings 
with up and down spins, respectively.
}
\label{fig:simulation_n2_non_spherical_monopole_vortex_01}
\end{center}
\end{figure}

A numerical simulation for the $N=2$ topological monopole-string composites is shown in
Fig.~\ref{fig:simulation_n2_non_spherical_monopole_vortex_01}.
The top-leftmost panel shows an initial configuration with random distributions of the monopoles and anti-monopoles,
and the $\mathbb{Z}_2^{(3)}$ charged strings.
We keep the initial number of monopoles to be the same order both for 
Figs.~\ref{fig:rtd_1monopole_1string} and
\ref{fig:simulation_n2_non_spherical_monopole_vortex_01}.
In comparison with Fig.~\ref{fig:rtd_1monopole_1string},
the life time of the (anti-)monopoles is relatively long. The life time of the strings is as long as that of monopoles.
This is due to the $\mathbb{Z}_2^{(3)}$ symmetry which ensures that the tensions of the strings with different 
$\mathbb{Z}_2^{(3)}$ charged are exactly equal. 
The cosmological monopole problem
may exist in this case.

\clearpage

\subsection{Topological monopole-wall composite  for $N=2$ with $\alpha < 0$}

%Next, we consider the potential (\ref{eq:V2}) with $\alpha_2 < 0$ that is the easy axis type. 
%There are two discrete vacua corresponding
%to the north and south poles of the $S^2$ at $\alpha_2 = 0$,
%see Fig.~\ref{fig:deformed_V_2}.
%The symmetry breaking pattern at the vacua reads
%\be
%O(2)^{(12)} \times \mathbb{Z}_2^{(3)} \to O(2)^{(12)}.
%\label{eq:SSB_negative_alpha2}
%\ee
%Compared this with Eq.~(\ref{eq:SSB_positive_alpha2}) for $\alpha_2 > 0$, the discrete symmetry is broken 
%which gives rise to  topological domain walls.
In this subsection, we study the 
$N=2$ cases with $\alpha < 0$ 
(the easy-axis potential). 

\subsubsection{The topological domain wall with $SO(2)^{(12)}$ moduli}
\label{sec:N=2,alpha<0}

Let us consider the Lagrangian with $V_2$ for $\alpha<0$.
As before, the $S^2$ structure remains as the quasi vacuum manifold of the potential in the weak coupling regime satisfying Eq.~(\ref{eq:wall_weak}).
The genuine vacuum manifold 
consists of the two points as shown in Fig.~\ref{fig:pot_N3} (the right panel with $\alpha < 0$).
As is summarized in Table~\ref{tab:2}, the discrete symmetry $\mathbb{Z}_2^{(3)}$ is spontaneously broken so that
there exist topologically stable domain walls.

If we impose $\phi_2 = 0$ which is consistent with EOMs, the three component field reduces 
to the two components as ${\bm \phi} \to (\phi_1,0,\phi_3)$. Then, the EOMs for $(\phi_1,\phi_3)$ are the same as those studied in Sec.~\ref{sec:tdw_z2} replacing $\phi_3$ by $\phi_2$. 
Hence, the domain wall solution 
shown in Fig.~\ref{fig:wall_n2} (a) 
in the linear $O(2)$ model 
can be embedded without any change.

\begin{figure}[tbp]
\begin{center}
\includegraphics[height=3cm]{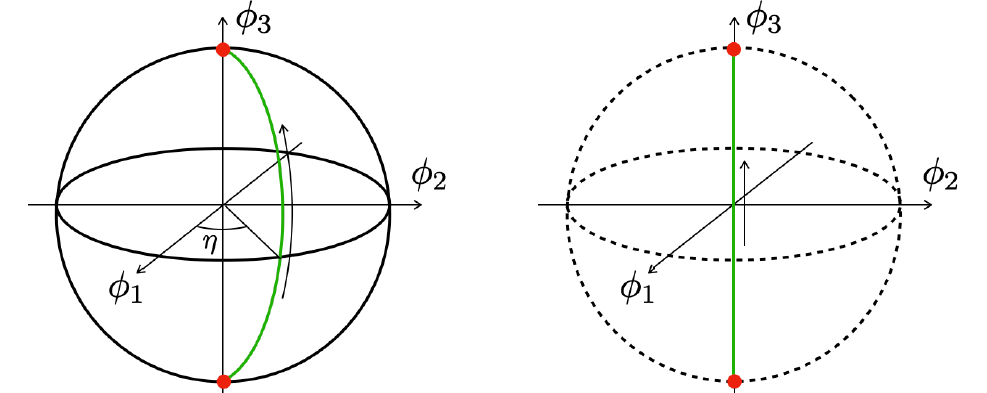}
\caption{The left figure shows a domain wall with $SO(2)$ moduli whereas the right one corresponds to the domain wall without
internal moduli.}
\label{fig:wall_orientation}
\end{center}
\end{figure}

There is an important difference between the $O(2)$ and $O(3)$ models.
In the former case, the spontaneously broken symmetry in  the presence of the domain wall is $\mathbb{Z}_2^{(2)}$,
whereas it is $O(2)^{(12)}$ which breaks down into $\mathbb{Z}_2^{(12)}$ in the latter as
\be
O(2)^{(12)} \times \mathbb{Z}_2^{(3)}
\xrightarrow[]{\ \text{vac}\ }
O(2)^{(12)} 
%\xrightarrow[]{\ \text{wall}\ }%(|\alpha| v^2/\lambda)^{1/4}\ }
\dotarrow{\ \text{wall}\ }
\mathbb{Z}_2^{(12)} .
\label{eq:ssb_wall}
\ee
Hence, the domain wall carries the continuous moduli $O(2)^{(12)}/\mathbb{Z}_2^{(12)} = SO(2)^{(12)} \simeq S^1$.
This is intuitively understood as follows. 
When one connects the two vacua,
the path passes through a point on the equator (great circle with $\phi_3=0$,
which is parameterized by an azimuth angle $\eta \in [0,2\pi)$ as shown in Fig.~\ref{fig:wall_orientation} (left).
$\eta$ is nothing but the Nambu-Goldstone mode localized on the domain wall.
\if0 %%%
\footnote{ 
In fact, this is a linear 
$O(3)$ model version 
of a domain wall in the nonlinear $O(3)$ model with an easy-axis potential, equivalent to
the massive ${\mathbb C}P^1$ model 
\cite{Abraham:1992vb,Abraham:1992qv,Arai:2002xa,Arai:2003es}.
}
\fi %%%

%%%%%%%%

On the contrary, the domain wall does not have any moduli in the strong coupling regime where the original $S^2$ structure completely disappears. The domain wall solution corresponds to a straight segment along the $\phi_3$ axis connecting the two vacua with $\phi_1+i\phi_2 = 0$ as illustrated
in Fig.~\ref{fig:wall_orientation} (right). This is neutral under $SO(2)^{(12)}$. 
The numerical solution is the same as those given in Fig.~\ref{fig:wall_n2} (b).

In the limit of  $\lambda \to \infty$, 
the model reduces to the $O(3)$ nonlinear sigma model 
with an easy-axis potential, 
equivalent to
the so-called massive ${\mathbb C}P^1$ model 
\cite{Abraham:1992vb,Abraham:1992qv,Arai:2002xa,Arai:2003es}.
It admits a domain wall with a $U(1)$ modulus.

%%%%%
\subsubsection{Topological monopole-wall composite: a vortex inside a domain wall}

Let us next study a global monopole 
in  the presence $V_2$ with negative $\alpha$ in the weak coupling regime.
As is given in Eq.~(\ref{eq:ssb_wall}), the $SO(2)^{(12)}$ symmetry of the vacuum is spontaneously broken
by the topological domain wall whose world volume is a $2+1$ dimensional  surface.
%Just as a usual $SO(2)$ breaking in $2+1$ dimensions, 
This SSB inside the domain wall gives rise to a topological vortex,
which is localized on the wall and looks a point-like object from the 3+1 dimensional bulk perspective.
This is nothing but a global monopole 
in the bulk. 
One can also regard this as the global monopole deformed by $V_2$ and immersed into a domain wall.
As before, the homotopy group  $\pi_2(S^2) \simeq {\mathbb Z}$ 
for monopoles is only approximately meaningful for characterizing the composite solitons. 
Alternatively, the combination of $\mathbb{Z}_2^{(3)}$ for the mother domain wall and
$\pi_1(SO(2)^{(12)}) = \mathbb{Z}$ for the vortices inside it
is more appropriate to characterize it.
This can be briefly summarized by 
a sequence of symmetry breaking of the system 
\begin{alignat}{3}
G_{\rm approx}=& O(3) \quad &\xrightarrow[\ \text{monopole}\ ]{\ \text{vac}\ } \quad &O(2) &\nonumber\\
& \cup & &\cup &\nonumber\\
G_{\rm exact}=&
O(2)^{(12)} \times \mathbb{Z}_2^{(3)}
\quad &\xrightarrow[\ \text{mother wall}\ ]{\ \text{vac}\ } \quad
&O(2)^{(12)} 
\quad&\xrightarrow[\ \text{vortex}\ ]{\ \text{mother wall}\ }\quad%(|\alpha| v^2/\lambda)^{1/4}\ }
1.
%\label{eq:ssb_wall}
\label{eq:SSB_sequence_monopole_wall}
\end{alignat}

%\be
%O(2)^{(12)} \times \mathbb{Z}_2^{(3)}
%\xrightarrow[\ \text{host wall}\ ]{\ \text{vac}\ }
%O(2)^{(12)} 
%\xrightarrow[\ \text{guest string (monopole)}\ ]{\ \text{host wall}\ }%%%%(|\alpha| v^2/\lambda)^{1/4}\ }
%\mathbb{Z}_2^{(12)} \changed{(I)?}.
%%\label{eq:ssb_wall}
%\label{eq:SSB_sequence_monopole_wall}
%\ee

%%%%%%%%
\begin{figure}[tbp]
\begin{center}
\includegraphics[width=12cm]{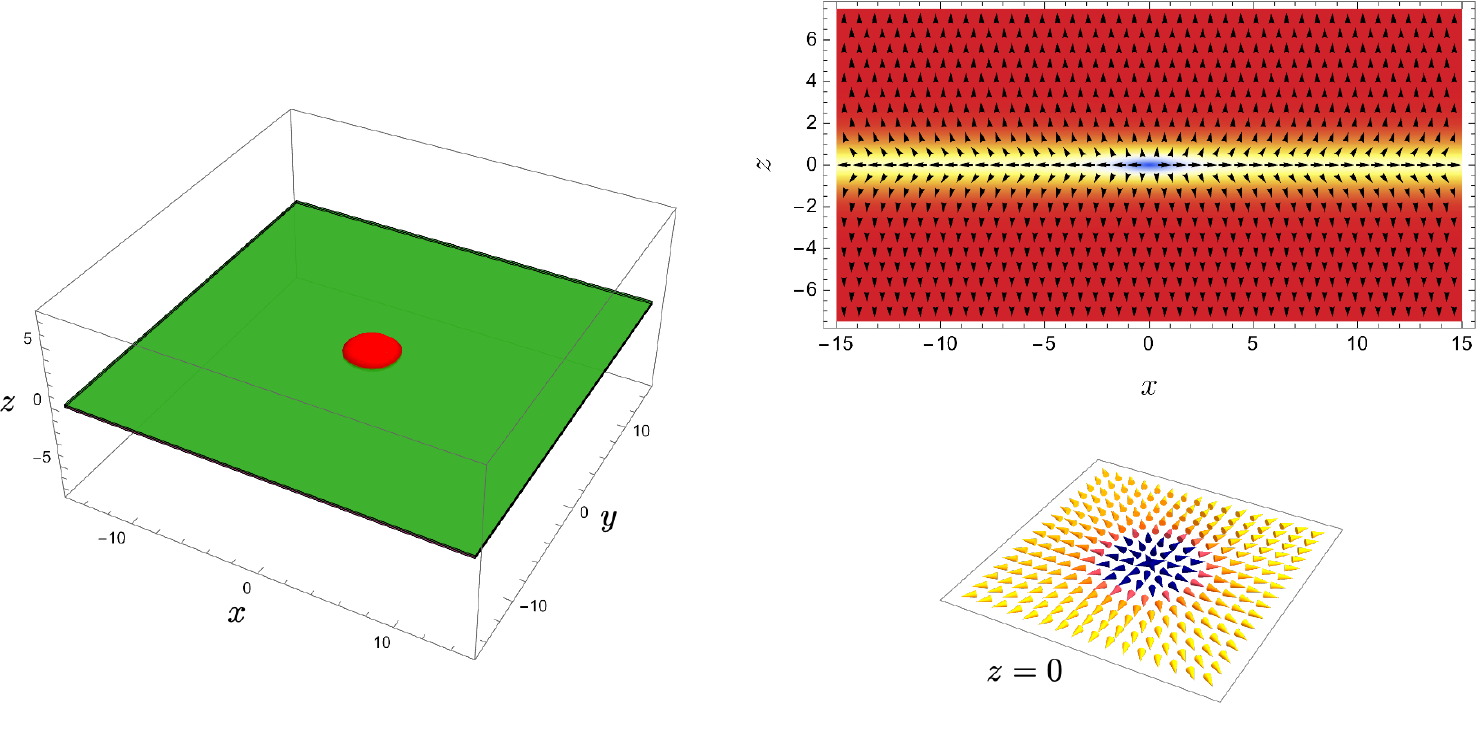}
\caption{A numerical solution of an $N=2$ topological monopole-wall composite for 
$(v,\lambda,\alpha) = (1,2,-1/2)$.
(left) The red surface corresponds to the monopole charge density ${\cal M}_0 = 1/24$.
The green (head) and purple (tail) surfaces corresponds to $0<\phi_3 < 0.15$ and $-0.15 < \phi_3 < 0$, respectively.
(up-right) The vector plot of ${\bm \phi}$ at the $y=0$ plane.
(lower-right) The vector plot of ${\bm \phi}$ on the domain wall at $z=0$ for $x \in [-15,15]$ and $y\in[-15,15]$.}
\label{fig:1monopole_domainwall}
\end{center}
\end{figure}
%%%%%%%%

With these observations at hand, we are lead to a product ansatz
\be
{\bm \phi} = v\left(
\begin{array}{c}
\frac{x}{\sqrt{\rho^2 + a^2}}\cos \Theta  \\
\frac{y}{\sqrt{\rho^2 + a^2}}\cos \Theta \\
c \sin \Theta 
\end{array}
\right),\quad
\Theta = 2 \arctan e^{b z} - \frac{\pi}{2},
\ee
with $a$, $b$ and $c$ are constants.
For large $\rho \gg 0$, we have
\be
{\bm \phi}\big|_{\rho \gg 0} = v\left(
\begin{array}{c}
\cos\theta \cos \Theta  \\
\sin\theta \cos \Theta \\
c \sin \Theta 
\end{array}
\right), \quad \tan\theta = \frac{y}{x},
\ee
which can be understood as the mother  domain wall with the orientational moduli $\theta$.
On the $z=0$ plane, this behaves as 
a vortex
\be
{\bm \phi}\big|_{z=0} = v\left(
\begin{array}{c}
\frac{x}{\sqrt{\rho^2 + a^2}}  \\
\frac{y}{\sqrt{\rho^2 + a^2}} \\
0
\end{array}
\right).
\ee
The 
$|{\bm \phi}|$ vanishes only at the origin, which implies the presence of the point-like defect, namely
a monopole, at the origin.

%Hence, it gives rises to global vortices on the topological domain wall.
%
%The unperturbed hedgehog vectors ${\bm \phi}$ tend to be directed to either the north or south poles.
%There are two possibilities: The one possibility is that the monopole configuration completely disappears and the topological domain
%wall remains. Namely, the ${\bm \phi} = (0,0,\tilde v)$ is continuously transformed to 
%${\bm \phi} = (0,0,-\tilde v)$ in such a way that there is a plane defect on which ${\bm \phi} = (0,0,0)$.
%For this $\phi_1$ and $\phi_2$ are everywhere zero.
%The other possibility is that the ${\bm \phi} = (0,0,\tilde v)$ transits to ${\bm \phi} = (0,0,-\tilde v)$ on a plane as the first case,
%but $\phi_1$ and $\phi_2$ are not zeros in general. Instead, the distribution of $(\phi_1,\phi_2)$ on the plane is like those 
%of the two dimensional hedgehog ${\bm \phi} \sim (\cos\theta,\sin\theta, 0)$.
%Namely, the monopole is pushed into the domain wall and it can be regarded as the vortex on the domain wall.
%Note that the vortex is a topological defect associated with the $SO(2)$ symmetry which is spontaneously broken only inside
%the domain wall. The $SO(2)$ symmetry is recovered at the vortex center inside the domain wall.
%Which, the pure domain wall without monopole or the domain wall with vortex, is realized depends on the parameter $\alpha_2$.
%If $|\alpha_2|$ is too large the former is realized. On the other hand, the latter should appear as long as
%the weak parameter condition (\ref{eq:smallness2}) is satisfied.

We can utilize the above product ansatz as an initial configuration for the relaxation scheme.
The numerical solution is shown in Fig.~\ref{fig:1monopole_domainwall}.
Note that the converged final configuration is independent of the initial configurations.
The monopole as a daughter is squashed, looking like a flat pancake inside the mother domain wall.

\begin{figure}[tbp]
\begin{center}
\includegraphics[width=12cm]{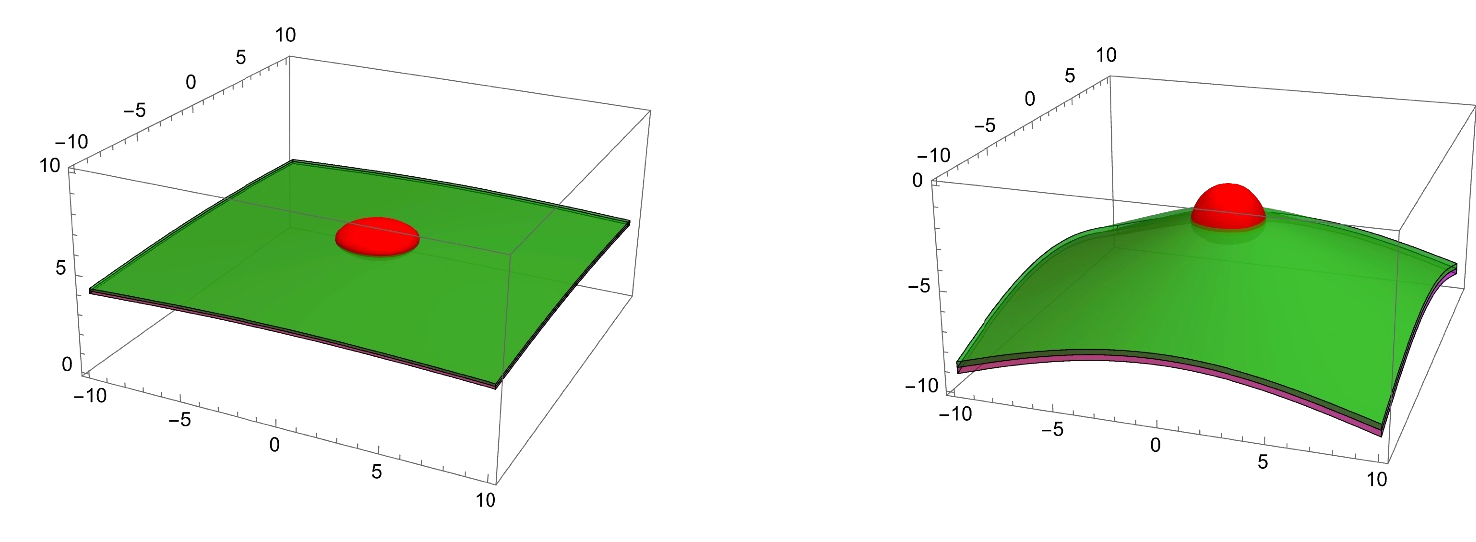}
\caption{Asymmetric monopole-wall composite.
The vacuum energies above and below the domain wall are different. This configuration is unstable since the vacuum with larger energy continuously push the domain wall. 
In the figure, the domain wall moves down, but the monopole behaves as an impurity.
The parameters are chosen as $(v,\lambda,\alpha_1,\alpha_2) = (1,2,-1/2,-1/4)$ for the left panel, and $(1,4,-1/4,-1/2)$ for the right panel.
}
\label{fig:dw_m}
\end{center}
\end{figure}

Let us discuss the case that two vacua are not degenerated, 
replacing the scalar potential $V_2$ by $V_{1+2}$ given in Eq.~(\ref{eq:V12}). Then,  the $\mathbb{Z}_2^{(3)}$ symmetry is only an approximate symmetry whereas the $O(2)^{(12)}$ symmetry remains as a genuine symmetry, leading to an imbalance between 
energies of vacua.
One remains as
a true vacuum 
and the other is lifted as a 
false vacuum. 
The domain wall 
connects the true and false vacua  and is no longer static. 
Nevertheless, once it is created, the $O(2)^{(12)}$ symmetry is spontaneously broken inside it, yielding daughter monopoles. We  provide a numerical simulation of an imbalanced monopole-wall composite in Fig.~\ref{fig:dw_m}.
One can see that the domain wall bends around the daughter monopole.
The domain wall
moves down in the figure but it is stacked by the monopole behaving as an impurity. 
This situation may affect bubble nucleation in the early universe.

%%%%%%
%%%%%%%

As mentioned in introduction,
a local version of this composite soliton is 
a local (`t Hooft-Polyakov) 
monopole immersed into a non-Abelian domain wall 
\cite{Nitta:2015mxa}.
A texture version of this 
composite configuration is a domain-wall Skyrmions 
in which Skyrmions are realized as  lumps (or baby Skyrmions) inside the domain wall \cite{
Nitta:2012wi,
Gudnason:2014nba,
Eto:2015uqa,
Eto:2023lyo}
(see also Ref.\cite{Kudryavtsev:1999zm}).

\subsubsection{Dynamical simulations for monopole-wall composites}

\begin{figure}[H]
\begin{center}
\includegraphics[width=10cm]{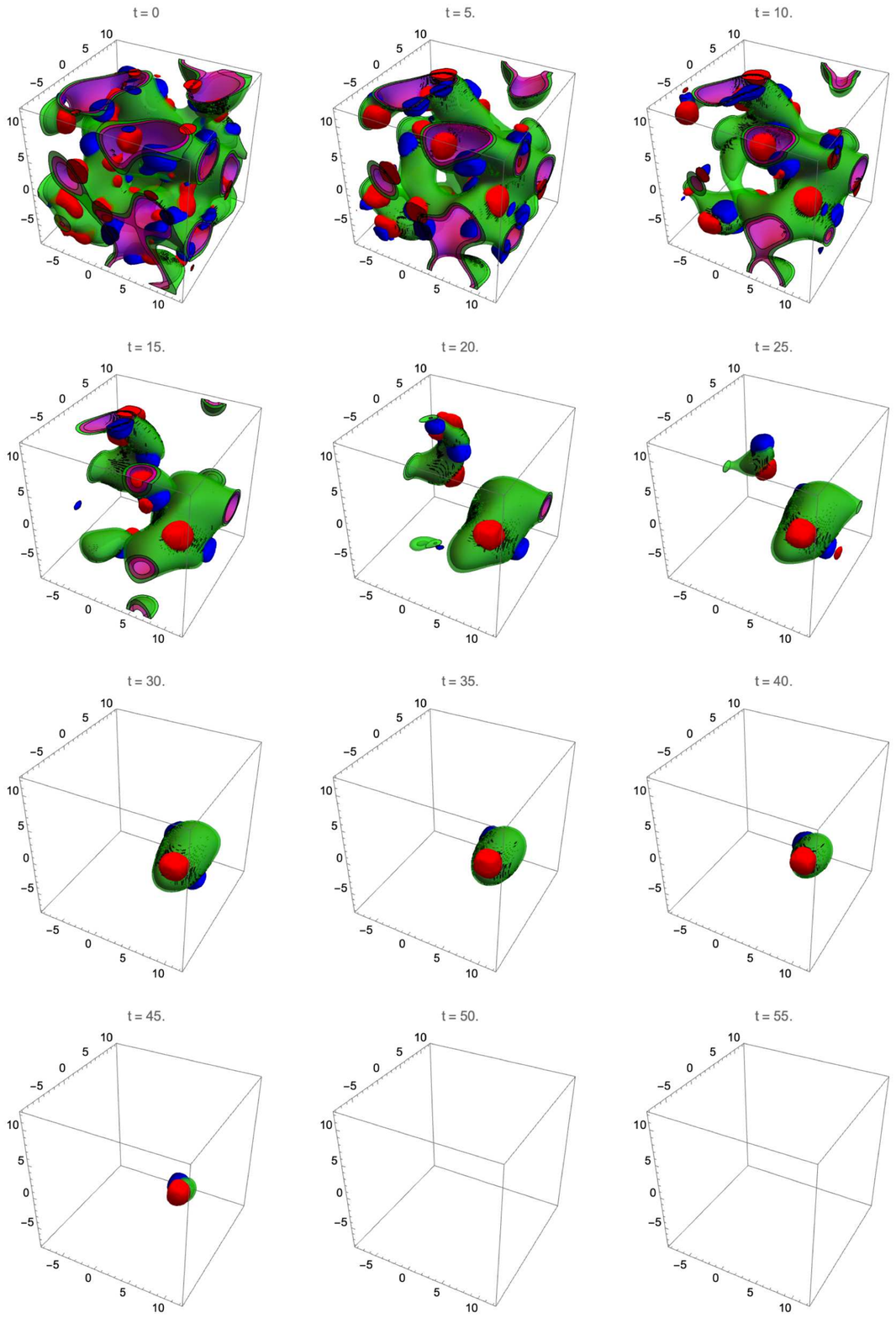}
\caption{$(v,\lambda,\alpha) = (1,4,-1/4)$.
The red (blue) surfaces are the isosurfaces of the monopole charge density ${\cal M}_0 = 1/24$ ($-1/24$), 
representing (anti-)monopoles.
The green (head) and purple (tail) surfaces correspond to $0<\phi_3 < 0.15$ and $-0.15 < \phi_3 < 0$, respectively, representing 
domain walls.
}
\label{fig:simulation_n2_non_spherical_monopole_wall_01}
\end{center}
\end{figure}

Fig.~\ref{fig:simulation_n2_non_spherical_monopole_wall_01} shows a numerical simulation for the real time dynamics evolving from a random and periodic initial configuration for  $(v,\lambda,\alpha) = (1,4,-1/4)$.
All the monopoles and anti-monopoles appear on domain walls.
The domain walls are in general closed surfaces under the periodic boundary condition,
%even though the $\mathbb{Z}_2$ symmetry is manifest, 
%the domain walls 
they shrink by their tension.
%, leaving either the north or south vacuum behind. 
Which of the north or south vacuum is left is determined by chance
with the probability $1/2$.
(In Fig.~\ref{fig:simulation_n2_non_spherical_monopole_wall_01}, the final vacuum is accidentally the north vacuum.)

%%%%%%%%%
%%%%%%%%%
%
%\clearpage
%
%On the other hand, Fig.\ref{fig:simulation_n2_asym_non_spherical_monopole_wall_01} shows a numerical simulation 
%for the $\mathbb{Z}_2$ asymmetric case. We chose $(v,\lambda,\alpha_1,\alpha_2) = (1,4,1/2,-1/4)$.
%Since $\alpha_1 > 0$, the true vacuum is the south pole. This can be seen in Fig.\ref{fig:simulation_n2_asym_non_spherical_monopole_wall_01}:
%The domain walls shrink by showing the purple surfaces outside while the green surfaces are closed inside the vacuum bubbles. 
%The potential bias accelerates the shrinking of the metastable vacuum bubble. Indeed, the life time of the $\mathbb{Z}_2$ asymmetric domain walls
%is much shorter than that of the $\mathbb{Z}_2$ symmetric one, compare Figs.~\ref{fig:simulation_n2_non_spherical_monopole_wall_01}
%and \ref{fig:simulation_n2_asym_non_spherical_monopole_wall_01}.
%\begin{figure}[h]
%\begin{center}
%\includegraphics[width=15cm]{simulation_n2_asym_non_spherical_monopole_wall_01}
%\caption{$(v,\lambda,\alpha_1,\alpha_2) = (1,4,1/2,-1/4)$.
%The red surface shows the isosurface of monopole charge density ${\cal M}_0^{(123)} = 1/24$.
%The yellow and cyan surfaces show the isosufaces of $|\phi_1+i\phi_2| = 1/2$ with $\phi_3 > 0$
%and $\phi_3 < 0$, respectively.}
%\label{fig:simulation_n2_asym_non_spherical_monopole_wall_01}
%\end{center}
%\end{figure}

Compared with stable monopoles on strings, one can observe that the life time is shorter 
in this case because the domain walls 
are long lived compared with strings.

\subsection{Relations to spherical monopoles}

\begin{figure}[tbp]
\centering
    \includegraphics[width=11cm]{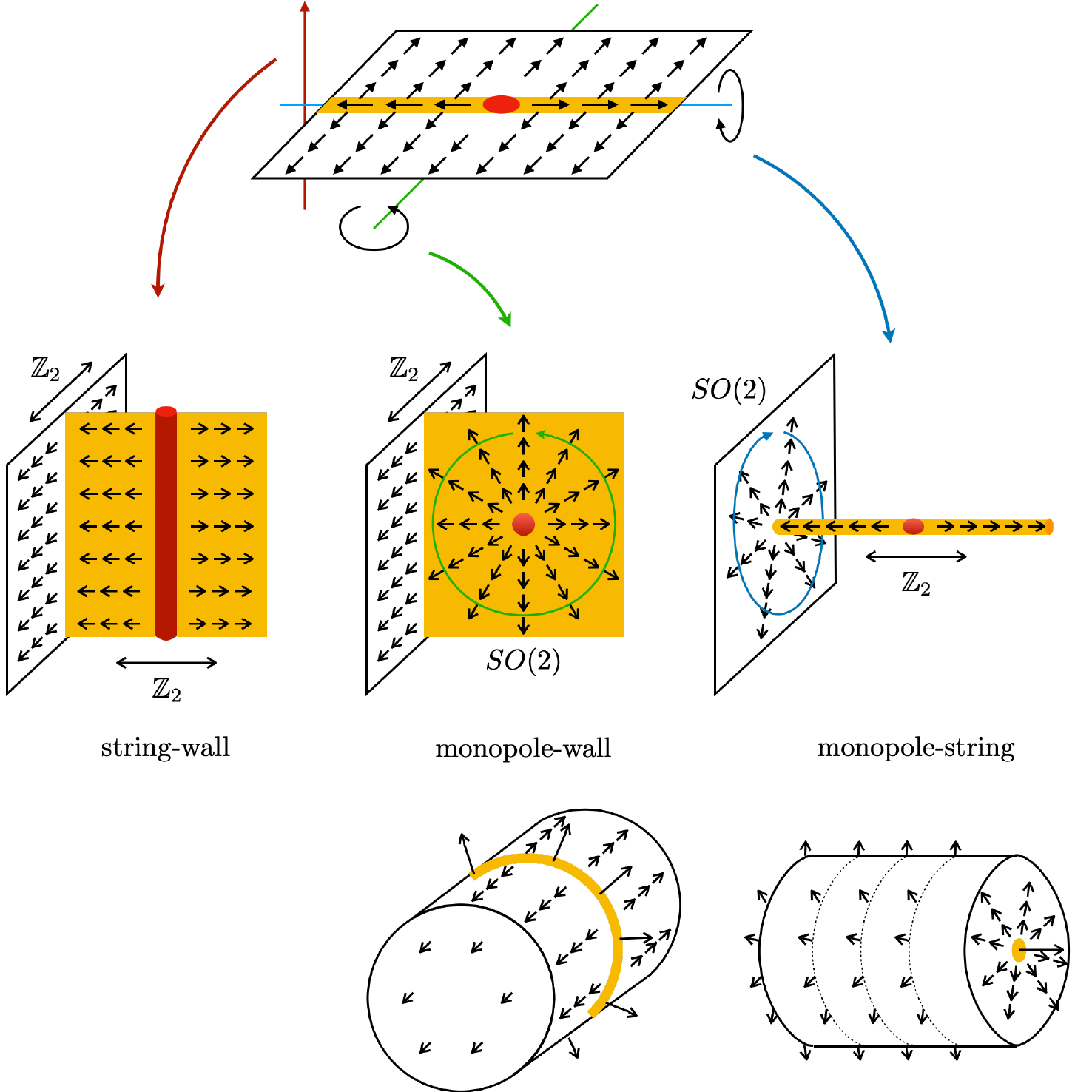}
    \caption{Schematic pictures of the composite solitons of two different solitons.
    The small arrows represent configurations of ${\bm \phi}$.
    The top one is the vortex-wall on two dimensional plane.
    In the second row,
    the string-wall (left) is made
    of the two-component scalar field whereas monopole-string (right) and 
    monopole-wall (middle) are made
    of the three-component scalar field.
    In the third row, the configuration of ${\bm \phi}$ on the cylindrical boundary is shown.}
    \label{fig:composite_2solitons}
\end{figure}

In the both  cases of $\alpha > 0$ and $\alpha <0$, 
the SSB pattern of the Lagrangian is the same, $SO(2) \times \mathbb{Z}_2 \to 1$.
However, the difference is the order of the SSBs, i.e., 
which of $SO(2)$ or $\mathbb{Z}_2$ is broken first, 
namely at high energy.
If $SO(2)$ is broken first, the mother soliton is 
the global string. On the other hand, the mother soliton is the domain wall
when $\mathbb{Z}_2$ is broken first. 
For the former case, the subsequent SSB of $\mathbb{Z}_2$ on the mother  string gives rise to
the inner kink which is identical to the monopole.
In the latter case, the $SO(2)$ SSB on the mother domain wall
leads to the inner vortex which is identical to the monopole.
These can be visually understood by drawing
the distributions of the $\bm{\phi}$ as in
the middle and right panels in the second row of Fig.~\ref{fig:composite_2solitons}.
For the both configurations, the $SO(2)$ transformation rotates the arrows and is associated with
the vortices whereas $\mathbb{Z}_2$ flips the arrows and is associated with
the domain walls.

%%%%%%%%%%%%%
%%%%%%%%%%%%

Another view point is that 
both the monopole-string and monopole-wall composites can be
constructed by a simultaneous rotation of the real space and internal space from the vortex-wall configuration in $\mathbb{R}^2$,
as illustrated in the first and second rows in Fig.~\ref{fig:composite_2solitons}.
If we rotate the two-dimensional vortex-wall composite along the axis
on the domain wall, we obtain the monopole-string composite in $\mathbb{R}^3$.
On the other hand, if we rotate it along the axis orthogonal to the domain wall,
we obtain the monopole-wall composite in $\mathbb{R}^3$.
Instead of the rotation, if we translate it to the vertical direction, 
we obtain the string-wall composite in $\mathbb{R}^3$ in the $O(2)$ model.

Let us compare these monopoles in the 
linear $O(3)$ model with the spherically symmetric one
associated with the usual SSB $SO(3) \to SO(2)$ whose
topology is characterized by $\pi_2(SO(3)/SO(2)) \simeq \mathbb{Z}$.
This counts how many times the hypersurface $\bm{\phi}(\bm{x})$ in the field internal space, 
whose domain is the spacial boundary $\{\bm{x} \mid \bm{x} \in S^2 = \p\mathbb{R}^3\}$,
wraps the vacuum manifold $S^2 \simeq SO(3)/SO(2)$.
In comparison, 
the relevant SSB, $SO(2) \times \mathbb{Z}_2 \to \mathbbm{1}$,
under the presence of the ESB  potential $V_2$
can be seen clearly on a cylindrical boundary as in the bottom row
in Fig.~\ref{fig:composite_2solitons}.
The distributions of ${\bm \phi}$ on the cylinder surfaces have distinct difference
between the monopole-string and monopole-wall. Nevertheless, for the both cases,
the hypersurface ${\bm \phi}$ on the boundary forms the sphere. 
Namely, they can be continuously deformed into the spherical hedgehog configuration. 
Thus, there should exist the monopole inside the cylinder.

%\textcolor{red}{This sentence is not needed $\rightarrow$ Of course, $\pi_2(S^2)$ is not the genuine topological characteristic for the monopole-string and monopole-wall composites, but we found $\pi_1(SO(2))$ together with the SSB of $\mathbb{Z}_2$ is a nice topological quantity to characterize them.}

%%%%%%%%%%%%%%%%%%%%%%%%%%%%%%%%%%%
\section{Monopole-string-wall composites}\label{sec:monopole-string-wall}

Finally, we consider the most generic quadratic potential
\be
V_2 =  (\phi_1)^2\cos\gamma + (\phi_2)^2 \cos\beta\sin\gamma + (\phi_3)^2 \sin\beta\sin\gamma.
\label{eq:V2_quadratic}
\ee
As before, we are interested in the weak coupling regime where 
the overall coefficient $\alpha$ satisfies the weak coupling condition (\ref{eq:wall_weak}).
Note that the above potential can be rewritten as
$
V_2 =  ({\bm \phi})^2\cos\gamma + (\phi_2)^2 \left(\cos\beta\sin\gamma-\cos\gamma\right) 
+ (\phi_3)^2 \left(\sin\beta\sin\gamma - \cos\gamma\right).
$
The first term can be absorbed into the quadratic potential $\frac{\lambda}{4}\left({\bm \phi}^2 - v^2\right)^2$.
It just shifts the VEV and potential energy by constants leading to no essential changes. 
Therefore, we can set $\gamma = \pi/2$ without loss of generality and we will adopt 
the potential with one parameter $\beta$
\be
V_2(\beta) =  (\phi_2)^2 \cos\beta + (\phi_3)^2 \sin\beta.
\ee
For generic $\beta$
the symmetry of the model is 
\be
G = \mathbb{Z}_2^{(1)} \times \mathbb{Z}_2^{(2)}  \times \mathbb{Z}_2^{(3)},
\ee
associated with the sign flip of $\phi_{1,2,3}$.
The accidental cases are $\beta = \pi/2$ and $3\pi/2$ for which $V_2$ has the single term $\pm (\phi_3)^2$, respectively. 
In such a case, 
the symmetry is enhanced as $G(\beta=\pi/2,3\pi/2) = O(2)^{(12)} \times \mathbb{Z}_2^{(3)}$ which we have studied in the previous sections.
Similarly, we have $G(\beta=0,\pi) = O(2)^{(13)} \times \mathbb{Z}_2^{(2)}$. In addition $\beta  = \pi/4, 5\pi/4$ belongs to the same class.
This can be understood by rewriting
$
V_2(\beta =\pi/4,5\pi/4) 
%= \pm \frac{(\phi_2)^2 + (\phi_3)^2}{\sqrt2}
= \pm \left({\bm \phi}^2 - (\phi_1)^2\right)/\sqrt2
%\frac{{\bm \phi}^2}{\sqrt2} \mp \frac{(\phi_1)^2}{\sqrt2}.
$.
Again, the first term can be absorbed into the quadratic potential $\frac{\lambda}{4}\left({\bm \phi}^2 - v^2\right)^2$, so that
the potential is essentially $V_2(\beta=\pi/4,5\pi/4) = \mp (\phi_1)^2/\sqrt2$. Therefore,
we have $G = O(2)^{(23)} \times \mathbb{Z}_2^{(1)}$.

\begin{figure}[ht]
\begin{center}
\includegraphics[width=12cm]{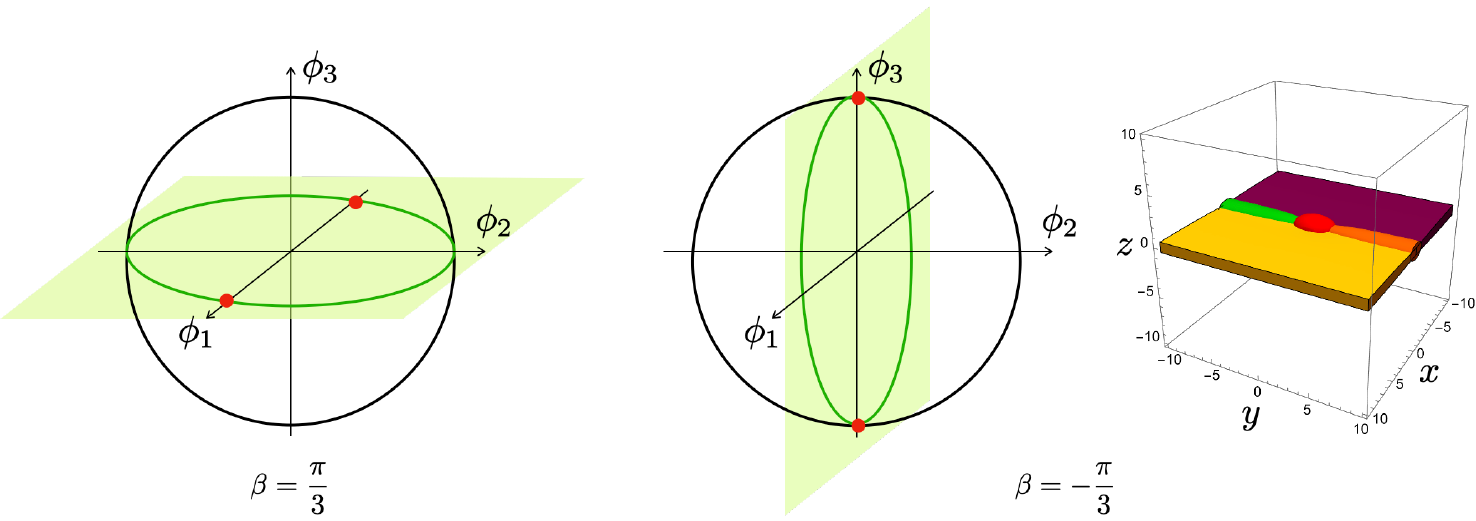}
\caption{The vacuum manifold under the presence of $V_2$ and a numerical solution for monopole-string-wall composite.}
\label{fig:deformed_V_2b}
\end{center}
\end{figure}

For illustration, let us take $\beta = \pm\pi/3$ which gives 
\be
V_2(\pm\pi/3) = \frac{1}{2}(\phi_2)^2 \pm \frac{\sqrt3}{2}(\phi_3)^2. \label{eq:pot_beta=pi/3}
\ee
For $\beta = \pi/3$,
the second term has a larger positive coefficient than the first one, so that it enforces $\phi_3=0$. Namely, the original vacuum manifold $S^2$ is
restricted to the great circle $S^1$ perpendicular to the $\phi_3$ axis. 
Subsequently, the first term enforces $\phi_2=0$.
Namely, the vacuum of the potential is understood by the following three steps;
the vacuum is $S^2$ when $\alpha=0$,
the second term in \eqref{eq:pot_beta=pi/3} deforms it so that the $S^1$ submanifold perpendicular to the $\phi_3$ axis is the lowest-energy one,
and the first term in \eqref{eq:pot_beta=pi/3} leaves two antipodal points (on the $\phi_1$ axis) among the $S^1$ manifold as the true vacua.
We call this potential structure the three-fold structure ($S^2 \supset S^1 \supset S^0$).
%the low-lying bottom of the potential has the three-fold structure, the first structure is $S^2$, the second one is the $S^1$ perpendicular to the $\phi_3$ axis, and the third one is the true vacua that are two antipodal points on the $\phi_1$ axis,
See Fig.~\ref{fig:deformed_V_2b} (left).

For $\beta = -\pi/3$, the second term chooses the two points $\phi_3 \sim \pm v$ as the true vacua.
In addition, the first term slightly reduces the potential energy on the great circle $S^1$ perpendicular to the $\phi_2$ axis.
The structure of the potential is similar to that with $\beta = \pi/3$,
and the two cases $\beta = \pm \pi/3$ are identical if we rotate in the internal space by $\pi/2$
as shown in Fig.~\ref{fig:deformed_V_2b} (middle). 
The generic $\beta$ can be straightforwardly understood in a similar manner. 
In summary, the parameter space $\beta \in [0,2\pi)$ is divided into 6 regions by the 6 boundaries
$\beta = 0,\pi/4,\pi/2,\pi,5\pi/4,3\pi/2$ where the symmetry is enhanced.

The three-fold structures $S^2$, $S^1$, and $S^0 ={\mathbb Z}_2$ of the potential give rise to the monopole, string, and domain wall, respectively.
%, at each step.
Hence, the resulting soliton is a monopole-string-wall composite.
This composite is topologically stable.
Note that 
characterizations 
of monopoles and strings by 
the homotopy group $\pi_2(S^2)$ and $\pi_1(S^1)$ are 
only approximate.
Instead, it is exactly characterized by the homotopy group associated with the spontaneous breaking of $G$.

To be specific, let us consider $\beta = -\pi/3$, see Fig.~\ref{fig:deformed_V_2b} (middle). 
The vacua are ${\bm \phi} = (0,0,\pm v)$.
There, the $\mathbb{Z}_2^{(3)}$ symmetry is spontaneously broken, and the mother domain wall is generated.
In this case, one has to choose 
a half circle on either positive or negative side of $\phi_1$ 
as the domain wall configuration 
in the broken phase.
Namely, the $\mathbb{Z}_2^{(1)}$ symmetry 
is spontaneously broken on the mother domain wall.
This gives rise to an inner domain wall inside the mother domain wall, namely a daughter string from the 3+1 dimensional bulk perspective.
In the daughter vortex core, $\phi_1=\phi_3=0$ and $\phi_2$ can be either positive or negative (since it does not lie on the origin $\bm{\phi}=0$ due to the quasi vacuum manifold $S^2$).
%and one has to choose again the 
%the system again has to choose a hemisphere either positive or negative side
%of $\phi_2$. 
Namely, the $\mathbb{Z}_2^{(2)}$ symmetry is spontaneously broken inside the daughter vortex localized inside the mother domain wall.
This gives rise to an inner kink 
inside the daughter vortex, 
which is nothing but a 
granddaughter monopole from the 3+1 dimensional bulk 
perspectives.
This can be briefly summarized by a sequence of the symmetry breakings of $G$ for $\beta = -\pi/3$
\begin{alignat}{2}
G_{\rm approx}&=O(3) &\xrightarrow[\text{monopole}]{\text{vac}}\qquad &O(2)\nonumber\\
&\cup & &\cup \nonumber\\
G_{\rm exact}&=
\mathbb{Z}_2^{(1)} \times &\mathbb{Z}_2^{(2)} \times  \mathbb{Z}_2^{(3)}
\xrightarrow[\text{mother wall}]{\text{vac}} \mathbb{Z}_2^{(1)} &\times \mathbb{Z}_2^{(2)}
\xrightarrow[\text{daughter vortex}]{\text{mother wall}} \mathbb{Z}_2^{(2)} 
\xrightarrow[\text{granddaughter monopole}]{\text{daughter vortex}} 1.
\label{eq:SSB_Z2^3}
\end{alignat}

%\be
%\mathbb{Z}_2^{(1)} \times \mathbb{Z}_2^{(2)} \times  %\mathbb{Z}_2^{(3)}
%\xrightarrow[\text{host wall}]{\text{vac}} %\mathbb{Z}_2^{(1)} \times \mathbb{Z}_2^{(2)}
%\xrightarrow[\text{guest wall (string)}]{\text{host wall}} \mathbb{Z}_2^{(2)} 
%\xrightarrow[\text{Jr.guest wall (monopole)}]{\text{guest wall}} 1.
%\label{eq:SSB_Z2^3}
%\ee

The composite soliton can be well captured by the following ansatz
for $\beta = -\pi/3$
\be
{\bm \phi} = v\left(
\begin{array}{c}
\cos \Theta \sin\Phi\\
\delta  \tanh \gamma y \cos \Theta  \cos\Phi\\
\kappa  \sin \Theta 
\end{array}
\right),\quad
\Theta = 2 \arctan e^{\xi z} - \frac{\pi}{2},\quad
\Phi = 2 \arctan e^{\chi x} - \frac{\pi}{2},
\ee
with parameters $\xi,\chi,\gamma,\delta$, and $\kappa$.
The presence of the mother wall can be seen by taking $\Phi = \pi/2$ for $x \gg 0$ and $\Phi = -\pi/2$ for $x \ll 0$
\be
{\bm \phi}\big|_{x\gg0} = v\left(
\begin{array}{c}
\cos \Theta\\
0\\
\kappa  \sin \Theta 
\end{array}
\right),\qquad
{\bm \phi}\big|_{x\ll0} = v\left(
\begin{array}{c}
-\cos \Theta\\
0\\
\kappa  \sin \Theta 
\end{array}
\right).
\ee
These are the $\mathbb{Z}_2^{(1)}$-charged domain wall perpendicular to the $z$-axis.
On the domain wall at $\Theta = 0$ ($z=0$), we have
%\be
%{\bm \phi} \to v \left(
%\begin{array}{c}
%0\\
%0\\
%\pm 1
%\end{array}
%\right),\quad z \to \pm \infty
%\ee
\be
{\bm \phi}\big|_{z=0} =
v\left(
\begin{array}{c}
\sin\Phi\\
\delta  \tanh \gamma y  \cos\Phi\\
0
\end{array}
\right).
%\to 
%v\left(
%\begin{array}{c}
%0\\
%\pm 1\\
%0
%\end{array}
%\right),\quad x \to \pm \infty.
\ee
This is essentially the same as Eq.~(\ref{eq:PA_sw}) for the string-wall composite in 
the $O(2)$ model.
We find the inner walls by taking $y \gg 0$ and $y \ll 0$ as
\be
{\bm \phi}\big|_{z=0,~y\gg0}
=
v\left(
\begin{array}{c}
\sin\Phi\\
\delta  \cos\Phi\\
0
\end{array}
\right),\qquad
{\bm \phi}\big|_{z=0,~y\ll0}
=
v\left(
\begin{array}{c}
\sin\Phi\\
-\delta   \cos\Phi\\
0
\end{array}
\right).
\ee
These are  $\mathbb{Z}_2^{(2)}$-charged domain walls
perpendicular to the $y$-axis.
Finally, the inner kink inside the daughter vortex 
(granddaughter monopole) can be found for $(\Theta,\Phi) = (0,0)$ at $z=x=0$ as
\be
{\bm \phi}\big|_{z=x=0}
=
v\left(
\begin{array}{c}
0\\
\delta \tanh \gamma y\\
0
\end{array}
\right).
\ee

We present a numerical solution for $\beta = -\pi/3$ in Fig.~\ref{fig:deformed_V_2b} (right)
which is obtained by evolving the above ansatz by the standard relaxation scheme.
Since the ansatz 
is appropriate as the initial configuration, the relaxation process quickly converges and the final state is independent of the parameters of the ansatz.
In Fig.~\ref{fig:deformed_V_2b} (right)
the mother domain walls are painted in yellow and purple reflecting the $\mathbb{Z}_2^{(1)}$ charge: 
the yellow corresponds to ${\bm \phi} \sim (1,0,0)$ while the purple indicates ${\bm \phi} \sim (-1,0,0)$.
The both connect two vacua 
${\bm \phi} \sim (0,0,1)$ in the upper bulk while ${\bm \phi} \sim (0,0,-1)$
in the lower bulk. 
Also, the daughter vortex strings are painted by orange and green reflecting the $\mathbb{Z}_2^{(2)}$ charge:
the orange indicates ${\bm \phi} \sim (0,1,0)$, and the green corresponds to ${\bm \phi} = (0,-1,0)$.

The field ${\bm \phi}$ vanishes at the center of the monopole (the red blob). 
Though the distribution of ${\bm \phi}$ is not a spherically symmetric hedgehog, 
the kink inside the daughter vortex can be regarded as a global monopole.
To see this, let us consider a plane $y=10$ in the right panel of Fig.~\ref{fig:deformed_V_2b}.
The field on the plane corresponds to a hemisphere with $\phi_2>0$ in the internal space (middle panel of Fig.~\ref{fig:deformed_V_2b}).
On the other hand, the field on a plane $y=-10$ corresponds to a hemisphere with $\phi_2<0$.
Since these two hemispheres are smoothly connected on $y=0$ on the boundary of the box, 
the field configuration wraps the quasi vacuum manifold $S^2$ 
representing a global monopole.
%Therefore, ${\bm \phi}$ of the monopole-string-wall composite is isomorphic to  the monopole.
%
%\be
%{\bm \phi}\big|_{z=0,x=0}
%=
%v\left(
%\begin{array}{c}
%\tanh \delta y \\
%0\\
%0
%\end{array}
%\right)
%\to 
%v\left(
%\begin{array}{c}
%\pm 1\\
%0\\
%0
%\end{array}
%\right),\quad y \to \pm \infty.
%\ee

\begin{figure}[tbp]
\begin{center}
\includegraphics[width=15cm]{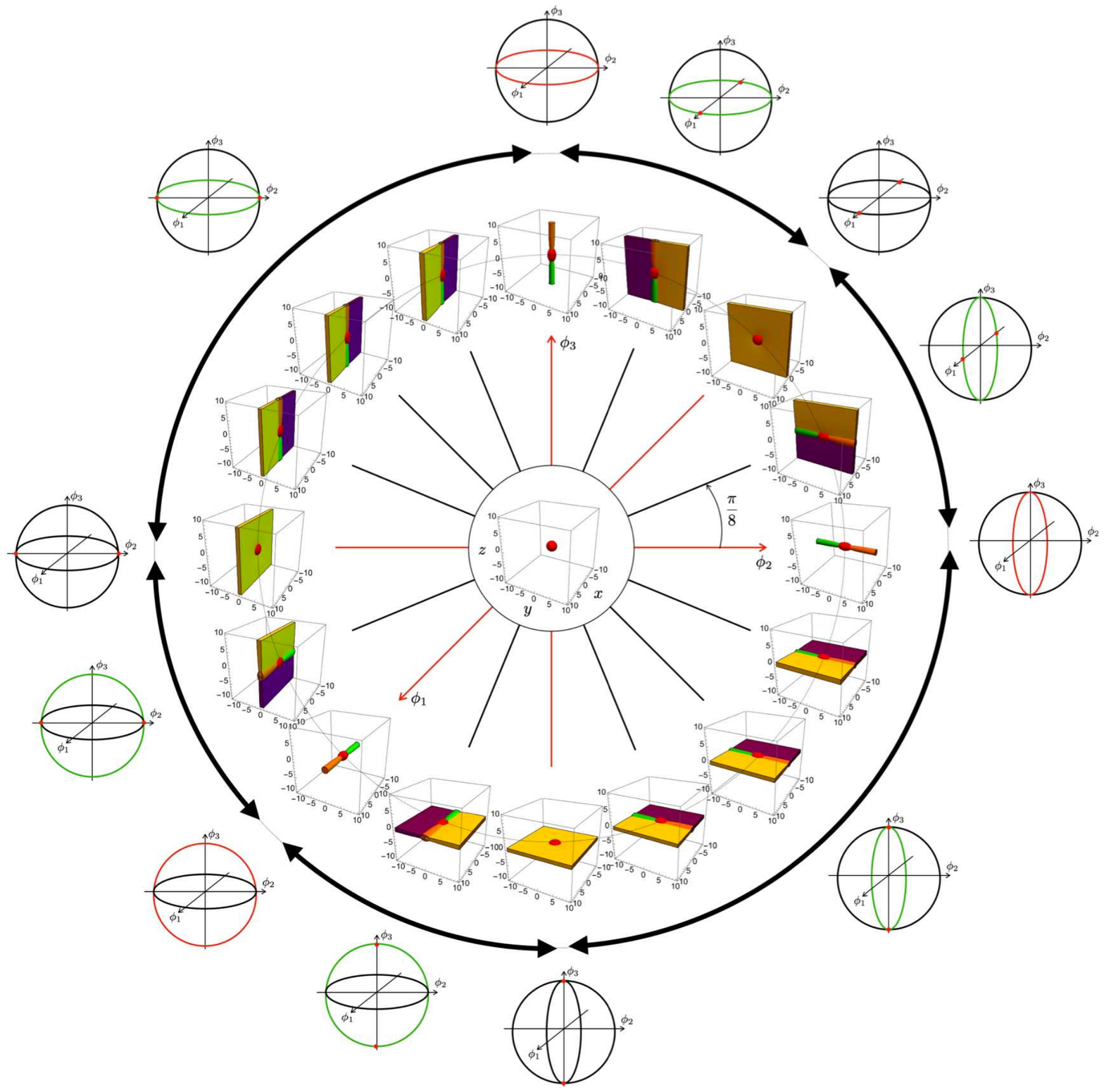}
\caption{The monopole-string-wall composites for $\beta = k\pi/8$ ($k=0,1,\cdots,15$). The right-most figure having a monopole-string corresponds to $\beta = 0$ ($k=0$). The others are placed counterclockwise by $\delta \beta = \pi/8$. The figure at the center shows the isolated global monopole for $\epsilon = 0$.
The schematic pictures at the outer circle showing $S^2$ in the internal space indicates the true vacua in red color and quasi vacua in green. }
\label{fig:msw}
\end{center}
\end{figure}

We show the numerical solutions of the monopole-string-wall composites for $\beta = k \pi/8$ with $k=0,1,\cdots,15$
in Fig.~\ref{fig:msw}. All the solutions are obtained from the unique initial configuration ${\bm \phi} \sim v \hat{\bm r}$ by the relaxation
scheme. Some of them are identical up to spacial rotations. Nevertheless we show all $\beta$ because it is useful to see which direction
the composite extends for a fixed $\beta$.

%%%%%%%%%%
%%%%%%%%%%%
\clearpage

%%%%%

\begin{figure}[tbp]
\begin{center}
    \includegraphics[width=10cm]{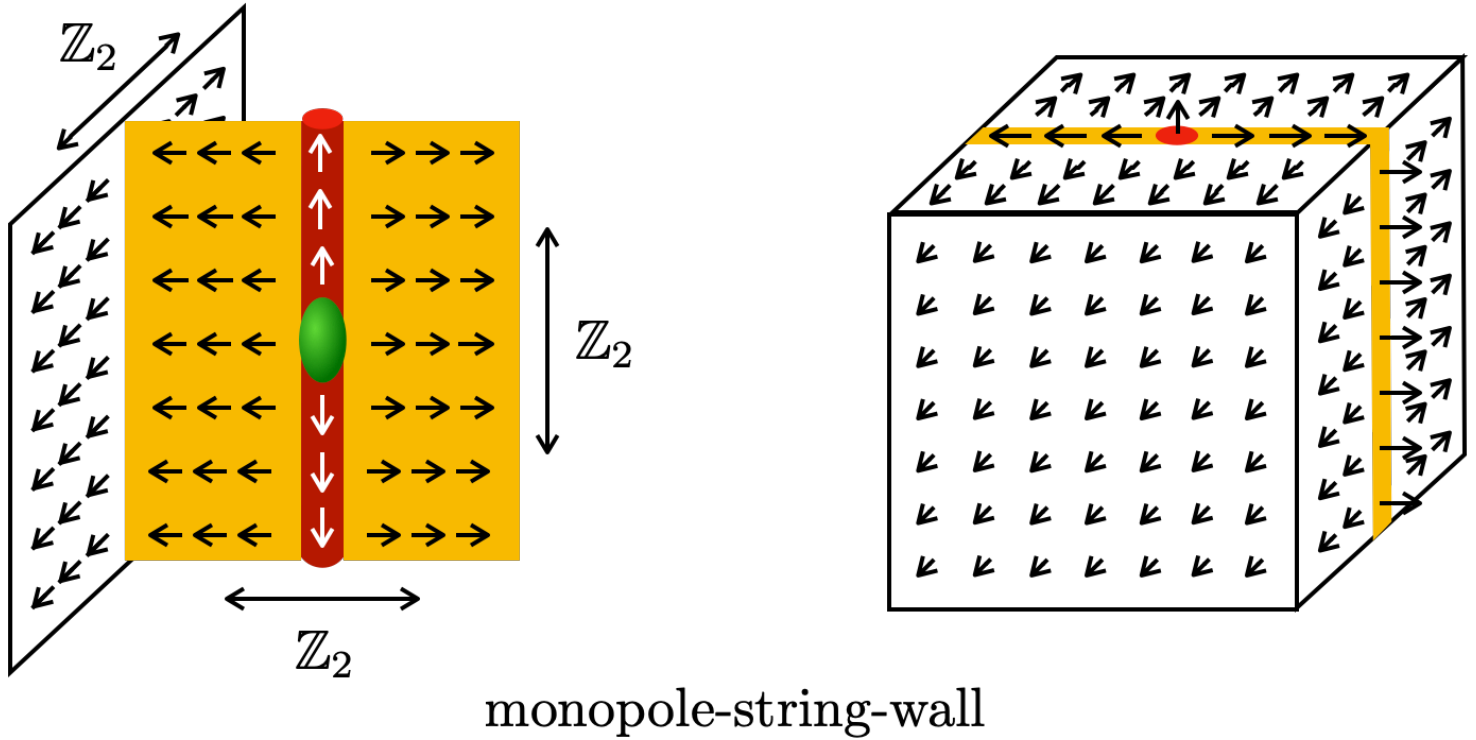}
    \caption{The composite of monopole-string-wall. The configuration of ${\bm \phi}$ is
    shown by the small arrows. }
    \label{fig:composite_msw}
\end{center}
\end{figure}

The profile of ${\bm \phi}$ is
shown in the left panel of Fig.~\ref{fig:composite_msw}.
The profile of ${\bm \phi}$ on the boundary cube is shown in the right panel of Fig.~\ref{fig:composite_msw}. It is very different from the spherical hedgehog 
distribution of the usual monopole. However,
indeed it can be continuously deformed into a spherical hedgehog, and therefore
there should exist the monopole inside the cube.
%%%%%%%%%%
%%%%%%%%%%

\begin{figure}[tbp]
\begin{center}
\includegraphics[width=12cm]{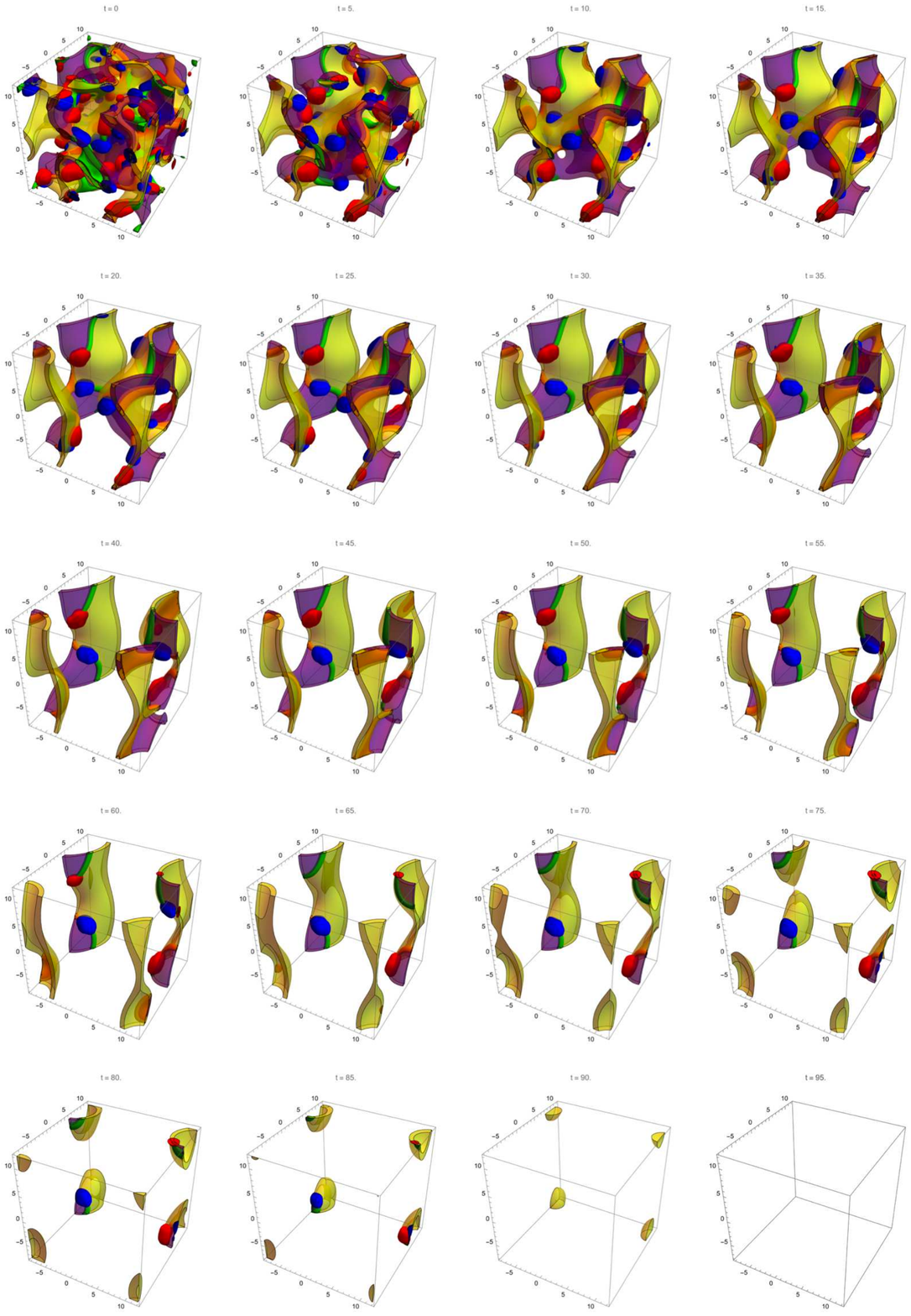}
\caption{The dynamical simulation of the monopole-string-wall composite for $(v,\lambda,\alpha) = (1,4,1/4)$ and $\beta = -\pi/3$.
The periodic boundary condition is imposed.}
\label{fig:msw_dynamics}
\end{center}
\end{figure}

%%%%%
We briefly present a result of dynamical simulation starting from a random and periodic initial configuration in Fig.~\ref{fig:msw_dynamics}.
%%%%%%%%%%
%%%%%%%%%

As a local-soliton counterpart 
of this configuration, 
a composite soliton of 
domain wall, local vortex, and local monopole was constructed in Ref.~\cite{Nitta:2015mxa} 
as a kink inside 
a non-Abelian sine-Gordon soliton 
\cite{Nitta:2014rxa,Eto:2015uqa} 
inside a non-Abelian domain wall \cite{Shifman:2003uh,Eto:2005cc,Eto:2008dm} 
(see also Ref.~\cite{Nitta:2022ahj}). 
This configuration 
is also similar to 
a topological Nambu monopole 
in two-Higgs doublet model
\cite{Eto:2019hhf,Eto:2020hjb,Eto:2020opf}. 
As a similar texture-type soliton, 
a Skyrmion as a kink inside a domain wall inside a domain wall was studied 
in Ref.~\cite{Nitta:2012rq}
in which it is called 
a matryoshka Skyrmion.

%%%%
\section{Summary and discussion}
\label{sec:summary}

In this work, we have studied various composite solitons of the global type
appearing in the $O(2)$ and $O(3)$ linear sigma models
${\bm \phi} = (\phi_1,\phi_2)$ and $(\phi_1,\phi_2,\phi_3)$ 
with perturbed ESB terms.

The $O(2)$ model 
with the ESB 
potential $V_N = \alpha (\phi_2)^N$ for $N=0,1,2$ 
is identical to the axion model with the domain-wall number $N_\mathrm{DW}=1,2$, 
admitting a composite of the global string and domain walls.
In the $N=0$ case, 
%the $O(2)$ symmetry is not explicitly broken 
%but is only spontaneously broken. As a consequence, 
the topological global string
exists as a stable soliton solution.
In the $N=1$ case, 
the $O(2)$ symmetry is explicitly broken
to $\mathbb{Z}_2$, 
the vacuum is a unique point, and there are no topologically
nontrivial solitons. Nevertheless, as long as the ESB term is sufficiently small, there are non-topological composite solitons: 
a string attached by a domain wall.
In the case of $N=2$, there exists a topological composite soliton
of the global string and the two domain walls. This is very well known but
we have given another picture on the string-wall composite.
From our viewpoint, the string-wall composite can be realized 
as an inner kink in the mother domain wall. The symmetry  behind this is
$G = \mathbb{Z}_2^{(1)} \times \mathbb{Z}_2^{(2)}$.
The $\mathbb{Z}_2^{(2)}$ is spontaneously broken in the vacuum.
This triggers the sequential spontaneous symmetry breaking and
generations of the mother domain wall and daughter vortex as given in Eq.~(\ref{eq:2comp_sequence}).
The mother domain wall has a $\mathbb{Z}_2^{(1)}$ charge and when
two mother domain walls with opposite $\mathbb{Z}_2^{(1)}$ charges
join, an inner kink is produced at the junction line.
It can be identified with a global vortex in the bulk viewpoint.

Next, 
we have studied
the $O(3)$ linear model 
with the small ESB potentials $V_N = \alpha (\phi_3)^N$ with $N=0,1,2$. 
In the $N=0$ case, the SSB is
$G = O(3) \to H=O(2)$ 
giving rise to the global
monopoles characterized by $\pi_2(G/H) \simeq \mathbb{Z}$.
In the $N=1$ case, we have $G = H = O(2)^{(12)}$. The vacuum is a unique point,
and there are no topologically nontrivial solitons.
Nevertheless, if the ESB potential is sufficiently weak,
there is a non-topological composite of the global monopole attached by a 
global string. 
%from one side.
%This is a counterpart of the non-topological string-wall composite in the
%two component model.
We have constructed the numerical solutions for
the global monopole attached by the string, and a pair of the monopole
and antimonopole confined by the string.
In the $N=2$ case, we have found that there are two branches corresponding to $\alpha > 0$ or $\alpha < 0$ (supposing small $|\alpha|$). 
The summary is given in Table \ref{tab:2}.
The symmetry of the Lagrangian is $G = O(2)^{(12)} \times \mathbb{Z}_2^{(3)}$ for both $\alpha > 0$ and $\alpha < 0$,
which is spontaneously broken into $H = \mathbb{Z}_2^{(12)} \times \mathbb{Z}_2^{(3)}$ for $\alpha > 0$ and
$H = O(2)^{(12)}$ for $\alpha < 0$.
The SSB $G \to H$ gives rise to the $\mathbb{Z}_2^{(3)}$ charged
string for $\alpha > 0$, and the $SO(2)^{(12)}$ charged domain wall for
$\alpha < 0$ in the vacuum. The production of these topological defects
triggers further SSB as given in Eqs.~(\ref{eq:SSB_sequence_monopole_string})
and (\ref{eq:SSB_sequence_monopole_wall}), resulting in the production of the monopoles inside the mother solitons.

The last composite soliton we have constructed in this paper 
is the composite of the monopole, string, and domain wall.
This is the most generic solitonic configuration in 
the presence of the ESB potential $V_2$ of
the quadratic terms of $\phi_{1,2,3}$ as Eq.~(\ref{eq:V2_quadratic}).
In general, the $O(3)$ symmetry is explicitly broken to 
$G=\mathbb{Z}_2 \times \mathbb{Z}_2 \times \mathbb{Z}_2$, and $G$
is spontaneously broken to $H = \mathbb{Z}_2\times \mathbb{Z}_2$
in the vacua. Then, the mother domain wall is generated, triggering the
subsequent SSB of the remaining $\mathbb{Z}_2 \times \mathbb{Z}_2$ as
Eq.~(\ref{eq:SSB_Z2^3}). On the mother domain wall,
the SSB $H \to \mathbb{Z}_2$ gives rise to the inner domain wall which is
nothing but the daughter string from the bulk viewpoint. Furthermore,
the SSB of the remaining $\mathbb{Z}_2$ takes place 
in the daughter string,
resulting in the production of the inner kink, that can be identified with 
a granddaughter monopole from the bulk viewpoint.

 \bigskip
Now, let us address several discussions and possible future directions.
In the case of the $O(3)$ model with 
the quadratic potential $V = \alpha \phi_3^2$, the $O(2)^{(12)}$ symmetry 
is exact. 
If we couple a $U(1)$ gauge field to $U(1) \subset O(2)^{(12)}$, 
solitons become partially local solitons. 
In the case of $\alpha>0$ 
discussed in  Subsec.~\ref{sec:N=2,alpha>0}, 
strings become 
local ANO strings with Ising spins  in their cores.
However,
a ${\mathbb Z}_2$ kink on a string 
remains as a global monopole in the bulk.\footnote{
In the limit of  $\lambda \to \infty$, 
the model reduces to a gauged $O(3)$ nonlinear sigma model 
with an easy-plane potential.
The model admits local string 
with an Ising spin, which carries 
a half-lump charge of $\pi_2(S^2) \simeq {\mathbb Z}$ \cite{Schroers:1995he,Schroers:1996zy,Nitta:2011um}. 
In this limit, this lump charge 
becomes exact in contrast to 
the linear model for which the lump 
charge is only approximate. 
On the other hand, a kink on the string becomes singular in this limit. }%%% end of footnote
In the case of $\alpha<0$
discussed in Subsec.~\ref{sec:N=2,alpha<0}, 
a domain wall becomes a superconducting domain wall, 
since the $U(1)$ gauge symmetry 
is spontaneously broken 
around the domain wall.
Consequently, 
a vortex on a domain wall becomes a local ANO vortex 
which is a global monopole 
in the bulk.
This  gives a field theoretical realization of 
a thin-film superconductor.

In addition, in the case of $\alpha<0$,
%discussed in Subsec.~\ref{sec:N=2,alpha>0}, 
global monopoles become global vortices on a domain wall.
It is known that 
the XY model in 2+1 dimensions 
exhibits 
the so-called Berezinskii-Kosterlitz-Thouless (BKT) transition
\cite{Berezinsky:1970fr,Berezinsky:1970-2,Kosterlitz:1972,Kosterlitz:1973xp} 
at finite temperature, 
due to a creation of a pair of 
vortex and anti-vortex. 
In our case, such a BKT transition 
could occur on the domain wall. 
Such a transition on a soliton has not been studied before. While it is interesting on its own, from the bulk point of view it implies 
a novel phenomenon that a BKT-like transition by monopoles.

When one of degenerate vacua is lifted, domain walls become unstable
as in Fig.~\ref{fig:dw_m}.
In such a case, 
a closed domain wall describes a bubble of the true vacua inside the false vacuum, 
which grows in time.
Even in such a case, topological solitons such as monopoles can appear in the domain wall worldvolume as in 
Fig.~\ref{fig:dw_m}.
While the production of topological solitons  by a collision of bubbles has been studied in the literature 
\cite{Hawking:1982ga,
Srivastava:1991sf,
Srivastava:1991nv,
Chakravarty:1992zp,
Copeland:1996jz}, 
topological solitons 
themselves 
on a bubble have not. 
The existence of such topological solitons  may affect 
dynamics and evolution of 
bubbles in the early universe.

As discussed in this paper, the $O(3)$ model with an easy axis potential yields 
a domain-wall monopole  
described by a global vortex in the domain wall world-volume, which is an $O(2)$ model or the XY model. 
On the other hand, the domain wall has also a translational modulus, 
that we have not discussed in this paper. 
By using the both $U(1)$ and translational moduli, 
one can construct 
a D-brane soliton, 
a lump string ending on the domain wall, 
at least in the limit 
$\lambda \to \infty$ of a nonlinear sigma model 
\cite{Gauntlett:2000de,Shifman:2002jm,Isozumi:2004vg,Eto:2008mf}. 
This is in fact a global vortex in the domain wall world-volume. 
Since the both configurations 
are global vortices on the domain wall, one may wonder a relation between them.
One can obtain a hint from 
Fig.~\ref{fig:dw_m}. 
In the heavy limit of monopoles $\lambda \to \infty$, bending of the domain wall would become that of the D-brane soliton.

\if0
\textcolor{red}{
Another interesting problem
is constructing composite counterparts in gauge theories. Many
composite solitons in gauge theories are well known but they are mostly
composite of two solitons. It is an interesting problem to seek
composite configurations of three or more solitons in gauge theories.
(I think MN gave many examples)
}
\fi

%%%%%%%%%%%%%%%%%%%%%%%%%%%%%%%%%%%%%%%%%%%%%%%%%%
\section*{Acknowledgements}
%%%%%%%%%%%%%%%%%%%%%%%%%%%%%%%%%%%%%%%%%%%%%%%%%%

% JSPS Grant-in-Aid for Scientific Research
This work is supported in part by JSPS Grant-in-Aid for
Scientific Research
KAKENHI Grant No.~JP22H01221 (M.\,E. and M.\,N),
Grant No.~JP19K03839 (M.\,E.),
Grant No.~JP21J01117 (Y.\ H.), and
Grant No.~JP18H01217 (M.\ N.).
% MEXT KAKENHI Grant-in-Aid for Scientific Research on Innovative Areas
The work of M.\ E. is also supported by the MEXT KAKENHI Grant-in-Aid for Scientific Research on Innovative Areas
``Discrete Geometric Analysis for Materials Design'' No.~JP17H06462 from the MEXT of Japan.
M.N. is also supported in part by the WPI program ``Sustainability with Knotted Chiral Meta Matter (SKCM$^2$)'' at Hiroshima University.

\appendix

%%%%%%%%%%%%%%%%%%%%%%%%%%%%%%%%%%%%%%%%%%%%%%%%%%%%%%%%%%%%%%%%
\section{Classifying composite solitons}
\label{sec:summary-composites}

We  classify composite solitons 
and present examples 
focusing on 
domain-wall vortex composites, 
monopole vortex composites, 
and monopole domain wall composites. 

\subsection{Classification}
Before doing so, 
we first proposed that  
composite solitons can be classified 
into the following two types:
\begin{enumerate}
    \item The SSB + ESB type, 

\item  The SSB + SSB
(hierarchical SSBs) type.
\end{enumerate}

The first type is focused in this paper. 
See also Ref.~\cite{Nitta:2022ahj} 
for detailed explanations.
In such a case, 
as explained in the main text, 
first, an SSB $G \to H$ 
occurs at high energy scale. 
Then, there appears a topological soliton
that often spontaneously 
breaks a symmetry in the vacuum 
$H$ to a subgroup $K$ 
in the vicinity of the soliton. 
Such a soliton (called a mother soliton) carries moduli $K/H$ 
corresponding to the locally happened SSB.
Subsequently, 
in the original theory, 
a small  ESB term is introduced 
at a low-energy scale 
that we can treat as a 
perturbation.
Then, it induces a potential term 
on the moduli of the mother soliton, 
triggering a further SSB on the mother soliton 
(or lifting up some part of the moduli).
This produces a daughter soliton inside the mother soliton.

The second type is not discussed in this paper 
but is also investigated in the literature, and so we mention it here. 
In this case, there are two scalar fields $\phi_1$ and $\phi_2$ belonging to 
irreducible representations 
of the symmetry group $G$. 
When these scalar fields develop VEVs  
in hierarchical energy scales,
there occur 
successive SSBs 
$G \to H \to L$. 
The first (second) SSB occurs 
by the VEV of the first 
(second) fields $\phi_1$ ($\phi_2$).
In this case, 
a daughter soliton 
$\pi_n (G/H)$
is 
created by the first SSB, 
and a mother soliton 
$\pi_m (H/L)$ with $m<n$
is created by the second SSB.
Note that the total SSB 
is $G \to K$, and thus
the daughter soliton is not  stable when 
$\pi_n (G/L) =0$ while 
the mother is not stable 
when $\pi_m (G/L) =0$. 

In the following, 
we summarize known examples 
for domain-wall vortex composites, 
monopole vortex composites, 
and monopole domain wall composites. 

%%%%%
\subsection{Domain-wall vortex composites}

Domain-wall vortex composites can be further 
classified into (1) confined or (2) deconfined vortices 
depending on the number of 
domain walls attached to the vortex.
A confined vortex is attached by a single domain wall. 
Thus, it is continuously pulled by the domain wall  
and so it is unstable.
A deconfined vortex is attached by two (or more) domain walls with the same tension, and is stable. 
However, it can be confined with its anti-soliton, 
for which they are connected by 
two (or more) domain walls.

In the confined case,
the wall is metastable and 
decays quantum mechanically 
by nucleation of holes 
bounded by a closed vortex string \cite{Preskill:1992ck}.

%%%%%%%
\subsubsection{The SSB + ESB type}

%{\bf (1) Confined vortices (domain wall bounded by vortices)}

This type of configurations was first studied in 
Refs.~\cite{Kibble:1982dd,Everett:1982nm}.

\begin{enumerate}

\item
{\it Axion strings.}
Axion strings with the domain wall number $N_{\rm DW}$ 
\cite{Vilenkin:1982ks,Kawasaki:2013ae}. 
The 
SSB is $U(1)_{\rm PQ} \to \{1\}$ and the ESB is 
$U(1)_{\rm PQ} \to 
{\mathbb Z}_{N_{\rm DW}}  
({\mathbb Z}_1 =\{1\})$. 
In the $N_{\rm DW}=1$ case (the confined case),
one axion string is attached by
one sine-Gordon soliton  
and is unstable.
In the general $N_{\rm DW}$ case (the deconfined case),
one axion string is attached by $N_{\rm DW}$ sine-Gordon domain walls of the same tension and is stable.
In this case, 
the axion string can be described by 
a kink inside a domain wall as shown in this paper.
In either case, 
a string and anti-string are connected 
by $N_{\rm DW}$ domain walls. 
There are many variations. 
One recent example is 
an axionic Alice string 
\cite{Sato:2018nqy,Chatterjee:2019rch}.
The
SSB is $U(1)_{\rm PQ} \times SU(2) \to 
({\mathbb Z}_4)_{{\rm PQ}+(2,3)} \ltimes U(1)_{1}$ and the
ESB is $U(1)_{\rm PQ} \to {\mathbb Z}_2$.
Although 
one axion string is attached by two domain walls ($N_{\rm DW}=2$),
it decays into two axionic Alice strings 
each of which is attached by one domain wall
(the Lazarides-Shafi mechanism \cite{Lazarides:1982tw}).

\item
{\it Two-Higgs doublet models (2HDMs)}. 
In this model,
there are non-Abelian fractional $Z$-strings 
\cite{Dvali:1993sg,Eto:2018hhg,Eto:2018tnk}. 
In addition to the standard 
electroweak symmetry breaking 
$U(1)_Y \times SU(2)_W 
\to U(1)_{\rm EM}$, 
the relative phase $U(1)_a$ between two Higgs doublets is also spontaneously broken: 
The SSB is $U(1)_a \times U(1)_Y \times SU(2)_W \to 
U(1)_{\rm EM}$
and the ESB is $U(1)_a \to \{1\}$ 
(or 
$\{{\mathbb Z}_2\}$), 
depending on the potential.
One non-Abelian fractional string is attached by one 
(or two) domain walls 
(depending on the interaction), because of 
an ESB term of relative phase $U(1)_a$.
A single $U(1)_a$ string is attached 
by two (or four) domain walls, 
$N_{\rm DW}=2 (4)$,
and decays into two non-Abelian fractional strings,
each of which is attached by 
one (or two) domain walls 
\cite{Eto:2018hhg,Eto:2018tnk}. 
If the number is two, the composite is stable.
On the other hand, a single $Z$-string can be decomposed into 
a pair of two non-Abelian fractional $Z$-strings 
connected by one (or two) domain wall(s), 
constituting a vortex molecule 
\cite{Eto:2021dca}.

\item
{\it Chiral symmetry breaking in QCD.} 
The SSB is that of chiral symmetry breaking 
$SU(3)_{\rm L} \times SU(3)_{\rm R} \times U(1)_A 
\to 
SU(3)_{\rm L+R}\times {\mathbb Z}_3$ 
and the ESB is 
$U(1)_{\rm A} \to 
{\mathbb Z}_3$ 
\cite{Eto:2009wu}.
One axial string winding around 
the axial $U(1)_{\rm A}$ symmetry 
is attached by three chiral domain walls ($N_{\rm DW}=3$)
because of 
an explicit $U(1)_{\rm A}$ symmetry breaking by anomaly
\cite{Balachandran:2001qn}.
It is unstable to decay into three non-Abelian global vortices 
each of which is attached by one chiral domain wall
\cite{Balachandran:2002je,
Eto:2013bxa,Eto:2013hoa}.

\item
{\it High density QCD.}
In high density limit of QCD, 
the ground state is the color-flavor locked (CFL) phase exhibiting color superconductivity \cite{Alford:2007xm}. 
It admits 
non-Abelian semi-superfluid vortices 
\cite{Balachandran:2005ev,Nakano:2007dr,Eto:2009kg,Eto:2009tr,Eto:2013hoa}. 
A single non-Abelian semi-superfluid vortex 
can be decomposed into a pair of 
chiral non-Abelian vortices 
connected by 
one (or two) non-Abelian chiral domain wall(s), 
depending on the interaction 
\cite{Eto:2021nle}.
A single chiral non-Abelian vortex 
is attached by one (or two) 
non-Abelian chiral domain wall(s)
 \cite{Eto:2021nle}.

\item
{\it Supersymmetric theories.}
A wall in a wall as a string in 
supersymmetric theories \cite{Ritz:2004mp,Eto:2005sw,Auzzi:2006ju,Bolognesi:2007bh}. 
If composite solitons are 
BPS, then a vortex is attached by two domain walls and they are stable.
In particular in the model in Ref.~\cite{Auzzi:2006ju}, when  
a single ANO string 
is absorbed into a domain wall, it splits into two half-quantized  strings. 
This situation is 
the same with  chiral $p$-wave superconductors 
(the item 9, below).

\item

{\it Domain-wall Skyrmions in field theory.} 
Domain-wall Skyrmions exist in the $U(1)$ gauge theory coupled with two complex scalar fields, or the ${\mathbb C}P^1$ model ($O(3)$ nonlinear sigma model)
with the easy-axis potential.
The SSB is $SU(2) \to U(1)$ 
and the ESB is $SU(2) \to U(1)$.
A domain wall connects vacua with a VEV of only one of $\phi_1$ and $\phi_2$.
A vortex in the bulk is absorbed into the wall 
and becomes a sine-Gordon soliton inside the wall \cite{Nitta:2012xq}.
In strong gauge coupling, the model reduces to 
the ${\mathbb C}P^1$ model.
A ${\mathbb C}P^1$ lump (baby Skyrmion) is 
absorbed into a ${\mathbb C}P^1$ wall 
and can be described as a sine-Gordon soliton in the domain wall 
\cite{Nitta:2012xq,Kobayashi:2013ju} 
(see also Refs.~\cite{Sutcliffe:1992he,
Stratopoulos:1992hq,Kudryavtsev:1997nw,Jennings:2013aea,Bychkov:2016cwc}). 
%This setup provides a physical proof for the lower dimensional analogue of the Atiyah-Manton construction
%\cite{Sutcliffe:1992he,Stratopoulos:1992hq}. 
%\item
This can be generalized to the ${\mathbb C}P^{N-1}$ model, for which
the SSB is $SU(N) \to SU(N-1)\times U(1)$ 
and the ESB is $SU(N) \to U(1)^{N-1}$.
The 
${\mathbb C}P^{N-1}$ 
model admits $N$ vacua and  $N-1$ parallel domain walls \cite{Gauntlett:2000ib,Tong:2002hi,Isozumi:2004jc,Isozumi:2004va,Eto:2004vy}. 
Then, 
${\mathbb C}P^{N-1}$ lumps 
absorbed into ${\mathbb C}P^{N-1}$ domain walls 
can be described by 
$U(1)^{N-1}$ coupled sine-Gordon solitons in 
the domain walls \cite{Fujimori:2016tmw}.
Non-Abelian vortices 
\cite{Hanany:2003hp,Auzzi:2003fs,Eto:2005yh,Eto:2006cx} absorbed into 
a non-Abelian domain wall \cite{Shifman:2003uh,Eto:2005cc,Eto:2008dm}  
%become non-Abelian Josephson vortices 
can be described as non-Abelian sine-Gordon solitons  \cite{Nitta:2014rxa,Eto:2015uqa} 
in the domain-wall effective theory which is a $U(N)$ chiral Lagrangian with a pion mass \cite{Nitta:2015mma,Nitta:2015mxa}.

\item
{\it Condensed matter example 1:
Domain-wall Skyrmions in chiral magnets} \cite{PhysRevB.99.184412,PhysRevB.102.094402,Nagase:2020imn,Yang:2021,Ross:2022vsa} 
(see also \cite{Kim:2017lsi}). 
This can be realized 
by adding the Dzyaloshinskii-Moriya interaction to the field theoretical 
${\mathbb C}P^1$  domain-wall Skyrmions 
mentioned above (the item 6).
Previously, a Bloch line in a Bloch wall in magnets was well known  \cite{Chen,Malozemoff}.

\item 
{\it Condensed matter example 2: 
Two-component miscible BECs} 
\cite{Son:2001td,Kasamatsu:2004tvg,Kasamatsu:2005,Cipriani:2013nya,
Tylutki:2016mgy,Eto:2017rfr,Kobayashi:2018ezm,Eto:2019uhe} 
{\it and 
two-gap superconductors} 
\cite{Babaev:2001hv,Goryo_2007,Crisan_2007,Tanaka_2007,PhysRevB.77.144518,TANAKA20176}.  
There are two complex scalar fields 
$\phi_1$ and $\phi_2$ 
describing 
condensates of two atoms 
of BECs or
gap functions 
of superconductors. 
In either case,
the SSB pattern is: 
$U(1)_1 \times U(1)_2 
\simeq [U(1)_g \times U(1)_r] /{\mathbb Z}_2
\to \{1\}$ 
and the ESB is   $U(1)_r \to \{1\}$, 
where the $U(1)_g$ is the overall phase rotation, which is gauged 
for superconductors but not 
for BECs. 
The ESB term is 
$\phi_1^*\phi_2 + {\rm c.c}$. 
called 
the Josephson interaction for superconductors,
or Rabi interaction for BECs.
Either of $\phi_1$ and $\phi_2$ can have a fractionally quantized 
vortex attached 
by a sine-Gordon soliton, 
like axion strings. 
A vortex winding around $\phi_1$ and 
the one around $\phi_2$  
are connected by a sine-Gordon soliton, 
constituting a vortex molecule. 
Generalizations to $N$-components were also studied for BECs \cite{Eto:2012rc,Eto:2013spa} 
and superconductors 
\cite{Nitta:2010yf}. 
The extension to more general charge assignments introduces various 
types of vortex molecules
\cite{Chatterjee:2019jez}

\item 
{\it Condensed matter example 3:
Chiral $p$-wave superconductors}. 
Similar to two-gap superconductors, 
there are two gap functions (complex scalar fields)
$\phi_1$ and $\phi_2$.
However, the ESB term in this case is 
$\phi_1^{*2}\phi_2^2 + {\rm c.c}$, 
in contrast to 
the conventional Josephson term 
$\phi_1^*\phi_2 + {\rm c.c}$ 
for two-gap superconductors. 
The ground states are either 
$(\phi_1,\phi_2)\sim (1,0)$ or $(0,1)$.
The SSB is 
$U(1)_1 \times U(1)_2 
\simeq [U(1)_g \times U(1)_r] /{\mathbb Z}_2
\to U(1)_1 =U(1)_{g-r}$ 
at $(\phi_1,\phi_2)\sim (1,0)$
or $U(1)_2 = U(1)_{g+r}$ at 
$(\phi_1,\phi_2)\sim (0,1)$,  
where 
$U(1)_g$ is the  overall gauge transformation 
and $U(1)_r$ is a relative phase rotation. 
The ESB is   $U(1)_r \to {\mathbb Z}_2$.
A chiral domain wall exists 
connecting the two ground states.
A half-quantum vortex can stably exist 
on the chiral domain wall 
\cite{Sigrist:1999,Etter_2020},
which can be identified with
a kink (sine-Gordon soliton with a half ($\pi$) period) inside the domain wall.
This situation is close to the above item 6 in which 
the kink is a sine-Gordon soliton with $2\pi$ period.
When a singly quantized vortex in the bulk is absorbed into the domain wall, 
it splits into two half-quantized vortices.
This situation is the same with supersymmetric theory \cite{Auzzi:2006ju}.
Apart from the domain wall, in one region, 
%a half-quantized vortex winding around either $\phi_1$ or $\phi_2$ is attached by two domain walls of the same tension \cite{PhysRevB.86.060514,Garaud:2015}.
a singly quantized vortex 
winding around $U(1)_g$ once
is split into 
two half-quantized vortices connected by two domain walls constituting a vortex molecule 
\cite{PhysRevB.86.060514,Garaud:2015}.
The other ground state appears inside the region between the domain walls.

\item
{\it Condensed matter example 3: 
$^3P_2$ neutron superfluids.} 
A singly quantized vortex 
is split into two half-quantized non-Abelian vortices 
connected by one or two solitons 
\cite{Masaki:2021hmk,Kobayashi:2022moc,Kobayashi:2022dae} 
as the case of chiral $p$-wave superconductors.

\item
{\it Condensed matter example 4: 
$^3$He superfluids.}
A spin-mass vortex is 
confined by a soliton in the B-phase, 
and 
a half-quantum vortex is confined by a soliton 
in polar and polar-distorted A phases 
\cite{
Volovik:2003fe,
PhysRevLett.117.255301,
Volovik:2019goo,
Volovik:2020zqc,
Makinen:2022}.
A Mermin-Ho vortex within a
domain wall exists 
\cite{
Volovik:2003fe}.

\end{enumerate}

%%%
\subsubsection{The SSB + SSB type}

\begin{enumerate}

\item
{\it Wall-vortex composites in 
GUTs} \cite{Kibble:1982dd,Everett:1982nm}.
%\textcolor{red}{(More?)}
The relevant sequential SSBs are 
$SO(10)\to SU(4)_c\times SU(2)_R\times SU(2)_L \times \mathbb{Z}_2^\mathrm{C}\to SU(3)_c \times SU(2)_L \times U(1)_Y$,
in which $\mathbb{Z}_2^\mathrm{C}$ corresponds to the charge conjugation.
The symmetry $SU(4)_c\times SU(2)_R\times SU(2)_L$ is the same as that of the Pati-Salam model~\cite{Pati:1974yy}.
At the first symmetry breaking, the $\mathbb{Z}_2$ string appears due to $\pi_1(SO(10)/ SU(4)_c\times SU(2)_R\times SU(2)_L \times \mathbb{Z}_2^\mathrm{C})\simeq \pi_0(\mathbb{Z}_2) = \mathbb{Z}_2$.
The second breaking gives rise to the domain wall separating two vacua which are interchanged by the charge conjugation.
Since $\mathbb{Z}_2^\mathrm{C}$ is embedded in the continuous $SO(10)$ symmetry,
the domain wall is not topologically stable but is bounded by the $\mathbb{Z}_2$ strings.
The domain walls collapse before they dominate the energy of the universe, so that there is no domain wall problem.

\item

{\it The Georgi-Machacek model.} 
The sequential 
SSBs are 
$U(1)_Y \times 
SU(2)_{\rm W} 
\to 
{\mathbb Z}_2
\times U(1)_{\rm EM} 
\to U(1)_{\rm EM}$
\cite{Chatterjee:2018znk}.
The first SSB 
due to 
VEVs of three $SU(2)$  triplet Higgs fields   
produces a 
${\mathbb Z}_2$ string, 
and the second SSB 
due to a VEV of 
an $SU(2)$ doublet Higgs field 
gives a domain wall 
attached to the string.

\item 
{\it Alice strings  attached by 
domain walls.} 
The sequential 
SSBs are
$U(1) \times SU(2) \to {\mathbb Z}_4 \ltimes U(1) \to 
\{1\}$ \cite{Chatterjee:2019zwx,Nitta:2020ggi}.
The first SSB 
due to 
the complex triplet  representation 
${\bf 3}$ 
of 
$SU(2)$ 
yields a (BPS) Alice strings 
\cite{Chatterjee:2017jsi,Chatterjee:2017hya},
and 
the second SSB 
occurring due to 
the fundamental 
representation 
of $SU(2)$ 
gives a domain wall
attached to each 
Alice string 
\cite{Chatterjee:2019zwx,Nitta:2020ggi}.
However, the existence of solitons at the second SSB does not rely on the usual
homotopy arguments, 
but on a nontrivial Aharonov-Bohm phase.
Such a soliton is called a 
Aharonov-Bohm defect.

\item
{\it Two-flavor dense QCD.}
The sequential 
SSBs are
$U(1)_{\rm B} \times SU(3)_{\rm C} \to {\mathbb Z}_6 \ltimes SO(3) \to 
{\mathbb Z}_3$. 
The first SSB 
due to 
the representation 
${\bf 6}$ 
(a $3 \times 3$ anti-symmetric tensor) of 
$SU(3)$ 
yields a non-Abelian Alice strings 
\cite{Fujimoto:2020dsa,Fujimoto:2021bes},
and 
the second SSB 
occurring due to 
the anti-fundamental representation 
$\bar {\bf 3}$ of 
$SU(3)$ 
gives solitons  
(Aharonov-Bohm defects) 
attached to the Alice strings 
\cite{Fujimoto:2021wsr}.
Thus, non-Abelian Alice string 
are  confined by 
solitons 
to constitute 
``mesonic or baryonic molecules".

\end{enumerate}

\subsubsection{The modified XY model}

The third type which is rather nontrivial is 
the modified XY model 
\cite{
Carpenter_1989}
or its Goldstone model extension
\cite{Kobayashi:2019sus,Kobayashi:2019npl}. 
In a certain parameter region, 
two hierarchical SSBs occur
$U(1) \to {\mathbb Z}_2 \to \{1\}$. 
In numerical simulations, a two-step phase transition was found. 
In the first step, vortices are created (BKT-type transition) and in the second step domain walls connecting them (Ising-type transition) are created 
\cite{Kobayashi:2019sus}.

\if0
{\bf (2) Deconfined vortices 
(Domain-wall vortices/
kinks on a domain wall)}
\fi

%%%%%
\subsection{Vortex monopole composites}

\if0
Vortex monopoles can be further 
classified into (1) confined or (2) deconfined monopoles.
A confined monopole is attached by a single vortex string. 
Thus, it is continuously pulled by the string 
and it is unstable.
On the other hand, 
a deconfined monopole is attached by two (or more) vortex strings with the same tension, 
and is stable.

Each case can be further 
classified into 
the SSB + ESB type 
and SSB$_1$ + SSB$_2$ 
(hierarchical SSBs) type.
The first type is the same with the configurations focused in this paper.

{\bf (1) Confined monopoles
(a vortex terminating on a monopole)}
\fi

\subsubsection{
The SSB + ESB type}

When a monopole is attached by two (or more) 
vortex strings of the same tension, 
the configuration is stable.
In this case, it is often the case 
that a monopole can be regarded as 
a kink on a vortex string.

\begin{enumerate}
    \item 
    {\it $1/4$ BPS vortex-monopoles in supersymmetric gauge theory.}
A $U(N)$ gauge theory with $N$ fundamental Higgs scalar fields 
with a common $U(1)$ charge 
admits 
a 't Hooft-Polyakov type monopole attached by two 
non-Abelian vortex strings of the same tension 
\cite{Tong:2003pz}.
The SSB is 
$U(N)_{\rm C} \times 
SU(N)_{\rm F} 
\to SU(N)_{\rm C+F}$ 
and the ESB is 
$SU(N)_{\rm F} \to U(1)^{N-1}$. 
The SSB without ESB 
yields 
a non-Abelian vortex carrying 
${\mathbb C}P^{N-1}$ 
moduli 
\cite{Hanany:2003hp,Auzzi:2003fs,Eto:2005yh,Eto:2006cx}, 
and the ESB induces 
a potential on the moduli. 
The monopole can be realized 
as a ${\mathbb C}P^{N-1}$ kink inside the non-Abelian vortex
world-sheet. 
In a supersymmetric extension, these are 
1/4 BPS states 
and are 
nonperturbatively stable.\footnote{
These can be obtained from vortex instantons 
(instanton immersed into a vortex)
\cite{Hanany:2004ea,Eto:2004rz,Fujimori:2008ee} 
by the Scherk-Schwarz dimensional reduction 
\cite{Scherk:1979zr}
or an $S^1$ compactification with a twisted boundary condition 
\cite{Eto:2004rz}.
}
Such a vortex monopole configuration 
%(as well as vortex instanton configuration given below)
yields 
the correspondence between 
quantum effects in the two dimensional ${\mathbb C}P^{N-1}$ model and four dimensional gauge theory \cite{Shifman:2004dr,Hanany:2004ea}.
Non-Abelian monopoles on multiple non-Abelian vortices 
can be also realized \cite{Nitta:2010nd}.
%giving rise to a physical realization  of the Donaldson's rational map \cite{Donaldson:1984ugy}. 
The cases of 
$SO, USp$ gauge groups were studied in Ref.~\cite{Eto:2011cv}.

\item
%[(B)SM]
{\it Topological Nambu monopole in 2HDMs} \cite{Eto:2019hhf,Eto:2020hjb,Eto:2020opf}.

The SSB is $SU(2)_{\rm W} 
\times SU(2)_{\rm cust.} 
\times U(1)_{a}
\to U(1)_{\rm em} \times SU(2)_{\rm cust.}$ 
when $\theta_W=0$.
Here $SU(2)_{\rm cust.}$ is called the Higgs custodial symmetry.
Turning on $\theta_W\neq 0$ or the ESB interaction in the Higgs potential,
the original symmetry is explicitly broken into
$SU(2)_{\rm W} 
\times  U(1)_Y\times \mathbb{Z}_2^\mathrm{CP}
\times U(1)_{a}$,
which is spontaneously broken into
$U(1)_{\rm em}\times \mathbb{Z}_2^\mathrm{CP}$.
The SSB without ESB yields a non-Abelian vortex carrying
${\mathbb C}P^{1}$ moduli, 
and the ESB induces a potential on the moduli. 
In contrast to a Nambu monopole in Standard Model(SM) 
which is attached by one $Z$-string as
 mentioned in Sec.~\ref{sec:Nambu}, there is a topological Nambu monopole in 2HDMs 
attached by the two non-Abelian topological $Z$-strings 
\cite{Eto:2019hhf,Eto:2020hjb,Eto:2020opf}. 
The $\mathbb{Z}_2^\mathrm{CP}$ symmetry ensures the degeneracy of the tensions of the two $Z$-strings.
If $\mathbb{Z}_2^\mathrm{CP}$ is also explicitly broken, 
the monopole is confined and is unstable~\cite{Eto:2020hjb}.
Otherwise the monopole is stable~\cite{Eto:2019hhf}. 
Although the two $Z$-strings correspond to two minima of the lifted potential of the ${\mathbb C}P^1$ moduli,
this monopole connecting the strings cannot be regarded as a kink on a string 
in the sense that magnetic fluxes 
spread out almost spherically but not confined. 
In spite of the trivial second homotopy group of the vacuum, the monopole has a non-trivial $U(1)_\mathrm{em}$ bundle~\cite{Eto:2020opf}.

\end{enumerate}

A texture version is 
Skyrmions on a vortex string:
vortex Skyrmions 
realized as sine-Gordon kinks 
inside a vortex string 
in a Skyrme model with a BEC-inspired potential
\cite{Gudnason:2014hsa,Gudnason:2014jga} 
or 
a simpler potential 
\cite{Gudnason:2016yix,
Nitta:2015tua}.

\subsubsection{
The SSB + SSB type
}
%\begin{itemize}

\begin{enumerate}

\item  
{\it Confinement phase of QCD}. 
Quarks are confined by color electric flux tubes in QCD, 
giving rise to a confinement.
As a dual superconductor description, 
monopoles are 
confined by 
color magnetic flux tubes
or vortices \cite{Nambu:1974zg,Mandelstam:1974vf,Mandelstam:1974pi},
where a monopole and anti-monopole 
is connected by a color magnetic flux tube.  
 
The generalization of this vortex-monopole composite  
to supersymmetric QCD was proposed to study the confinement problem 
\cite{Auzzi:2003em,Auzzi:2004yg,Eto:2006dx,Cipriani:2011xp,Chatterjee:2014rqa} 
in which hierarchical SSBs was studied: 
$SU(N+1) \times SU(N)_{\rm F} 
\to 
U(N)_{\rm C} \times 
SU(N)_{\rm F} 
\to SU(N)_{\rm C+F}$.
In the first SSB a 
't Hooft-Polyakov monopole is created, and in the second SSB 
it is attached by a non-Abelian vortex string.

\item
{\it High density QCD.}
In high density QCD,
a monopole and anti-monopole are confined 
by non-Abelian vortex strings \cite{Gorsky:2011hd,Eto:2011mk}, 
explaining 
a quark-hadron continuity or duality 
\cite{Eto:2011mk}.

\item
%[GUT]
{\it Grand unified theory(GUT)}.
There is a cosmological 
monopole problem 
in GUTs: 
monopoles are created 
at GUT phase transition from a unified gauge group 
$G$ to that of the standard model but no monopoles are observed \cite{Dokos:1979vu,Lazarides:1980va,Preskill:1979zi}.
To avoid this problem, it was proposed that 
in a certain period of early universe, 
the electric magnetic $U(1)$ symmetry was 
spontaneously broken. 
$G =SU(5), SO(10) \to SU(3) \times 
SU(2) \times U(1) \to SU(3) \times 
SU(2) $.
Thus 
all (anti-)monopoles are confined 
to enhance pair annihilations of monopole and anti-monopole 
(the Langacker-Pi mechanism~\cite{Langacker:1980kd}).

\item
{\it Beads on a string.}
Monopoles on a string exist in an $SU(2)$ gauge theory 
coupled to two adjoint real Higgs scalar fields 
\cite{Hindmarsh:1985xc,Ng:2008mp,
Kibble:2015twa,Hindmarsh:2016dha}.
In this case, hierarchical symmetry breakings are 
considered as $SU(2) \to U(1) \to {\mathbb Z}_2$: 
in the first breaking a monopole is created, 
and in the second breaking ${\mathbb Z}_2$ strings 
attached to the monopole are created.
This composite is sometimes called a necklace.
This can be regarded as a kink on a string. 
The gauge kinetic mixing with the SM $U(1)_\mathrm{em}$ field was discussed in Refs.~\cite{Hiramatsu:2021kvu,Chitose:2023vae}. 
See Refs.~\cite{Kleihaus:2003nj,Kleihaus:2003xz} for similar objects.

\item
{\it Alice-string monopole composites.} 
Alice-string monopole composites
are considered 
\cite{Nitta:2020ggi}. 
One monopole is confined but two monopoles are not.

\item
{\it Condensed matter example 1: 
Chiral liquid crystals
 and 
chiral magnets} 
\cite{Ackerman:2017b,
RevModPhys.84.497,Smalyukh:2020zin,Smalyukh:2022}.
A toron is a pair of monopole and anti-monopole 
connected by lump (baby Skyrmion) string.

\item
{\it Condensed matter example 2:
$^3$He superfluids} \cite{Volovik:2003fe}. 
This is called 
a nexus soliton.

\end{enumerate}

\subsubsection{Exception}\label{sec:Nambu}

{\it Nambu monopole in the Standard Model (SM)}~
\cite{Nambu:1977ag,Achucarro:1999it,Eto:2012kb}.
The fact that the vacuum manifold of the SM is $S^3$ leads to no stable monopole nor strings ($\pi_1(S^3)=\pi_2(S^3)=\{1\}$).
However, a monopole attached by a single $Z$-string, called the Nambu monopole, exists and has a non-zero magnetic $U(1)_{\rm em}$ flux.
In contrast to the Nambu monopole in 2HDMs,
the monopole in the SM cannot be stable but is confined.
In addition, the monopole cannot have any non-trivial $U(1)_{\rm em}$ bundle around it because $U(1)_{\rm em}$ cannot be defined at the core of the $Z$-string~\cite{Eto:2020opf}.

%%%%%%
\subsection{Domain-wall monopoles}

\subsubsection{The SSB + ESB type}
Compared with vortex monopoles, 
there are not many studies for domain-wall monopoles~\footnote{
There is an example of unstable composite of a domain wall and monopole in a GUT. 
The SSB is
$SU(5) \times {\mathbb Z}_2 \to 
[SU(3)_{\rm C}\times SU(2)_{\rm L}\times
U(1)_{\rm Y}]/
{\mathbb Z}_6$ 
with a global symmetry 
${\mathbb Z}_2$.
A domain wall and monopoles coexist. 
However, after the monopoles 
are absorbed into the domain wall, they are unstable to decay 
\cite{Dvali:1997sa,Pogosian:1999zi,Brush:2015vda}. 
This was proposed as 
a monopole eraser 
to solve the cosmological monopole problem.}
except for the following examples.

\begin{enumerate}
    \item 
{\it Local monopoles in a non-Abelian domain wall.}
$SU(N)$ 't Hooft-Polyakov monopoles are 
$U(1)^{N-1}$ global vortices 
\cite{Nitta:2015mxa} inside a non-Abelian domain wall  
\cite{Shifman:2003uh,Eto:2005cc,Eto:2008dm}.\footnote{
These can be obtained from domain-wall instantons 
\cite{Eto:2005cc,Nitta:2013vaa,Nitta:2015mxa}
by the Scherk-Schwarz dimensional reduction 
\cite{Scherk:1979zr}
or an $S^1$ compactification with a twisted boundary condition.
}
A global analogue of this composite 
is studied in the main text of this paper. 
A texture version of this 
composite configuration is a domain-wall Skyrmions 
where Skyrmions are realized as  lumps  inside the domain wall \cite{Nitta:2012wi,
Gudnason:2014nba,
Eto:2015uqa,
Eto:2023lyo} 
(see also Ref.\cite{Kudryavtsev:1999zm}).

\if0
\item
\textcolor{red}{
{\it GUT.}
In a GUT, 
the SSB is
$SU(5) \times {\mathbb Z}_2 \to 
SU(3)_{\rm C}\times SU(2)_{\rm L}\times
U(1)_{\rm Y}]/
{\mathbb Z}_6$ 
and the ESB is ${\mathbb Z}_2 \to \{1\}$ \textcolor{red}{(?)}
A domain wall and monopoles coexist. 
However, after the monopoles 
are absorbed into the domain wall, they are unstable to decay 
\cite{Dvali:1997sa,Pogosian:1999zi,Brush:2015vda}. 
This was proposed as 
a monopole eraser.}
\yh{It seems that they do not consider ESB of $\mathbb{Z}_2$...}
\fi

\end{enumerate}

\subsubsection{The SSB + SSB type}
As far as we know, 
there are no examples of this type.

\bigskip

\section{$N=2$ axion string-wall composite: An effective Lagrangian approach}
\label{sec:appendix}

In this appendix, we show the validity of the ansatz in Eq.~(\ref{eq:kink-on-string}) 
for a vortex-monopole composite, 
i.~e. a kink inside a vortex, in the $O(2)$ model ($N=2$) 
from the view point of the moduli approximation  
\cite{Manton:1981mp,Eto:2006uw}.
We consider the two component scalar fields ${\bm \phi} = (\phi_1,\phi_2)$ 
with 
the Lagrangian (\ref{eq:O3_lag}) with the additional potential (\ref{eq:VN}) for  $N=2$:
\be
{\cal L}_2 = \frac{1}{2}(\p_\mu{\bm \phi})^2 - \frac{\lambda}{4}\left({\bm \phi}^2 - v^2\right)^2 +  \tilde \alpha (\phi_2)^2.
\ee
We assume $\tilde \alpha > 0$ hereafter (in the main text we have used $\alpha = -\tilde \alpha$), and the weak coupling regime is 
\be
\frac{\lambda v^4}{4} \gg \tilde \alpha v^2.
\ee
The vacua can be obtained by solving
\be
\frac{\delta V}{\delta \phi_1} = \lambda \phi_1({\bm \phi}^2 - v^2) = 0, \quad
\frac{\delta V}{\delta \phi_2} = \lambda \phi_2({\bm \phi}^2 - v^2) - 2\tilde\alpha \phi_2 = 0.
\ee
There are two solutions
\be
\phi_1 = 0, \qquad
%\quad
%\lambda (\phi_2^2 - v^2) - 2\tilde\alpha  = 0
%\quad \rightarrow \quad
\phi_2 = \pm v \sqrt{1 + \frac{2\tilde\alpha}{\lambda v^2}}.
% \simeq \pm \left(v + \frac{\tilde\alpha}{\lambda v^2}\right)
\ee
Let us perturb ${\bm \phi}$ around one of the vacua as
\be
\phi_1 = v \delta_1,\quad \phi_2 = v \sqrt{1 + \frac{2\tilde\alpha}{\lambda v^2}} + v\delta_2.
\ee
Plugging these into the potential and keeping only quadratic terms of $\delta_{1,2}$, we have
\be
V &=& \frac{\lambda}{4}\left[ v^2 \delta_1^2 + v^2 \left( \sqrt{1 + \frac{2\tilde\alpha}{\lambda v^2}}+\delta_2\right)^2 - v^2\right]^2
- \tilde\alpha v^2 \left( \sqrt{1 + \frac{2\tilde\alpha}{\lambda v^2}}+\delta_2\right)^2 \nonumber\\
&=& \frac{\lambda v^4}{4}\left[\delta_1^2 + 1 + \frac{2\tilde\alpha}{\lambda v^2} + 2 \sqrt{1+\frac{2\tilde\alpha}{\lambda v^2}} \delta_2 + \delta_2^2 -1\right]^2
- \tilde \alpha v^2\left(1 + \frac{2\tilde\alpha}{\lambda v^2} + 2 \sqrt{1+\frac{2\tilde\alpha}{\lambda v^2}} \delta_2 + \delta_2^2\right)
\nonumber\\
&=& \frac{\lambda v^4}{4}\left[\frac{2\tilde\alpha}{\lambda v^2} + 2 \sqrt{1+\frac{2\tilde\alpha}{\lambda v^2}} \delta_2 + \delta_1^2 + \delta_2^2 \right]^2
- \tilde \alpha v^2\left(1 + \frac{2\tilde\alpha}{\lambda v^2} + 2 \sqrt{1+\frac{2\tilde\alpha}{\lambda v^2}} \delta_2 + \delta_2^2\right)
\nonumber\\
&\simeq&
 \frac{\lambda v^4}{4}\left[ \left( \frac{2\tilde\alpha}{\lambda v^2}\right)^2 
 + 2 \frac{2\tilde\alpha}{\lambda v^2} 2 \sqrt{1+\frac{2\tilde\alpha}{\lambda v^2}} \delta_2
 + 2 \frac{2\tilde\alpha}{\lambda v^2}(\delta_1^2 + \delta_2^2)
 + 4 \left(1+\frac{2\tilde\alpha}{\lambda v^2}\right)\delta_2^2
  \right] \nonumber\\
  && -~ \tilde \alpha v^2\left(1 + \frac{2\tilde\alpha}{\lambda v^2} + 2 \sqrt{1+\frac{2\tilde\alpha}{\lambda v^2}} \delta_2 + \delta_2^2\right) \nonumber\\
 &\simeq& 
 \frac{\lambda v^4}{4}\left[ 
 2 \frac{2\tilde\alpha}{\lambda v^2}(\delta_1^2 + \delta_2^2)
 + 4 \left(1+\frac{2\tilde\alpha}{\lambda v^2}\right)\delta_2^2
 \right] - \tilde \alpha v^2 \delta_2^2 \nonumber\\
 &\simeq& \tilde \alpha v^2 \delta_1^2 + \lambda v^4 \left( 1 + \frac{2\tilde\alpha}{\lambda v^2}\right) \delta_2^2.
\ee
Hence, there are two typical masses
\be
m_1 = \sqrt{2 \tilde \alpha},\qquad
m_2 = \sqrt{2\lambda \left( 1 + \frac{2\tilde\alpha}{\lambda v^2}\right)}v \simeq \sqrt{2\lambda}v, 
\ee
where $m_1$ is the mass of the pseudo-NG mode while $m_2$ is that of the Higgs mode. The weak coupling regime requires
\be
m_1 \ll m_2.
\ee

The $\mathbb{Z}_2^{(2)}$ is spontaneously broken in the above vacua, and it gives rises to a mother domain wall.
In order to estimate energy scales, let us consider 
the $\mathbb{Z}_2^{(1)}$ charged mother domain wall by fixing the amplitude $|{\bm \phi}|= v$ as
\be
{\bm \phi} = v \left(\pm \cos\Theta,\sin\Theta\right),
\ee
the $\pm$ signature is responsible for the $\mathbb{Z}_2^{(1)}$ charge.
Then, the Lagrangian reduces to the sine-Gordon model with a half period as
\be
{\cal L}_2 = \frac{v^2}{2}\p_\mu\Theta\p^\mu\Theta - v^2 \tilde\alpha \sin^2\Theta.
\ee
The mother domain wall solution can be approximated by 
a sine-Gordon soliton 
with a half period
\be
\Theta_0(x) = 2 \arctan e^{m_1 x} + \frac{\pi}{2},
\ee
with the wall width  $m_1^{-1}$.
Then, the tension (energy per unit area) of the mother domain wall can be approximated by
\be
\sigma = 2 v^2 m_1.
\ee
The mother domain wall is produced at the mass scale $v$, but its tension $v^2 m_1$ is much smaller than the typical scale $v^3$.

As soon as the mother domain wall is generated, it breaks the $\mathbb{Z}_2^{(1)}$ symmetry, and 
a daughter domain wall can be produced in general.
To estimate the energy scales,
let us next take the product ansatz
\be
{\bm \phi} = v \left(\varphi(y) \cos\Theta_0(x),\sin\Theta_0(x)\right),
\label{eq:product_ansatz}
\ee
where $y$ is the 
world-volume coordinates 
of the mother domain wall. %except for $x$.
Plugging this into the Lagrangian and integrate it over $x$, we get 
the effective Lagrangian 
\be
L_2 = \int dx~{\cal L}_2 = \frac{v^2}{m_1} \p_{\mu'}\varphi\p^{\mu'}\varphi
- \frac{\lambda v^4}{3m_1}\left(\varphi^2 - 1\right)^2 - \frac{m_1 v^2}{3}\varphi^2,
\label{eq:effective}
\ee
where $\mu' = 0,2,3$. There is a  $\mathbb{Z}_2^{(1)}$ symmetry and it is spontaneously broken in the vacua $\varphi = \pm 1$ of the effective Lagrangian 
(\ref{eq:effective}).
The inner kink is obtained as 
\be
\varphi %= \frac{\sqrt{2\lambda  v^2-2\beta }}{\sqrt{2\lambda } v} \tanh \left(\frac{\sqrt{2\lambda  v^2-2\beta }}{\sqrt{6}} y \right)
\simeq \frac{\sqrt{m_2^2-m_1^2 }}{m_2} \tanh \left(\frac{\sqrt{m_2^2-m_1^2}}{\sqrt{6}} y \right),
\ee
and its tension (energy per unit length) is given by
\be
\mu \simeq \frac{4 \sqrt{2} \left(m_2^2-m_1^2 \right)^\frac{3}{2}}{3 \sqrt{3} m_1 \lambda}
\simeq \frac{4\sqrt2}{3\sqrt3} \frac{m_2^3}{\lambda m_1}
= \frac{8\sqrt2}{3\sqrt3} \frac{m_2}{m_1}v^2.
\ee
Thus, the inner kink, 
generated at the same instant as the mother domain wall produced at the energy scale $v$, has
the tension $\mu \simeq (m_2/m_1) v^2$ and the width is $\sqrt{6} m_2^{-1}$.  
This inner kink is nothing but a vortex from the bulk point of view, as discussed in the main text.
We thus have obtained the ansatz in Eq.~(\ref{eq:kink-on-string}).

\bibliographystyle{jhep}
%\begin{thebibliography}{99}
%\end{thebibliography}

\bibliography{references}
\end{document}